%% file: main.tex
\documentclass[prb,twocolumn,superscriptaddress,float,aps]{revtex4-2}
\usepackage{bm,color,amsmath,amssymb,mathrsfs,latexsym,graphicx,psfrag,tabularx}

\usepackage[hidelinks,colorlinks,linkcolor=blue,
citecolor=blue,urlcolor=blue]{hyperref}

\newcommand{\beq}{\begin{equation}}
\newcommand{\eneq}{\end{equation}}

\def\qq{\mathbf{q}}
\def\kk{\mathbf{k}}
\def\KK{\mathbf{K}}
\def\DKK{\Delta\mathbf{K}}
\def\pp{\mathbf{p}}
\def\RR{\mathbf{R}}
\def\tt{\mathbf{t}}
\def\rr{\mathbf{r}}
\def\GG{\mathbf{G}}
\def\QQ{\mathbf{Q}}
\def\aa{\mathbf{a}}
\def\bb{\mathbf{b}}
\def\uu{\mathbf{u}}

\def\KK{\mathbf{K}}
\def\qq{\mathbf{q}}
\def\pp{\mathbf{p}}
\def\pp{\mathbf{p}}
\def\GG{\mathbf{G}}
\def\QQ{\mathbf{Q}}
\def\RR{\mathbf{R}}
\def\tt{\mathbf{t}}
\def\dd{\mathbf{d}}
\def\aa{\mathbf{a}}
\def\bb{\mathbf{b}}
\def\mm{\mathbf{m}}
\def\ee{\epsilon}
\def\CC{\mathcal{P}}

\def\BZ{{\rm BZ}}
\def\mS{{\mathcal{S}}}

\def\BZ{{\rm BZ}}

\def\spin{{\varsigma}}

\def\spin{{\varsigma}}

\def\hH{{ \hat{H} }}
\def\hrho{ \hat{\rho} }
\def\hg{\hat{g}}
\def\hS{\hat{S}}

\def\mG{{\mathcal{G}}}

\def\UC{{\hat{\Theta}}}
\def\UF{{\hat{\Sigma}}}

\def\mK{{\mathcal{K}}}
\def\pr{\prime}
\def\mJ{{\mathcal{J}}}
\def\mK{{\mathcal{K}}}

\def\ie{{\it i.e.},\ }
\def\eg{{\it e.g.}\ }
\def\ea{{\it et al.}}

\usepackage{etoolbox}
\makeatletter
\patchcmd{\@maketitle}{\@author}{\@author\show\@thanks}{}{}
\makeatother

\usepackage{dsfont}

\begin{document}

\title{Surface Phonon Hall Viscosity Induced Phonon Chirality and Nonreciprocity in Magnetic Topological Insulator Films}

\author{Abhinava Chatterjee$^{1}$, and Chao-Xing Liu$^{*}$}
\affiliation{%
   Department of Physics, The Pennsylvania State University, University Park, Pennsylvania 16802, USA
}

\begin{abstract} 
The surface half-quantum Hall effect, a hallmark consequence of axion electrodynamics, can be induced by gapping out the surface states of topological insulators through surface magnetization, and has led to a variety of topological response phenomena observed in experiments. In this work, we investigate phonon dynamics originating from an acoustic analog - the surface phonon Hall viscosity - that can also occur at the surface of magnetic topological insulators. This surface phonon Hall viscosity stems from the Nieh-Yan action in the strain response of topological insulators, where strain acts as the effective vierbein field for the bulk low-energy massive Dirac fermions. Crucially, this viscosity term entangles phonon dynamics with surface magnetization.  In magnetic topological insulator films, we find that this interaction causes acoustic phonons to become chiral when the magnetization at the top and bottom surfaces is parallel, and nonreciprocal when it is anti-parallel. We further discuss potential experimental signatures of phonon dynamics induced by surface phonon Hall viscosity, specifically the phonon thermal Hall effect and magnon-polarons.
\end{abstract}

\date{\today}


\maketitle

\section{Introduction}

Topological insulator (TI) \cite{hasan2010colloquium,qi2011topological,moore2010birth}  represents a unique class of quantum materials characterized by an insulating bulk band gap and gapless metallic surface states protected by nontrivial bulk topological invariants. In three-dimensional (3D) TIs, this nontrivial topology manifests in an unconventional electromagnetic response described by a topological $\theta$ term, $\theta \frac{e^2}{2\pi h} {\bf E\cdot B}$  \cite{qi2008topological,essin2009magnetoelectric,sekine2021axion,nenno2020axion}, in addition to the ordinary Maxwell's terms, where ${\bf E}$ and ${\bf B}$ are the electromagnetic fields and $\theta$ is a pseudo-scalar quantized to $\pi$ due to time-reversal or inversion inside TIs. This term is known as axion electrodynamics due to its similar mathematical structure as the axion field originally proposed in particle physics \cite{peccei1977cp,peccei1977constraints,weinberg1978new,wilczek1978problem}. As a consequence, integrating magnetism into TIs to gap out surface states could generate a variety of topological quantum states, including quantum anomalous Hall insulators \cite{qi2008topological,yu2010quantized,liu2016quantum,chang2023colloquium}, axion insulators \cite{wilczek1987two,qi2008topological,mong2010antiferromagnetic,morimoto2015,wang2015quantized} and magnetic Weyl semimetals \cite{zyuzin2012topological,son2012berry,grushin2012consequences,wang2013chiral,goswami2013axionic}. These states, in turn, host a wealth of physical phenomena including the surface half-quantum Hall effect \cite{qi2008topological,liu2009magnetic,nomura2011surface,abanin2011ordering} (also referred to as the parity anomaly \cite{redlich1984gauge,redlich1984parity,ryu2012electromagnetic}), topological Faraday and Kerr rotation \cite{qi2008topological,maciejko2010topological,tse2010giant,wu2016quantized,okada2016terahertz,dziom2017observation}, the quantized magnetoelectric effect \cite{qi2008topological,essin2009magnetoelectric}, axion optical induction \cite{ahn2022theory,qiu2023axion} and dynamical axion quasiparticles \cite{li2010dynamical,sekine2016chiral,ooguri2012instability,taguchi2018electromagnetic,qiu2025observation}. The quantum anomalous Hall effect has been observed in magnetically doped TIs \cite{chang2013experimental,checkelsky2014trajectory}, MnBi$_2$Te$_4$
films \cite{deng2020quantum,liu2020robust}, and moir\'e materials \cite{serlin2020intrinsic,sharpe2019emergent,chen2020tunable} when there is ferromagnetism, while the axion insulator has been shown in a variety of antiferromagnetic TIs, including magnetically doped TIs \cite{mogi2017magnetic,mogi2017tailoring,xiao2018realization}, MnBi$_2$Te$_4$ films \cite{otrokov2019prediction,zhang2019topological,li2019intrinsic,gong2019experimental,liu2020robust}, and other candidates like EuIn$_2$As$_2$ \cite{sato2020signature} and EuSn$_2$As$_2$ \cite{li2019dirac}. Axion electrodynamics has been convincingly demonstrated in magnetic TIs.

\begin{figure*}
\includegraphics[width=\textwidth]{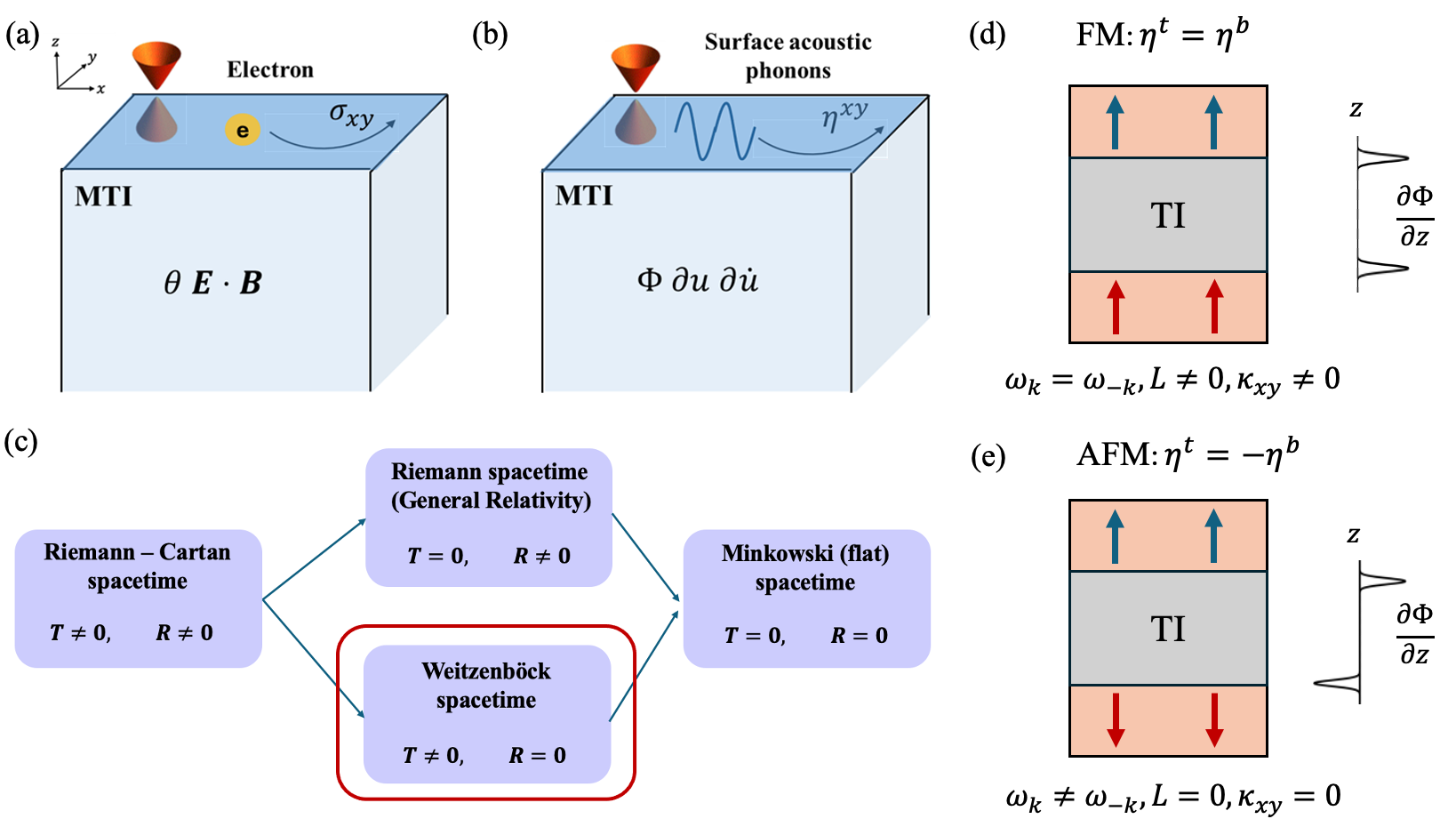}
\caption{(a) Axion electrodynamics in 3D magnetic TI can result in surface Hall conductivity $\sigma_{xy}= e^2/2h$ when surface Dirac cone is gapped. (b) The Nieh-Yan action for the strain response of magnetic TIs can lead to surface PHV $\eta_{xy}$.  (c) The connection between different classes of gravity theories. Non-zero torsion and Riemann curvature ($T, R\neq 0$) ensues Riemann-Cartan spacetime. In ordinary general relativity, the torsion is zero ($T=0$), whereas the Riemann curvature is zero in Weitzenböck spacetime (\(R=0\)). When both $T, R = 0$, it is flat Minkowski spacetime. (d) Magnetic TI sandwiches with surface FM configuration have equal values of surface PHV, $\eta_t = \eta_b$, at the top and bottom surfaces. Acoustic phonon modes are chiral ($L\neq 0$), but reciprocal ($\omega_\kk = \omega_{-\kk}$).  (e) The surface AFM configuration can lead to opposite surface PHV, $\eta_t = -\eta_b$, at two surfaces, and the acoustic phonons become non-reciprocal ($\omega_\kk \neq \omega_{-\kk}$) but also non-chiral ($L=0$). }
 \label{fig:Setup}
 \end{figure*}

Beyond the well-established axion electrodynamics, theoretical studies have also predicted that a distinct topological term — the Nieh‑Yan action \cite{nieh1982identity,nieh1982quantized,chandia1997topological,nieh2007torsional} — can arise in the viscoelastic response of magnetic TIs \cite{hughes2011torsional,hughes2013torsional,parrikar2014torsion,shapourian2015viscoelastic}. This action originates from the interaction between electrons and strain in TIs, which can be described as an effective coupling between Dirac fermions and the vierbein fields of the curved spacetime \cite{chandia1997topological,weinberg2013gravitation}. Such a mechanism can induce a surface phonon Hall viscosity (PHV) in magnetic TIs, serving as the precise acoustic phonon counterpart to the electronic surface half-quantum Hall effect. This analogy between axion electrodynamics and Nieh-Yan action, as well as the corresponding physical consequence, is illustrated in Fig. \ref{fig:Setup}(a) and (b). Despite early theoretical predictions\cite{shapourian2015viscoelastic,tuegel2017hall,barkeshli2012dissipationless}, the experimental demonstration of surface PHV has yet to be achieved. 

In this work, we investigate the phonon dynamics induced by surface PHV derived from a prototype model describing the electron-phonon interaction in magnetic TI materials, such as magnetically doped TI sandwiches (e.g. Cr doped TI/pure TI/V doped TI sandwiches) \cite{yu2010quantized,chang2013experimental,zhang2012tailoring,morimoto2015,xiao2018realization,mogi2017magnetic,mogi2017tailoring,zhuo2023axion,jiang2020concurrence,kou2015magnetic} and MnBi$_2$Te$_4$ \cite{deng2020quantum,liu2020robust,otrokov2019prediction,zhang2019topological,li2019intrinsic,gong2019experimental,liu2020robust}. We demonstrate that surface acoustic phonon modes at one surface of magnetic TIs are both chiral with a  finite angular momentum and non-reciprocal due to surface PHV. Consequently, the phonon dynamics in magnetic TI films with finite thickness is strongly entangled with the surface magnetization configuration. In the surface ferromagnetic (FM) configuration, where magnetization is parallel at the top and bottom surfaces, acoustic phonons are chiral; conversely, in the surface antiferromagnetic (AFM) configuration with anti-parallel magnetization, they become non-reciprocal. We further examine the phonon thermal Hall effect in magnetic TI films, which reveals a strong contribution for the surface FM configuration but becomes zero for the surface AFM configuration. As bulk phonon modes do not contribute to thermal Hall effect, the low-temperature thermal Hall conductivity only originates from surface phonon modes and thus always scales as $T^2$, distinguishing it from the $T^3$ dependence typical of three‑dimensional magnetic materials \cite{qin2012berry,ye2021phonon}. The thermal Hall conductivity is further enhanced by the hybridization of acoustic phonons and surface magnons, forming collective excitations known as "magnon-polarons". These predicted properties not only constitute compelling evidence for surface PHV — pointing to its deeper origin in the bulk Nieh–Yan topological term — but also provide a novel mechanism for controlling phonon dynamics via surface magnetization.

\section{Surface Phonon Hall viscosity in Magnetic Topological Insulators} \label{sec_M:Surface PHV}
We start with introducing a coupled electron-phonon model for magnetic TIs. The crystal structure of TI materials, such as (Bi,Sb)$_2$(Te,Se)$_3$ and MnBi$_2$Te$_4$, belongs to the $R\Bar{3}m$ space group. At the $\Gamma$ point, the corresponding point group symmetry is $D_{3d}$, generated by a three-fold rotation $C_{3z}$, a two-fold rotation $C_{2x}$ and spatial inversion $\mathcal{P}$. We consider magnetic doping in magnetic TI sandwiches only at the surface layers (Fig.\ref{fig:Setup}(d) and (e)), and thus time reversal $\mathcal{T}$ appears in pure TI layers. On the other hand, MnBi$_2$Te$_4$ exhibits an antiferromagnetic ground state at a low temperature. Although this magnetic ordering breaks $\mathcal{T}$, it preserves an antiunitary symmetry $\mathcal{S} = \mathcal{T}\tau_{1/2}$ that combines time-reversal with a half-unit-cell translation $\tau_{1/2}$ along the $z$-axis. The $\mathcal{S}$ symmetry plays the same role as the $\mathcal{T}$ in magnetic TI sandwiches for the low energy effective theory. Thus, our theory below can be directly applied to both magnetic TI sandwiches and MnBi$_2$Te$_4$. The bulk low-energy electron Hamiltonian is given by
\begin{eqnarray} \label{eq:He}
    H_e =H_0 + H_{ep}. 
\end{eqnarray}
Based on $D_{3d}$ crystal symmetry and antiunitary $\mathcal{T/S}$ symmetry, $H_0$ describes 3D massive Dirac electrons in magnetic TIs \cite{zhang2019topological,wang2020dynamical} and is given by
\begin{equation} \label{eq_M:H0}
    H_{0} = m \Gamma_0 + v \left( k_x \Gamma_1 + k_y \Gamma_2 \right) + v_z k_z \Gamma_3,
\end{equation}
up to linear $k$ order in the basis $\{|P1^{'+}_{z},\uparrow \rangle,|P2^{'-}_{z},\uparrow \rangle, |P1^{'+}_{z},\downarrow \rangle, |P2^{'-}_{z},\downarrow \rangle\}$ with $\pm$ labelling inversion parity and $\uparrow, \downarrow$ labelling spin, where $\kk$ is the crystal momentum,  $m$ is the Dirac mass and $v$ and $v_z$ are in-plane and out-of-plane Fermi velocities \cite{SM}.  The Dirac matrices $\Gamma$ are defined as $\Gamma_0 = \sigma_3, \Gamma_1 = -\sigma_1 \otimes s_2, \Gamma_2 =  \sigma_1 \otimes s_1, \Gamma_3 = \sigma_2, \Gamma_4 = \sigma_1 \otimes s_3 $, where $\sigma, s$ correspond to the orbital and spin degrees of freedom, respectively. The electron-phonon interaction can also be derived from the $D_{3d}$ group and anti-unitary $\mathcal{T}$/$\mathcal{S}$ symmetry and takes the form
\begin{equation} \label{eq_M:Hep}
    H_{ep} = C(\textbf{u}) I +  M(\textbf{u})  \Gamma_0 +  k_j \Delta_a^j(\textbf{u}) \Gamma_a ,  
\end{equation}
with $ C(\textbf{u}) = C_1 u_{yy} + C_2 \left( u_{xx} + u_{yy} \right), \quad M(\textbf{u}) = M_1 u_{zz} + M_2 \left( u_{xx} + u_{yy} \right) $, where $a=0,1,2,3; j=1,2,3$, $k_{1,2,3} = k_{x,y,z}$ and $C_{1,2},M_{1,2}$ are material dependent parameters \cite{SM}. $\Delta^j_a$ is a function of the strain tensor $u_{ij} = \frac{1}{2} \left( \partial_i u_j + \partial_j u_i \right)$, with $u_i$ being the $i^{th}$ component of the phonon displacement field $\textbf{u}$. The explicit form of $\Delta^j_a$ is given in Sec. S1.B of Supplementary Materials (SM)\cite{SM}. The Einstein summation convention is implied hereafter. Based on Eqs.(\ref{eq_M:H0}) and (\ref{eq_M:Hep}), we can define a tensor field $e_a^j = \delta_a^j + \Delta_a^j$, which serves as the frame field (inverse of the vierbein) within the framework of Einstein-Cartan gravity\cite{chandia1997topological,hughes2011torsional, hughes2013torsional, parrikar2014torsion, weinberg2013gravitation}. When comparing this to the general form of the Dirac equation in Riemann--Cartan spacetime, we note that the electron-phonon interaction generates a non-trivial frame field but does not induce a spin connection. Consequently, the Hamiltonian $H_e$ in Eq.(\ref{eq:He}) effectively describes a Dirac fermion in Weitzenb\"ock spacetime, a specific subset of Riemann--Cartan spacetime\cite{Hehl1976} defined by vanishing Riemann curvature but non-zero torsion\cite{AldrovandiPereira2013,nieh1982identity, nieh1982quantized, nieh2007torsional,Shapiro2002,HayashiShirafuji1979,HayashiShirafuji1980}. The distinctions and relationships between Riemann–Cartan, Weitzenb\"ock, and the Riemannian spacetime of general relativity, as well as Minkowski spacetime, are illustrated in Fig.\ref{fig:Setup}(c). 

\begin{figure*}
\includegraphics[width=\textwidth]{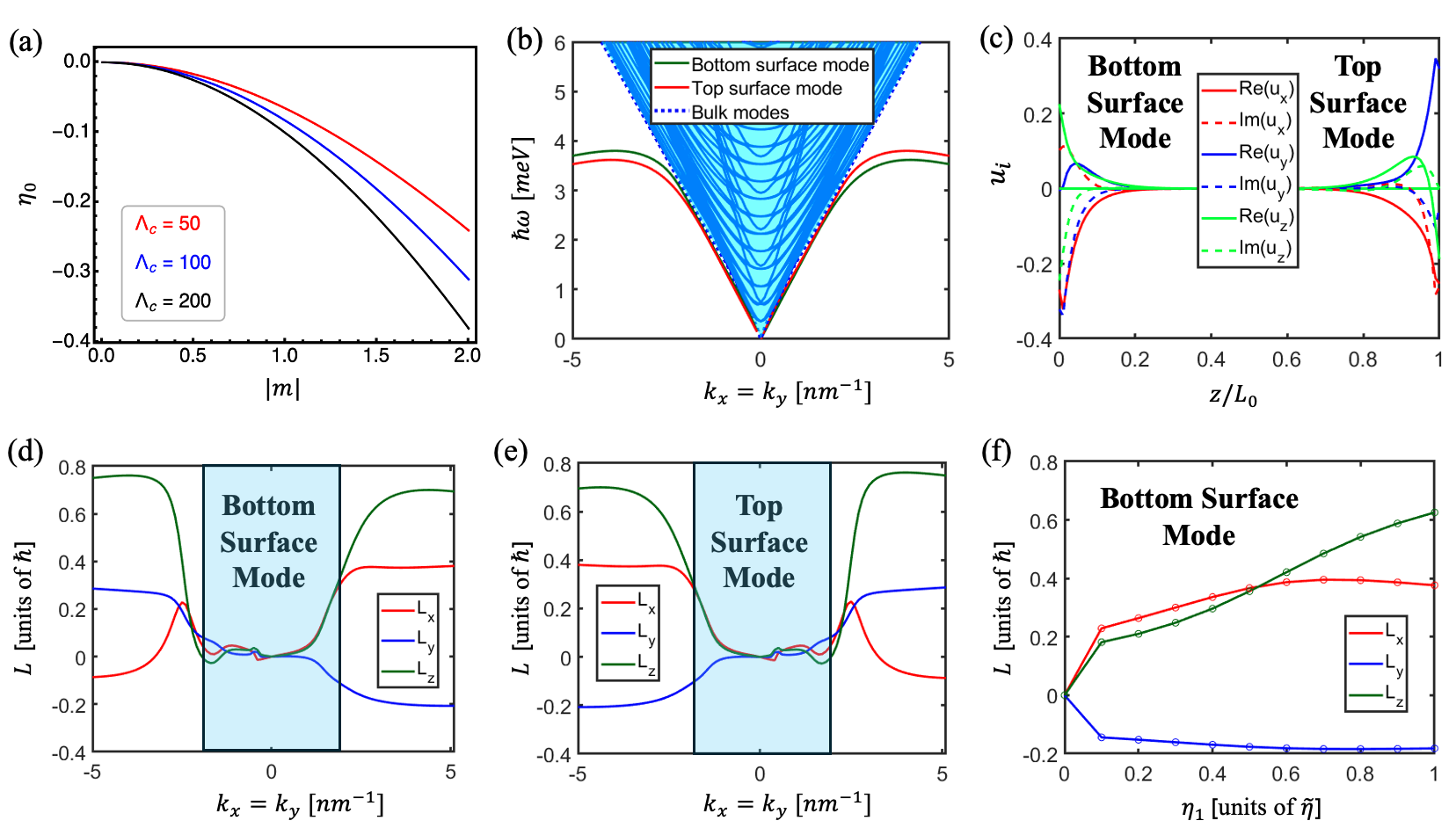}
 \caption{ (a) Dependence of the Nieh–Yan coefficient $\eta_0$ on the Dirac mass $|m|$ for different momentum cutoff $\Lambda_c$. (b) The surface phonon modes (green and red) with frequency below the bulk  phonon mode frequencies (blue region) in the FM configuration. (c) The spatial distribution of the displacement fields for the top and bottom surface phonon modes at $k_x = k_y = 2.9$ nm$^{-1}$. (d) and (e) shows the angular momentum $L_{x,y,z}$ as a function of momentum $\kk$ for (d) the bottom surface mode and (e) top surface mode in the FM configuration, respectively. The light blue shaded region corresponds to the momenta where the surface modes are indistinguishable from the bulk modes due to finite size effects in our numerical calculations. (f) The phonon angular momentum $L_{x,y,z}$ of the bottom surface phonon mode as a function of $\eta_1$ in units of $\Tilde{\eta} = 8  \hbar/nm^{2}$. For (a)-(e), we use $\eta_1 =  \Tilde{\eta}$.
 }
 \label{fig:Nonreciprocal phonon}
 \end{figure*}

Since we are interested in the phonon dynamics that is directly connected to the strain field in the above Hamiltonian, we integrate out the Dirac electrons and the corresponding effective action is given by (See SM Sec. S1.C for derivations \cite{SM})
\begin{equation} \label{eq_M:SNY}
     S_{NY} =  \eta_0 \delta_{ij} \int d t d^3 r \Phi \epsilon^{alb} \partial_l \Delta_a^i \partial_t \Delta_b^j, 
\end{equation}
where $a,l,b=1,2,3=x,y,z$, $i,j=1,2,3$ and $\eta_0$ is a material dependent parameter determined by electron-phonon coupling, treated here as a tuning parameter. $\Phi$ represents the phase of a complex Dirac mass term, which arises because the effective action is intended to describe the surface or interface region of magnetic TI, where both inversion $\mathcal{P}$ and anti-unitary symmetry $\mathcal{T}$/$\mathcal{S}$ are broken (See SM Sec. S1.C \cite{SM}). This action is known as the Nieh-Yan action \cite{nieh1982identity,nieh1982quantized,chandia1997topological}, which can be viewed as the gravitational analog to axion electrodynamics. We note that the coefficient $\eta_0$ is quadratic in the Dirac mass $m$ and depends on the momentum cutoff, as shown in Fig.\ref{fig:Nonreciprocal phonon}(a). This stands in sharp contrast to the coefficient of the axion term, which takes a quantized value, independent of Dirac mass $m$ and momentum cutoff (See SM Sec. S1.D \cite{SM}). 

Since the axion term is a total derivative, the main physical consequence of axion electrodynamics is the surface half-quantum Hall effect when the surface states of TIs is gapped by magnetism. Similarly, the Nieh-Yan action (\ref{eq_M:SNY}) is also a total derivative and leads to PHV \cite{avron1995viscosity,avron1998odd,barkeshli2012dissipationless,shapourian2015viscoelastic} at the surface or interface of magnetic TIs. To see that, we consider an interface to be the $xy$-plane with the $\Phi$ field only depending on the z direction, e.g. $\Phi = \Phi(z)$, and Eq.(\ref{eq_M:SNY}) can be rewritten as
\begin{eqnarray} \label{eq_M:SPHV}
    S_{PHV} = - \int dt d^3 r \frac{\partial \Phi}{\partial z} \eta_{ijmn} u_{ij} \Dot{u}_{mn},
\end{eqnarray}
where we have substituted $\Delta^i_a$ with $u_{ij}$ given in SM Sec.S1.B and absorbed all parameters into $\eta_{ijmn}$. $\eta_{ijmn}$ is the Hall viscosity tensor \cite{avron1995viscosity,avron1998odd} which satisfies $\eta_{ijmn} = \eta_{jimn} = \eta_{ijnm} = -\eta_{mnij} $, and its explicit form is given in SM Sec.S1.E \cite{SM}. 
In the specific context of acoustic phonons, it is referred to as phonon Hall viscosity \cite{barkeshli2012dissipationless, shapourian2015viscoelastic,guo2021extrinsic,chen2020enhanced,cortijo2015elastic,Heidari2019optical,ye2020phonon,huang2021electron,zhang2021phonon}. 
Due to anti-symmetric properties of the PHV and the $D_{3d}$ symmetry, we find three independent PHV coefficients $\eta_1, \eta_2, \eta_3$ defined as $\eta_1 \equiv \eta_{xxxy}, \eta_2 \equiv \eta_{xzyz}, \eta_3 \equiv \eta_{xyyz}/2$, which are related to the material parameters derived in SM Sec.S1.E\cite{SM}. The isotropic $\eta_1$ term has been derived in Refs. \cite{barkeshli2012dissipationless} and  \cite{shapourian2015viscoelastic} for the 2D and 3D isotropic systems, while the $\eta_2$ and $\eta_3$ terms are new independent PHV coefficients for materials with the $D_{3d}$ group symmetry. The full action is given by $S = S_0 + S_{\text{PHV}}$ where $S_0$ is the bulk phonon action,  $S_0 = \int dt d^3r \left( \frac{1}{2} \rho \dot{u}^2_i  -  \frac{1}{2} \lambda_{ijkl} u_{ij} u_{kl} \right)$  where $\lambda_{ijkl}$ are the elastic modulii of the $D_{3d}$ crystal group in SM Sec.S1.A\cite{SM} and $S_{\text{PHV}}$ is given by Eq.(\ref{eq_M:SPHV}). From the action $S$, the equation of motion can be derived from $\frac{\delta S}{\delta u_i} = 0$ as
\begin{equation}\label{eq_M:EOMBulk}
    \rho \ddot{u_i}  = \partial_j \left( \lambda_{ijkl} u_{kl} - 2 \eta_{ijkl} \frac{\partial \Phi}{\partial z} \dot{u}_{kl} \right) .
\end{equation} 
Compared to the standard elastic wave equation, Eq.(\ref{eq_M:EOMBulk}) includes an additional surface terms from PHV in Eq.(\ref{eq_M:SPHV}) and forms the basis for the discussion of the phonon dynamics in magnetic TI materials.

\section{Chiral and Nonreciprocal Phonon dynamics in Magnetic Topological Insulator Films}


\subsection{Chiral and Nonreciprocal Surface Acoustic Phonons} \label{sec_M:Chiral}

Chiral phonons describe lattice vibrations with non-zero angular momentum, manifested as circular or elliptically polarized displacement fields \cite{schaack1976observation,rebane1983faraday,lin1985study,mclellan1988angular,bermudez2008chirality,kagan2008anomalous,jones1973asymmetric,zhang2014angular,zhang2015chiral,chen2018chiral,hamada2018phonon,chen2019chiral,chen2021propagating,ren2021phonon,saparov2022lattice,zhang2023gate,wang2024chiral,juraschek2025chiral,shabala2025axial,wang2025ab,wang2025alteraxial}, and have been realized in TMD WSe$_2$ \cite{zhu2018observation}, bulk 3D chiral crystals \cite{ishito2023chiral}, Tellurium single-crystals \cite{zhang2025measurement}, ferromagnetic Kagome Weyl semimetals \cite{che2025magnetic} and polar crystals \cite{ueda2025chiral}. Nonreciprocal phonons describe phonon waves traveling at different speeds in opposite directions, and have been predicted and experimentally demonstrated in magnetic materials
\cite{heil1982nonreciprocal,sasaki2017nonreciprocal,xu2020nonreciprocal,chakraborty2023nonreciprocal,liao2024nonreciprocal,yu2020nonreciprocal,yang2020nonreciprocal,nomura2019phonon,sengupta2020magnetochiral,atzori2021magneto,nomura2023nonreciprocal,li2012colloquium,shan2023nonreciprocal,sengupta2025nonreciprocal,hirokane2020nonreciprocal,huang2025non,liao2023valley,liu2020observation,ren2025nonreciprocal,tokura2018nonreciprocal,cheong2018broken}, anomalous Hall crystals \cite{dong2025phonons} and quasicrystals \cite{matsuura2024singular}. Both phonon chirality and nonreciprocity are closely related to the symmetries. A finite phonon angular momentum requires the breaking of {\it either} inversion $\mathcal{P}$ or time-reversal $\mathcal{T}$ (or $\mathcal{S}$),  while nonreciprocity requires the breaking of {\it both} $\mathcal{P}$ and $\mathcal{T}$/$\mathcal{S}$. Thus, the bulk phonon modes of magnetic TI sandwiches or MnBi$_2$Te$_4$ are non-chiral and reciprocal due to the presence of both $\mathcal{P}$ and anti-unitary $\mathcal{T}$/$\mathcal{S}$. In contrast, the surface layer of magnetic TIs always breaks inversion $\mathcal{P}$. Furthermore, $\mathcal{T}$ in magnetic TI sandwiches is broken by surface magnetic doping, and the $\mathcal{S}$ symmetry is broken by the A-type of anti-ferromagnetism at the surface of MnBi$_2$Te$_4$. Thus, we would expect that phonon chirality and nonrecprocity may appear only at the surface of magnetic TI sandwiches or MnBi$_2$Te$_4$. 

To confirm the above symmetry argument, we solve Eq.(\ref{eq_M:EOMBulk}) in a thick slab configuration with  thickness $L=N a_0$ , where $N=100$ is the layer number and $a_0$ is the lattice constant. A thick slab configuration allows us to distinguish surface phonons from bulk phonons and decouple the surface phonon modes at the top and bottom surfaces. We choose stress free boundary conditions on both surfaces \cite{landau2012theory,benedek2013surface} and consider the ferromagnetic (FM) alignment of magnetization between the top and bottom surfaces such that $\eta^t_i = \eta^b_i $, where $i=1, 2, 3$ and the superscripts $t, b$ label the top and bottom surfaces, in Fig.\ref{fig:Setup}(d). The computed phonon dispersion is shown in Fig.\ref{fig:Nonreciprocal phonon}(b), from which we find two surface phonon modes \cite{landau2012theory} with their frequency (green and red curves) below the frequency of bulk phonon modes (the blue color region). These two surface phonon modes are located at the $z=0$ (bottom) and the $z=L_0$ (top) surfaces, respectively, as demonstrated via examining the displacement fields ($u_{x,y,z}$) at $k_x=k_y = 2.9$nm$^{-1}$ in Fig.\ref{fig:Nonreciprocal phonon}(c). For the surface FM configuration, these two surface phonon modes at opposite momenta are related by inversion $\mathcal{P}$. On each surface, we note that the phonon dispersion is nonreciprocal, $\omega^{\text{bottom (top)}}(+{\bf k}) \neq \omega^{\text{bottom (top)}}(-{\bf k})$, in Fig. \ref{fig:Nonreciprocal phonon}(b), which is consistent with the above symmetry argument since a single surface breaks both $\mathcal{P}$ and $\mathcal{T}$/$\mathcal{S}$. In addition, phonon nonreciprocity also requires to break two-fold rotation about $z$-axis, which is absent for the $D_{3d}$ group of our systems. 

Next we examine the angular momentum of phonon modes, which is defined by $L_i(\textbf{k}) = \textbf{u}_0^\dagger M_i \textbf{u}_0$ \cite{zhang2014angular,hamada2018phonon,liu2022probing,kishine2020chirality}, where the eigenmode $\textbf{u}_0 =(u_x,u_y,u_z)^T$ and the matrix $(M_i)_{mn} = -i \epsilon_{imn}$. In Fig. \ref{fig:Nonreciprocal phonon}(d) and (e), we plot $L_i$ ($i=x, y, z$) along $k_x = k_y$ for the surface modes at the top (red) and bottom (green) surfaces, respectively. In the blue color regions, the surface modes are not well separated from the bulk mode due to finite size effects in our numerical calculation. Thus, we focus on the angular momentum in the large momentum regions. We note that $L_z$ generally keeps its sign for opposite momenta, while $L_x$ and $L_y$ reveal a more complex behavior with sign changes. The phonon angular momentum at two surfaces can also be related by inversion $\mathcal{P}$, $L^{\text{bottom}}_i ({\bf k}) = L^{\text{top}}_i (-{\bf k})$. Fig. \ref{fig:Nonreciprocal phonon}(f) shows the dependence of phonon angular momentum $L_{x,y,z}$ on the dimensionless parameter $\eta_0$ in Eq(\ref{eq_M:SPHV}) at $k_x=k_y = 2.9$ nm$^{-1}$. $L_{x,y,z}$ is zero for $\eta_0 =0$, increases rapidly with increasing $\eta_0$, and gradually saturates for a large $\eta_0$, which clearly demonstrates the surface PHV origin of phonon chirality in magnetic TIs. Thus, we have established that the surface phonon mode is nonreciprocal and chiral at each individual surface due to surface PHV.



\begin{figure*}
\includegraphics[width=\textwidth]{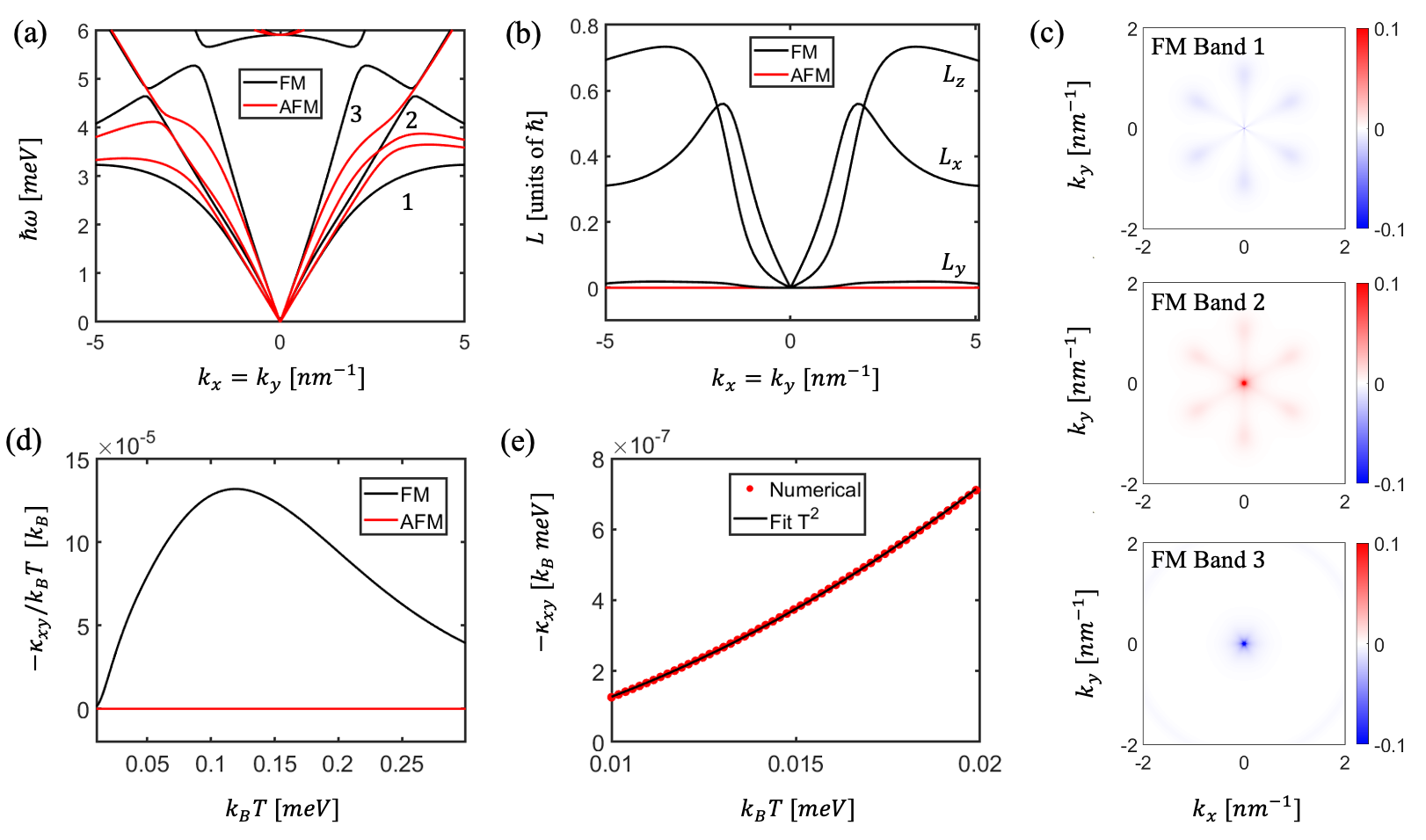}
\caption{ (a) The phonon dispersion of the thin film slab model (N=6 sites) for the FM case (black lines) and AFM case (red lines). (b) The angular momentum $L_{x,y,z}$ of the lowest frequency acoustic phonon modes for the FM case (black lines) and the AFM case (red lines). (c) The phonon Berry curvature distribution in the $k_x-k_y$ plane for the three acoustic phonon modes in the FM configuration. (d) The phonon thermal Hall conductivity $\kappa_{xy}$ as a function of temperature $k_B T$ for the FM (black) and AFM (red) configurations. (e) The low temperature behavior of $\kappa_{xy}$ from the numerical calculations (red points) can be well described by the $T^2$ fitting (black curve). }
 \label{fig:Phonon thermal Hall}
 \end{figure*}

\subsection{Chiral and Non-reciprocal Acoustic Phonon Modes in magnetic TI films} \label{Sec_M:nonreciprocal phonn}
We next consider the phonon dynamics in magnetic TI thin films. The thickness of a typical magnetic TI film grown by MBE is on the order of tens or hundreds of nanometers \cite{chang2013experimental,zhang2012tailoring,morimoto2015,xiao2018realization,mogi2017magnetic,mogi2017tailoring,zhuo2023axion,jiang2020concurrence,kou2015magnetic,deng2020quantum,liu2020robust,otrokov2019prediction,zhang2019topological,li2019intrinsic,gong2019experimental,liu2020robust}, which is much smaller than the penetration depth of a surface acoustic wave, which is usually around tens of microns \cite{landau2012theory,gerus1974amplification,gerus1975amplification}. Consequently, the surface acoustic phonon modes on two opposite surfaces will strongly hybridize; therefore, the phonon dynamics can be treated as effectively two-dimensional. We solve the phonon dynamics for a slab model with N=6 sites and stress-free boundary conditions on both surfaces. Particularly, we consider two magnetization configurations in Eq.(\ref{eq_M:EOMBulk}): the FM configurations with the PHV $\eta^t_i = \eta^b_i $ and the antiferromagnetic (AFM) configuration with $\eta^t_i = -\eta^b_i$. Both FM and AFM configurations can be realized in magnetic TI sandwiches, e.g. the Cr doped TI/pure TI/V doped TI sandwiches \cite{yu2010quantized,chang2013experimental,zhang2012tailoring,morimoto2015,xiao2018realization,mogi2017magnetic,mogi2017tailoring,zhuo2023axion,jiang2020concurrence,kou2015magnetic}, and controlled by tuning external magnetic fields. In MnBi$_2$Te$_4$, the FM and AFM configurations correspond to the odd and even setuple layers (SLs) of the film thickness \cite{chen2024even,zhao2021even,li2024fabrication,yang2024intrinsic,mei2024electrically,lin2022direct,ovchinnikov2021intertwined}. Fig.\ref{fig:Phonon thermal Hall}(a) depicts the phonon dispersions for the FM (black curves) and AFM (red curves) configurations, and one can see a sharp contrast between the FM and AFM configurations. The phonon dispersion of the FM configuration is symmetric between positive $\kk$ and negative $-\kk$, thus reciprocal, while that of the AFM configuration exhibits asymmetry, giving rise to nonrecriprocal phonons. We further examine the phonon angular momenta $L_{x,y,z}$ for the FM and AFM configurations of magnetic TI films in Fig.\ref{fig:Phonon thermal Hall}(b), which show finite values for the FM configuration and zero value for the AFM configuration. Combining Fig.\ref{fig:Phonon thermal Hall}(a) and (b), we have demonstrated that acoustic phonon dynamics in magnetic TI thin films is strikingly correlated with the surface magnetization configuration: the acoustic phonons are chiral for the FM configuration and non-reciprocal for the AFM configuration. Experimentally, the surface magnetization configuration can be controlled by tuning external magnetic fields for the Cr doped TI/pure TI/V doped TI sandwiches or fabricating even and odd SLs of MnBi$_2$Te$_4$. 

This unique property of phonon dynamics in our system can be understood from the symmetry aspect. In the FM configuration, the inversion $\mathcal{P}$ is preserved and guarantees the reciprocity of phonons. Due to the breaking of $\mathcal{T}$/$\mathcal{S}$, phonon angular momentum is allowed and generated by the surface PHV at both the top and bottom surfaces. In contrast, in the AFM configuration, although both $\mathcal{P}$ and $\mathcal{T}$/$\mathcal{S}$ are broken, the combined symmetry $\mathcal{P}\mathcal{T}$/$\mathcal{P}\mathcal{S}$ is preserved and forbids the phonon angular momentum at any momentum. Since $\mathcal{P}\mathcal{T}$/$\mathcal{P}\mathcal{S}$ does not change momentum, it does not provide any constraint on the phonon nonreciprocity. 

While the PHV has been previously proposed to generate chiral phonons \cite{barkeshli2012dissipationless,shapourian2015viscoelastic,guo2021extrinsic,zhang2021phonon}, the mechanism of inducing nonreciprocal phonons purely from surface PHV has not been considered. As nonreciprocal phonons require to break both $\mathcal{P}$ and $\mathcal{T}$ while the PHV action itself only breaks $\mathcal{T}$ but preserves $\mathcal{P}$, the early literature proposed to include additional symmetry-breaking terms in the phonon equation of motion like the flexo-elastic modulus \cite{ren2025nonreciprocal}, or the kineo-elastic modulus \cite{dong2025phonons}. In our mechanism, the inversion $\mathcal{P}$ is broken because of the opposite values of surface PHV, $\eta^t_i = -\eta^b_i$, in the AFM configuration. 

Experimentally, the phonon chirality can lead to phonon thermal Hall effect, as discussed below in Sec. \ref{sec_M:thermal Hall}, while longitudinal thermal conductivity measurements with opposite directions of the thermal gradient ($\kappa_{xx} (+\nabla T) \neq \kappa_{xx} (-\nabla T)$) can probe phonon nonreciprocity \cite{seif2018thermal,schmotz2011thermal,hirokane2020nonreciprocal,vu2023magnon}. 

\begin{figure*}
    \centering
\includegraphics[width=\textwidth]{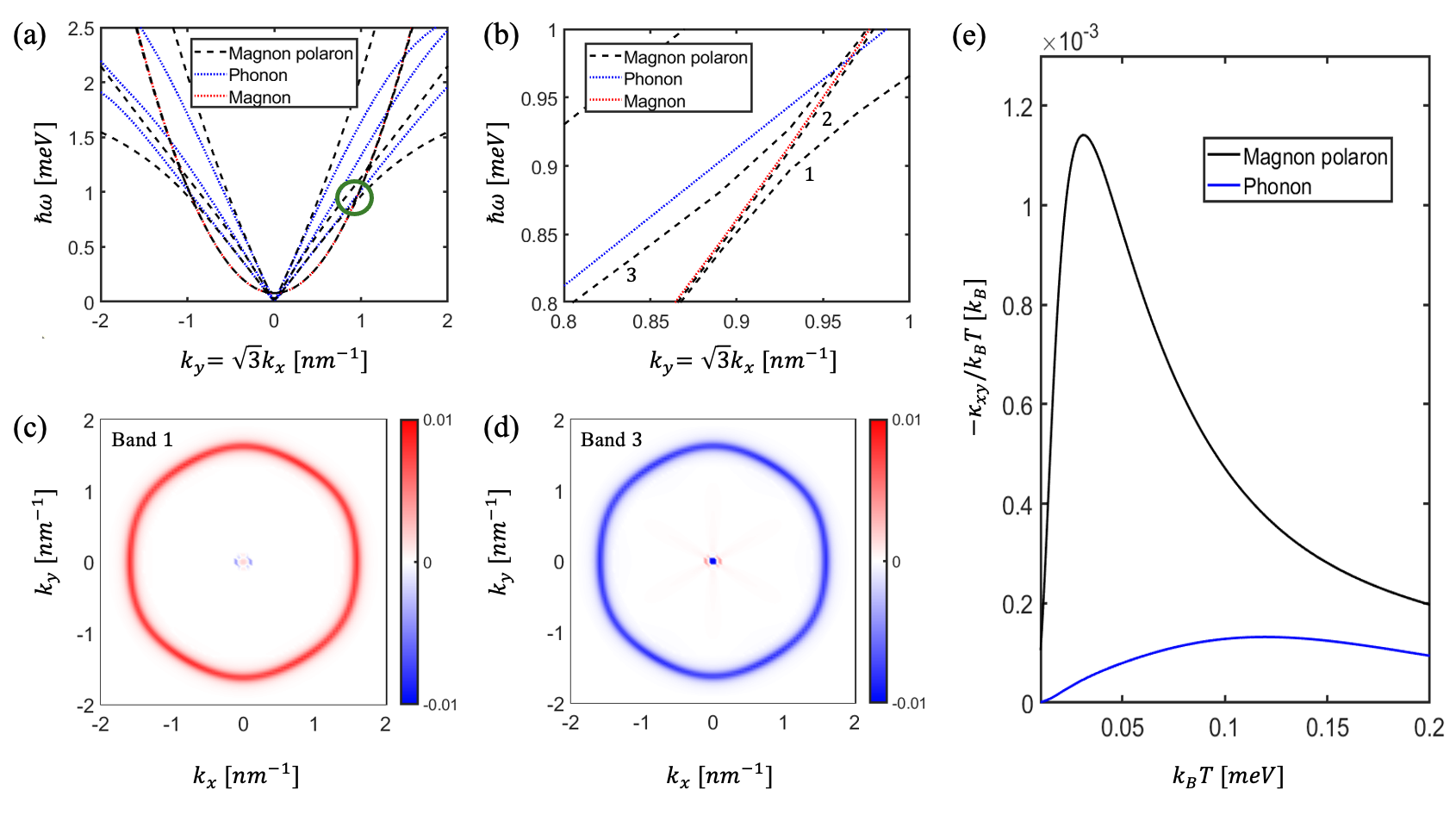}
\caption{  (a) The dispersion of magnon-polaron modes (black), bare phonon modes (blue) and bare magnon modes (red) in the FM configuration.  (b) The zoom-in for the region indicated by the green circle in (a). The Berry curvature distribution of magnon-polaron (c)  branch 1 and (d) branch 3. (e) The surface thermal Hall conductivity $\kappa_{xy}$ as a function of temperature $k_B T$ for magnon-polarons (black) and bare phonons (blue). }
 \label{fig:Magnon polaron}
 \end{figure*}

\subsection{Phonon thermal Hall effect in magnetic TI films} \label{sec_M:thermal Hall}
One physical consequence of the surface PHV is the phonon thermal Hall effect \cite{zhang2009Kubo,zhang2010topologicalphononHall,zhang2011phonon}, which describes the x-directional thermal current induced by y-directional temperature gradient. The thermal Hall effect is characterized by thermal Hall conductivity, which can be evaluated by 
\cite{qin2011energy,qin2012berry,ye2021phonon}
\begin{equation}\label{eq_M:kappaxy}
   \kappa_{xy}^{tr} = - \frac{1}{T} \int d \epsilon \epsilon^2 \sigma_{xy} (\epsilon) \frac{d n (\epsilon)}{d \epsilon}  
\end{equation}
where 
\begin{equation} \label{eq_M: sigmaxyZph}
    \sigma_{xy} (\epsilon) = -\frac{1}{V \hbar} \sum_{\hbar \omega_{\textbf{k},n} \leq \epsilon} \Omega^z_{\textbf{k},n}. 
\end{equation}
Here $T$ is the temperature, $V$ is the volume of the system, $\Omega_{\textbf{k}n} = -\text{Im}[ \frac{\partial \Bar{\psi}_{\textbf{k}n}}{\partial \textbf{k}} \times \frac{\partial \psi_{\textbf{k}n}}{\partial \textbf{k}} ]$ is the phonon Berry curvature distribution of band $n$ and $n(\epsilon) = \frac{1}{e^{\beta \epsilon} -1}$ is the Bose distribution. $\Bar{\psi}_{{\bf k},n}$ and $\psi_{\textbf{k}n}$ are the left and right eigenstates of the non-Hermitian Hamiltonian that describes phonon eigen-problem and is derived from the phonon equation of motion in Eq.(\ref{eq_M:EOMBulk}), as discussed in Sec. S3.B of SM \cite{SM}. Fig. \ref{fig:Phonon thermal Hall}(c) depicts the phonon Berry curvature distribution of three acoustic phonon modes in the $k_x-k_y$ plane for the FM configuration. The absolute values of phonon Berry curvatures are peaked around $\Gamma$, e.g. $k_x=k_y=0$, for all three acoustic phonon modes, and decay quickly with increasing $\kk$, satisfying three-fold rotation symmetry $C_{3z}$. For acoustic phonon modes 1 and 2, small peaks/dips appear around six momenta $|{\bf k}| = 2.5 $ nm$^{-1}$ at the angles $\{\pi/6,\pi/2,5 \pi/6, 7\pi/6, 3 \pi/2, 11 \pi/6 \}$ with respect to $k_x$, which is due to the anti-crossings between these two modes. The Berry curvature distribution at these six momenta are equal in magnitude and have opposite sign for acoustic phonon modes 1 and 2. Similarly, the Berry curvature distributions of acoustic phonon modes 2 and 3 at $\Gamma$ cancel each other. However, the 3 acoustic phonon modes appear at different energies, and they are filled differently in accordance to the Bose distribution, $n(\epsilon)$, giving rise to the thermal Hall conductivity. Based on the phonon Berry curvature distribution in Fig.\ref{fig:Phonon thermal Hall}(c), we can evaluate the phonon thermal Hall conductivity $\kappa_{xy}^{tr}$ with Eqs.(\ref{eq_M:kappaxy}) and (\ref{eq_M: sigmaxyZph}), which is shown as a function of temperature $k_B T$ (black curve) in Fig.\ref{fig:Phonon thermal Hall}(d). 
For the FM configuration, $\kappa_{xy}^{tr}$ increases rapidly at a small $T$, but decreases at a large $T$, thus revealing a peak at $k_B T \sim 0.1$ meV for the current parameters. This peak behavior of the temperature dependence is similar as the previous studies on bulk phonon thermal Hall effect\cite{ye2021phonon}. However, we note the low temperature behavior of $\kappa_{xy}$ can be fitted well by $\kappa_{xy} \sim T^2$, as shown in Fig.\ref{fig:Phonon thermal Hall}(e), in sharp contrast to the $\kappa_{xy} \sim T^3$ dependence shown in Refs. \cite{qin2012berry,ye2021phonon}. This difference is because the acoustic phonon modes that contribute to $\kappa_{xy}$ are of 2D nature, which is different from 3D acoustic phonon modes in Refs. \cite{qin2012berry,ye2021phonon}. We emphasize that even with increasing the film thickness to the bulk limit, this $T^2$ behavior remains valid since only the surface phonon modes originated from surface PHV can contribute to the thermal Hall effect, while the bulk phonons in magnetic TIs give no contribution due to the antiunitary $\mathcal{T}$ or $\mathcal{S}$ symmetry. We developed a 2D effective model for magnetic TI sandwiches with the FM configuration to analytically show this $T^2$ behavior at low temperatures in Sec. S3.C of SM \cite{SM} for acoustic phonon modes. 

As the AFM configuration preserves $\mathcal{PT}$ symmetry, the acoustic phonon modes must be non-chiral with zero phonon angular momentum (red curve in Fig.\ref{fig:Phonon thermal Hall}(b)), and the phonon Berry curvatures for all acoustic phonon modes vanish for any ${\bf k}$, and thus we always find $\kappa_{xy}=0$, as shown by the red line in Fig.\ref{fig:Phonon thermal Hall}(d).

\subsection{Magnon-polarons}
In magnetic TI films, magnon excitations can also exist, in addition to phonons, and couple to surface electrons. Therefore, acoustic phonons will couple to magnons via surface electrons, forming phonon-magnon hybrid excitations known as "magnon-polarons" \cite{kamra2015coherent,shen2015laser}, in magnetic TI films. In this section, we will show that magnon-polarons can significantly enhance the thermal Hall effect. 

To extract the coupling between acoustic phonons and magnons, we consider an effective model with the surface electrons at the top and bottom surfaces couple to both the strain field and surface magnons, as shown in SM Sec. S3.E \cite{SM}. After integrating out the surface electrons, we obtain a coupled phonon-magnon effective Lagrangian with the form 
\begin{eqnarray}\label{eq:Lagrangian_2D_full}
    \mathcal{L}_{2D} = \mathcal{L}^{ph}_{2D} + \sum_a (\mathcal{L}_{2D,a}^{m_0} +  \mathcal{L}_{2D,a}^{m-ph}),
\end{eqnarray}
where $a=t,b$ denotes the top and bottom surfaces. $\mathcal{L}^{ph}_{2D}$ is the phonon Lagrangian in 2D, which is derived from symmetry consideration in SM Sec. S3.B \cite{SM}, which can also be obtained by projecting $S_0 + S_{\text{PHV}}$ given in Sec. \ref{sec_M:Surface PHV} onto the 2D phonon displacements $u_{x,y}$. We consider magnetic TI sandwiches in Fig.\ref{fig:Setup}(d) and (e) and assume magnons in the top and bottom layers are decoupled. The free magnon Lagrangian $\mathcal{L}_{2D,a}^{m_0}$ on one surface ($a=t, b$) is given by \cite{herring1951theory,kittel1958interaction}
\begin{align}\label{eq_M:Lm0}
    \mathcal{L}^{m_0}_{2D,a} = &\frac{m_x^a \dot{m}^a_y}{\omega_s} - \frac{\gamma_A}{2} \left( (\nabla m_x^a)^2 + (\nabla m_y^a)^2  \right) \nonumber \\ &- \frac{\gamma_K}{2} \left( (m_x^{a})^2 + (m_y^{a})^2 \right),
\end{align}
where $\omega_s = \gamma M_s$, $\gamma$ is the gyromagnetic ratio and $M_s$ is the saturation magnetization, $\gamma_A,\gamma_K$ are the material parameters, $m^a_{x,y}$ denotes the magnetic fluctuation around the overall magnetization along the $z$-axis. The acoustic phonon-magnon interaction is given by
\begin{align}
    &\mathcal{L}_{2D,a}^{m-ph} = \zeta^a_1\Big[ \left(  u^a_{xx} - u^a_{yy} \right) \dot{m}^a_x - m^a_x \left( \dot{u}^a_{xx} - \dot{u}^a_{yy} \right) + m^a_y 2 \dot{u}^a_{xy} \nonumber \\
    &  - 2 u^a_{xy} \dot{m}^a_y \Big]  + 2 \zeta^a_2 \Big[ \left( u^a_{xx} + u^a_{yy} \right) \left( \partial_x m^a_x + \partial_y m^a_y\right) \Big] ,
\end{align}
on the surface $a=t, b$, where $\uu^a$ are the strain field and $\zeta^a_1 , \zeta^a_2 $ determine the magnon-phonon coupling strengths. The relative signs between $\zeta^a_1$ and $\zeta^a_2$ can be fixed by $\mathcal{P}$ for the FM configuration and $\mathcal{P}\mathcal{T}$ for the AFM configuration, respectively. 
$\zeta^a_1 \sim \text{sgn}(m_z^a)/(v_f^a)^2, \zeta^a_2 \sim \text{sgn}(m_z^a)/v_f^a$, where $m_z^a$ is the out-of-plane magnetization on surface $a=t,b$ and $v_f^a$ is the Fermi velocity for the Dirac. The exact expression for $\zeta^a_{1,2}$ is given in SM Sec S3.E. \cite{SM}. The surface magnetizations are related by $m_z^t = \pm m_z^b$ for the FM and AFM configurations respectively and the Fermi velocities are $v_f^t = - v_f^b$ for both configurations. The FM configuration corresponds to the $\mathcal{P}$ symmetric case, $\zeta^t_1 = \zeta^b_1, \zeta^t_2 = -\zeta^b_2$ and the AFM configuration corresponds to the $\mathcal{PT}$ symmetric case, $\zeta^t_1 = \zeta^b_1, \zeta^t_2 = \zeta^b_2$. The coupled equations of motion for strain field $\uu$ and magnetic fluctuation $\mm$ in the slab model can be derived from the Lagrangian (\ref{eq:Lagrangian_2D_full}) and the details of the derivation is discussed in Sec. S3.E of SM \cite{SM}. 

Fig. \ref{fig:Magnon polaron}(a) depicts the dispersions for the bare phonons (blue dotted curves) and magnons (red dotted curves), comparing them with the magnon-polaron dispersion (black dashed curves) in magnetic TI sandwiches for the FM configuration. There are three acoustic phonon modes and two degenerate magnon modes, which are located separately on the top and bottom surfaces. We note that the magnon-phonon coupling induces the anti-crossing between magnon and phonon bands, leading to the formation of magnon-polaron bands, as shown in Fig.\ref{fig:Magnon polaron}(b) as the zoom-in of the green circle region in Fig.\ref{fig:Magnon polaron}(a). The anti-crossing in Fig. \ref{fig:Magnon polaron}(b) occurs between two magnon polaron bands, namely bands 1 and 3. Within the anti-crossing gap of magnon-polaron, there is an additional magnon dispersion branch, band 2, which remains unaffected. In order to obtain the Berry curvature of the two magnon polarons, we project our model onto the bands 1 and 3. Fig. \ref{fig:Magnon polaron}(c) and (d) show the Berry curvature distribution of two branches of magnon-polaron modes. 
A large phonon Berry curvature is distributed in a near circle, with strong peaks near the anti-crossing regions of magnon-polarons, which leads to a significant enhancement of the phonon thermal Hall conductivity. Fig. \ref{fig:Magnon polaron}(e) shows a comparison of the thermal Hall conductivity $\kappa_{xy}$ between magnon-polarons (black curve) and acoustic phonons (blue curve). While the line shape of thermal Hall conductivity as a function of temperature remains the same, its magnitude is increased by almost one order, which is attributed to the large Berry curvature around the magnon-polaron anti-crossing regions. 

Magnon-polarons have been experimentally observed using neutron spectroscopy\cite{bao2023direct}, magneto-Raman spectroscopy\cite{liu2021direct}, and thermally driven spin transport measurements, including the spin Seebeck effect\cite{li2020observation,kikkawa2016magnon}. Their contribution to the thermal Hall effect has also been predicted and observed  \cite{ideue2017giant,nawwar2025large,takahashi2016berry,zhang2019thermal}. Similar experimental methods can be applied to magnetic TI sandwiches or MnBi$_2$Te$_4$ films to characterize the magnon-polaron spectrum and elucidate the unique magnon-polaron mechanism responsible for the enhanced thermal Hall effect in these systems.



\section{Conclusion and Discussion}

In this work, we have theoretically studied the phonon dynamics induced by surface PHV in magnetic TI materials, including magnetically doped TI sandwiches and MnBi$_2$Te$_4$. We show that the phonon dynamics can be controlled by the relative magnetization orientations of the top and bottom surfaces in magnetic TI films. For the FM configuration, we demonstrate the emergence of  chiral phonon modes carrying finite angular momentum. In contrast, for the AFM configuration, we predict the existence of nonreciprocal phonon propagation. These results establish a surface-magnetization-dependent switching between chiral phonons and nonreciprocal phonons, highlighting the tunability of acoustic responses in magnetic TI films. 

We further demonstrate the thermal Hall effect can arise from surface PHV in the FM configuration - vanishing in the bulk or the AFM configuration — and is further enhanced by coupling between acoustic phonons and magnons. Thus, the experiment measurements of thermal Hall effect provide a direct experimental probe of surface PHV in magnetic TIs. We find due to magnon-polaron effect, the thermal Hall conductivity $\kappa_{xy} \sim 10^{-3} k_B^2 T/\hbar$ in SI units, which is one order larger than the estimated thermal Hall conductivity per unit layer in bulk magnetic insulators \cite{ye2021phonon}. At 100mK, the thermal Hall conductivity is estimated as $\kappa_{xy} \sim 10^{-10} $ W/K. The thermal Hall conductivities due to electrons in 2D antiferromagnets\cite{wang2025layer}, altermagnets \cite{zhou2024crystal} and noncollinear antiferromagnets\cite{xu2020finite} are of a similar order of $\sim 10^{-10}$ W/K at 300 K. At the charge neutral point with the Fermi energy within the magnetic gap, the thermal Hall conductivity due to electrons is almost negligible \cite{wang2025layer}. Thus, we expect that the major contribution of the low temperature thermal Hall effect in magnetic TI films comes from phonons and magnons. It has been suggested that phonon scattering from impurities could enhance the thermal Hall conductivity \cite{guo2021extrinsic,chen2020enhanced}.Notably, the thermal Hall conductivity in magnetic TIs follows a $T^2$ law, distinguishing it from the conventional $T^3$ scaling found in 3D magnetic insulators. This deviation reflects the fact that the thermal Hall conductivity arises exclusively from surface PHV. Together, these predicted signatures offer compelling evidence for the existence of surface PHV in magnetic TI materials.

Magnetic TIs have long been recognized as a compelling platform for studying electronic topological phenomena, including the quantum anomalous Hall effect and axion electrodynamics \cite{qi2008topological,chang2013experimental,chang2023colloquium,zhang2019topological}. Our work demonstrates that magnetic TI films can also serve as an effective platform for controlling phonon dynamics with potential relevance to phononic applications. More broadly, our results clarify the interplay of topological electronic states, magnetization, and lattice dynamics in magnetic TIs, paving the way for potential devices that integrate spintronics and phononics \cite{vangessel2018review,maldovan2013sound,li2012colloquium,balandin2012phononics,liu2020topological}.

\section{Acknowledgments}
We thank C.Z. Chang, X.L. Qi, H. Nieh and B.H. Yan for helpful discussions. This project is mainly supported by the ONR Award (N000142412133). A.C. and C.-X.L. also acknowledge support from NSF grant via the grant number DMR-2241327.

\bibliographystyle{apsrev4-2}
\bibliography{ref}

\clearpage
\onecolumngrid    

\input{SM_ArXiv}

\end{document}

%% file: SM_ArXiv.tex




\clearpage
\onecolumngrid

\section*{Supplementary material for ``Surface Phonon Hall Viscosity Induced Phonon Chirality and Nonreciprocity in Magnetic Topological Insulator Films"}

\setcounter{section}{0}
\setcounter{figure}{0}
\setcounter{table}{0}
\setcounter{equation}{0}

\renewcommand{\thefigure}{S\arabic{figure}}
\renewcommand{\thesection}{S\arabic{section}}
\renewcommand{\thetable}{S\arabic{table}}

\numberwithin{equation}{section}
\newcommand{\eqsref}[1]{Eq(S\ref{#1})}

\def\qq{\mathbf{q}}
\def\kk{\mathbf{k}}
\def\KK{\mathbf{K}}
\def\DKK{\Delta\mathbf{K}}
\def\pp{\mathbf{p}}
\def\RR{\mathbf{R}}
\def\tt{\mathbf{t}}
\def\rr{\mathbf{r}}
\def\GG{\mathbf{G}}
\def\QQ{\mathbf{Q}}
\def\aa{\mathbf{a}}
\def\bb{\mathbf{b}}
\def\uu{\mathbf{u}}
\def\AA{\mathbf{A}}
\def\ff{\mathbf{f}}
\def\mm{\mathbf{m}}

\def\KK{\mathbf{K}}
\def\qq{\mathbf{q}}
\def\pp{\mathbf{p}}
\def\pp{\mathbf{p}}
\def\GG{\mathbf{G}}
\def\QQ{\mathbf{Q}}
\def\RR{\mathbf{R}}
\def\tt{\mathbf{t}}
\def\dd{\mathbf{d}}
\def\aa{\mathbf{a}}
\def\bb{\mathbf{b}}
\def\ee{\epsilon}
\def\CC{\mathcal{P}}
\def\UU{\mathbf{U}}

\def\BZ{{\rm BZ}}
\def\mS{{\mathcal{S}}}

\def\spin{{\varsigma}}

\def\hH{{ \hat{H} }}
\def\hrho{ \hat{\rho} }
\def\hg{\hat{g}}
\def\hS{\hat{S}}

\def\mG{{\mathcal{G}}}

\def\UC{{\hat{\Theta}}}
\def\UF{{\hat{\Sigma}}}

\def\mK{{\mathcal{K}}}
\def\pr{\prime}
\def\mJ{{\mathcal{J}}}
\def\mK{{\mathcal{K}}}

\def\ie{{\it i.e.},\ }
\def\eg{{\it e.g.}\ }
\def\ea{{\it et al.}}

%
%

\pagenumbering{arabic}

\section{Effective theory of magnetic topological insulators}

\subsection{Symmetry properties of $D_{3d}$ group for magnetic topological insulators} \label{sec:d3d}

In this section, we will first discuss the symmetry property of magnetic topological insulator (TI) films. Here we consider magnetically doped TI sandwiches (e.g. Cr doped (Bi,Sb)$_2$(Te,Se)$_3$/(Bi,Sb)$_2$(Te,Se)$_3$/V doped (Bi,Sb)$_2$(Te,Se)$_3$) \cite{yu2010quantized,chang2013experimental,zhang2012tailoring,morimoto2015,xiao2018realization,mogi2017magnetic,mogi2017tailoring,zhuo2023axion,jiang2020concurrence,kou2015magnetic}, and MnBi$_2$Te$_4$ \cite{deng2020quantum,liu2020robust,otrokov2019prediction,zhang2019topological,li2019intrinsic,gong2019experimental,liu2020robust}, which is an antiferromagnetic TI\cite{mong2010antiferromagnetic,li2010dynamical,zhang2019topological,wang2020dynamical,zhang2020mobius}. The low-energy effective theory of (Bi,Sb)$_2$(Te,Se)$_3$ and MnBi$_2$Te$_4$ is around the $\Gamma$ point in the Brillouin zone and the corresponding little group at $\Gamma$ is the $D_{3d}$ group, generated by three-fold rotation $\Hat{C}_{3}$ about the $z$ axis, two-fold rotation $\Hat{C}_{2x}$ about the $x$ axis and inversion $\Hat{P}$. The character table of $D_{3d}$ group generators is given in Table \ref{table:d3d character}. 

For magnetic TI sandwiches, since magnetic doping only exists at two surface layers, the bulk state of (Bi,Sb)$_2$(Te,Se)$_3$ preserves time reversal $\Hat{T}$. For MnBi$_2$Te$_4$, the antiferromagnetic ground state breaks time reversal symmetry $\Hat{T}$, but preserves the combined time reversal and lattice half-translation labelled by $\Hat{S} = \Hat{T} \Hat{\tau}_{1/2}$, where $\Hat{\tau}_{1/2}$ is the translation by half of the unit cell along the z direction. 

\begin{table}[h]
\centering
\begin{tabular}{||c || c c c c c c||} 
 \hline
  $D_{3d}$ & $E$ & $P$ & $m$ & $C_{3z}$ & $C_{2x}$ & $S_6$ \\ [0.5ex] 
 \hline\hline
 $A_{1g}$ & +1 & +1 & +1 & +1 & +1 & +1\\ 
 \hline
 $A_{2g}$ & +1 & +1 & $-$1 & +1 & $-$1 & +1\\ 
 \hline
 $E_{g}$ & +2 & +2 & 0 & $-$1 & 0 & $-$1 \\ 
 \hline
 $A_{1u}$ & +1 & $-$1 & $-$1 & +1 & +1 & $-$1 \\ 
 \hline
 $A_{2u}$ & +1 & $-$1 & +1 & +1 & $-$1 & $-$1\\ 
 \hline
 $E_{u}$ & +2 & $-$2 & 0 & $-$1 & 0 & +1\\ 
 \hline
\end{tabular}
\caption{Character Table of $D_{3d}$ group}
\label{table:d3d character}
\end{table}

We focus on the four bands around the band gap in magnetic TI sandwiches or MnBi$_2$Te$_4$, and introduce the four-by-four Dirac matrices, defined as 
\begin{equation}
\begin{split}
    &\Gamma_0 = \sigma_3, \Gamma_1 = -\sigma_1 \otimes s_2, \Gamma_2 =  \sigma_1 \otimes s_1, \Gamma_3 = \sigma_2, \Gamma_4 = \sigma_1 \otimes s_3 \\ &\Gamma_{01} = \sigma_2 \otimes s_2, \Gamma_{02} = -\sigma_2 \otimes s_1, \Gamma_{03} = -\sigma_1, \Gamma_{04} = \sigma_2 \otimes s_3, \Gamma_{12} = s_3 \\ & \Gamma_{13} = \sigma_3 \otimes s_2, \Gamma_{14} = s_1, \Gamma_{23} = -\sigma_3 \otimes s_1, \Gamma_{24} = s_2, \Gamma_{34} - \sigma_3 \otimes s_3  
\end{split}
\end{equation}
where $\sigma$, $s$ are two sets of Pauli matrices representing orbital and spin degrees of freedoms, respectively. The basis wavefunctions at the $\Gamma$ point are given by the bonding states of four Bi atomic layers  $\{|P1^{'+}_{z},\uparrow \rangle, |P1^{'+}_{z},\downarrow \rangle\}$, and the antibonding states of four Te atomic layers $\{|P2^{'-}_{z},\uparrow \rangle, |P2^{'-}_{z},\downarrow \rangle\}$ in one unit cell \cite{zhang2019topological,wang2020dynamical,zhang2020mobius}. The symmetry generators of the $D_{3d}$ group and the anti-unitary operator $\hat{S}$ on the basis $\{|P1^{'+}_{z},\uparrow \rangle,|P2^{'-}_{z},\uparrow \rangle, |P1^{'+}_{z},\downarrow \rangle, |P2^{'-}_{z},\downarrow \rangle\}$ are given by
\begin{eqnarray}
    \Hat{C}_{3z} = e^{i \frac{\pi}{3} s_3}, \quad \Hat{P} = \sigma_3, \quad \Hat{C}_{2x} =  i s_1 \sigma_3, \quad \Hat{S} = i s_2 K,
\end{eqnarray}
where $K$ is the complex conjugate. The representations of the Dirac $\Gamma$-matrices, the polynomials of the momentum $\boldsymbol{k}$ and the strain tensors, as well as their behavior under the anti-unitary operator $\Hat{S}$ are given in Table \ref{table:Reps}. The direct products of irreducible representation (irreps) can be found in Table \ref{table:direct product}. The above irreps can be even or odd $S=\pm$ under $\Hat{S}$, which is called $S$-parity below. The $S$-parity for the direct product of the irreps follows the simple multiplication rules, e.g. $E_{g}(S=+) \times E_u (S=-) = A_{1u}(S=-) \bigoplus A_{2u}(S=-) \bigoplus E_{u}(S=-)$.

\begin{table}[h!]
\centering
\begin{tabular}{||c c c||} 
 \hline
  & Representation & S  \\ [0.5ex] 
 \hline\hline
 $I, \Gamma_0$ & $A_{1g}$ & + \\ \hline
 $\{\Gamma_1,\Gamma_2\}$ & $E_u$ & \textemdash\\ 
 \hline
 $\Gamma_3$ & $A_{2u}$ & \textemdash\\ 
 \hline
 $\Gamma_4$ & $A_{1u}$ & \textemdash\\ 
 \hline
 $\{ \Gamma_{01}, \Gamma_{02} \}$ & $E_u$ & + \\ \hline
 $\Gamma_{03}$ & $A_{2u}$ & +\\ 
 \hline
 $\Gamma_{04}$ & $A_{1u}$ & +\\ 
 \hline
 $\Gamma_{12}, \Gamma_{34}$ & $A_{2g}$ & \textemdash \\ \hline
 $\{ \Gamma_{13}, \Gamma_{23} \}, \{ \Gamma_{24},\Gamma_{14} \}$ & $E_g$ & \textemdash \\ \hline
 $\{k_x,k_y\}$ & $E_u$ & \textemdash\\
 \hline $k_z$ & $A_{2u}$ & \textemdash \\
 \hline$\{u_{xz},u_{yz}\}, \{2 u_{xy}, u_{xx} - u_{yy} \}$ & $E_g$ & +\\
 \hline$u_{zz},u_{xx}+u_{yy}$ & $A_{1g}$ & +\\
 \hline
\end{tabular}
\caption{Representations of $\Gamma$, \textbf{k} and $u_{ij}$} 
\label{table:Reps}
\end{table}

\begin{table}[h]
\centering
\begin{tabular}{|c || c | c | c | c | c | c|} 
 \hline
   & $A_{1g}$ & $A_{2g}$ & $E_g$ & $A_{1u}$ & $A_{2u}$ & $E_u$ \\ [0.5ex] 
 \hline\hline
 $A_{1g}$ & $A_{1g}$ & $A_{2g}$ & $E_g$ & $A_{1u}$ & $A_{2u}$ & $E_u$ \\  
 \hline
 $A_{2g}$ & $A_{2g}$ & $A_{1g}$ & $E_g$ & $A_{2u}$ & $A_{1u}$ & $E_u$ \\  
 \hline
 $E_{g}$ & $E_{g}$ & $E_{g}$ & $A_{1g} \bigoplus A_{2g} \bigoplus E_g $ & $E_u$ & $E_{u}$ & $A_{1u} \bigoplus A_{2u} \bigoplus E_u$ \\  
 \hline
 $A_{1u}$ & $A_{1u}$ & $A_{2u}$ & $E_u$ & $A_{1g}$ & $A_{2g}$ & $E_g$ \\  
 \hline
 $A_{2u}$ & $A_{2u}$ & $A_{1u}$ & $E_u$ & $A_{2g}$ & $A_{1g}$ & $E_g$ \\  
 \hline
 $E_{u}$ & $E_{u}$ & $E_u$ & $A_{1u} \bigoplus A_{2u} \bigoplus E_u$ & $E_g$ & $E_g$ & $A_{1g} \bigoplus A_{2g} \bigoplus E_g$  \\  
 \hline
\end{tabular}
\caption{Direct products of the irreducible representations of the $D_{3d}$ group}
\label{table:direct product}
\end{table}

\subsection{Low-energy Effective Hamiltonian}\label{sec:effHSuppl}

The low-energy effective model of magnetic TI sandwiches and MnBi$_2$Te$_4$ is constructed using the symmetries $\Hat{C}_{3z}, \Hat{P}, \Hat{C}_{2x}$ and $\hat{T}/\Hat{S}$, which takes the form\cite{zhang2009topological,liu2010model}
\begin{equation}
    H_{0} = \left( m + m_1 \left(k_x^2 + k_y^2 \right) + m_2 k_z^2 \right) \Gamma_0 + v \left( k_x \Gamma_1 + k_y \Gamma_2 \right) + v_z k_z \Gamma_3,
\end{equation}
up to $k^2$ order in the basis $\{|P1^{'+}_{z},\uparrow \rangle,|P2^{'-}_{z},\uparrow \rangle, |P1^{'+}_{z},\downarrow \rangle, |P2^{'-}_{z},\downarrow \rangle\}$, where $\kk$ is the crystal momentum, and $m$, $m_1$, $m_2$, $v$ and $v_z$ are material dependent parameters. The quadratic term $m_1 (k_x^2 + k_y^2)$ sets the momentum cut-off $\Tilde{k}_c$, at which the linear term $v \Tilde{k}_c$ is of the same order as the quadratic term $m_1 \Tilde{k}_c^2$, e.g. $\Tilde{k}_c \sim v/m_1$. Here we have assumed that $m_2$ ($v_z$) is of the same order as $m_1$ ($v$). Therefore, when the momentum amplitude is smaller than $\Tilde{k}_c$, e.g. $|\kk|<\Tilde{k}_c$, we neglect quadratic and higher-order $\kk$ terms in the effective model below, so the effective Hamiltonian is of Dirac type.


\begin{table} [h]
\centering
\begin{tabular}{| c c | c c | c c |}
\hline
     $v$ & $3.1964 eV \mathring{A} $ & $a$ & $ 11 m^2/s^2$  & $d$ & $ 3 m^2/s^2$  \\ \hline  $v_3$ &
     $2.7023 eV \mathring{A}$ & $b$ & $ 3 m^2/s^2$  & $f$ & $ 13 m^2/s^2$  \\ \hline $m$ & $-0.1165 eV$ & $c$ & $ 8 m^2/s^2$  & $g$ & $ 0.7 m^2/s^2$  \\ 
     \hline
\end{tabular}
\caption{Material parameters}
\label{table:material}
\end{table}

Using Tables \ref{table:Reps} and \ref{table:direct product}, we write down all possible terms belonging to the identity irrep of the $D_{3d}$ group (invariant under the symmetry generators of $D_{3d}$ and the anti-unitary symmetry $\hat{T}/\hat{S}$) in the presence of electron - strain interactions as
\begin{eqnarray} \label{eqn:HLconstruct}
    H_{\Gamma} &  = & m \Gamma_0 + v \left( k_x \Gamma_1 + k_y \Gamma_2 \right) + v_3 k_z \Gamma_3 + C(\textbf{u}) I + M(\textbf{u}) \Gamma_0 + A_1 u_{zz} \left( k_x \Gamma_1 + k_y \Gamma_2 \right) + A_2 \left(u_{xx} + u_{yy}\right) \left( k_x \Gamma_1 + k_y \Gamma_2 \right)  \nonumber \\ & + & A_3 u_{zz} k_z \Gamma_3 + A_4 \left(u_{xx} + u_{yy}\right) k_z \Gamma_3 + A_5 \left( u_{xz} k_y - u_{yz} k_x \right) \Gamma_4 + A_6 \left( 2 u_{xy} k_y - \left( u_{xx} - u_{yy} \right) k_x \right) \Gamma_4  \nonumber \\ & + & B_1 \left( u_{xz} k_x + u_{yz} k_y  \right) \Gamma_3 + B_2 \left( 2 u_{xy} k_x + \left( u_{xx} - u_{yy} \right) k_y \right) \Gamma_3 + B_3 k_z \left( u_{xz} \Gamma_1 + u_{yz} \Gamma_2 \right)  \nonumber \\ & + & B_4 k_z \left( 2 u_{xy} \Gamma_1 + \left( u_{xx} - u_{yy} \right) \Gamma_2 \right)  + D_1 \Big[ u_{yz} \left( k_x \Gamma_1 - k_y \Gamma_2 \right) + u_{xz} \left( k_x \Gamma_2 + k_y \Gamma_1 \right) \Big] \nonumber \\
    & + & D_2 \Big[ \left( u_{xx} - u_{yy} \right) \left( k_x \Gamma_1 - k_y \Gamma_2 \right) + 2 u_{xy} \left( k_x \Gamma_2 + k_y \Gamma_1 \right) \Big]. 
\end{eqnarray}
We neglect higher order terms like $\mathcal{O}(k^2), \mathcal{O}(u_{ij}^2)$, etc., and by rearranging terms, the minimal Hamiltonian can be re-written as
\begin{equation} \label{eqn:HL}
    H_{\Gamma} = C(\textbf{u}) I + \left( m + M(\textbf{u}) \right) \Gamma_0 + v \left( k_x \Gamma_1 + k_y \Gamma_2 \right) + v_3 k_z \Gamma_3 +  k_j \Delta_a^j(\textbf{u}) \Gamma_a, \quad a=0,1,2,3; j=1,2,3 
\end{equation}
with
\begin{equation}\label{eq_SM:C_M_functions}
    C(\textbf{u}) = C_1 u_{yy} + C_2 \left( u_{xx} + u_{yy} \right), \quad M(\textbf{u}) = M_1 u_{zz} + M_2 \left( u_{xx} + u_{yy} \right)
\end{equation}
 where $k_{1,2,3} = k_{x,y,z}$ and $\Delta^j_a$ is a function of the strain tensor $u_{ij} = \frac{1}{2} \left( \partial_i u_j + \partial_j u_i \right)$, with $u_i$ being the $i^{th}$ component of the phonon displacement field $\textbf{u}$. In the language of general relativity, the $\Delta^j_a$ field is called the frame field (inverse of the triad) \cite{chandia1997topological,hughes2011torsional, hughes2013torsional, parrikar2014torsion}.  We have assumed Einstein's summation convention, which will be used for the remainder of the paper. The explicit form of the frame field is given by 
\begin{align}\label{eq_SM:Delta_functions}
    & \Delta^1_1 = A_1 u_{zz} + A_2 \left( u_{xx} + u_{yy} \right) + D_1 u_{yz} + D_2 \left(u_{xx} - u_{yy} \right), \nonumber \\ &\Delta^2_1 = D_1 u_{xz} + D_2 u_{xy}, \nonumber \\ &\Delta_1^3 = B_3 u_{xz} + B_4 u_{xy} \nonumber \\
    & \Delta^1_2 = D_1 u_{xz} + D_2 u_{xy}, \nonumber \\ & \Delta^2_2 = A_1 u_{zz} + A_2 \left( u_{xx} + u_{yy} \right) - D_1 u_{yz} - D_2 \left( u_{xx} - u_{yy} \right),\nonumber \\ & \Delta^3_2 = B_3 u_{yz} + B_4 \left( u_{xx} - u_{yy} \right) \nonumber \\
    & \Delta^1_3 = B_1 u_{xz} + B_2 u_{xy}, \nonumber \\ &\Delta^2_3 = B_1 u_{yz} + B_2 \left( u_{xx} - u_{yy} \right), \nonumber \\ & \Delta^3_3 = A_3 u_{zz} + A_4 (u_{xx} + u_{yy}) 
\end{align}
where $A_{1,2,3,4},B_{1,2,3,4},D_{1,2},M_{1,2}$ depend on material parameters. 

\begin{figure}[h]
 \centering
 \includegraphics[width=0.5\textwidth]{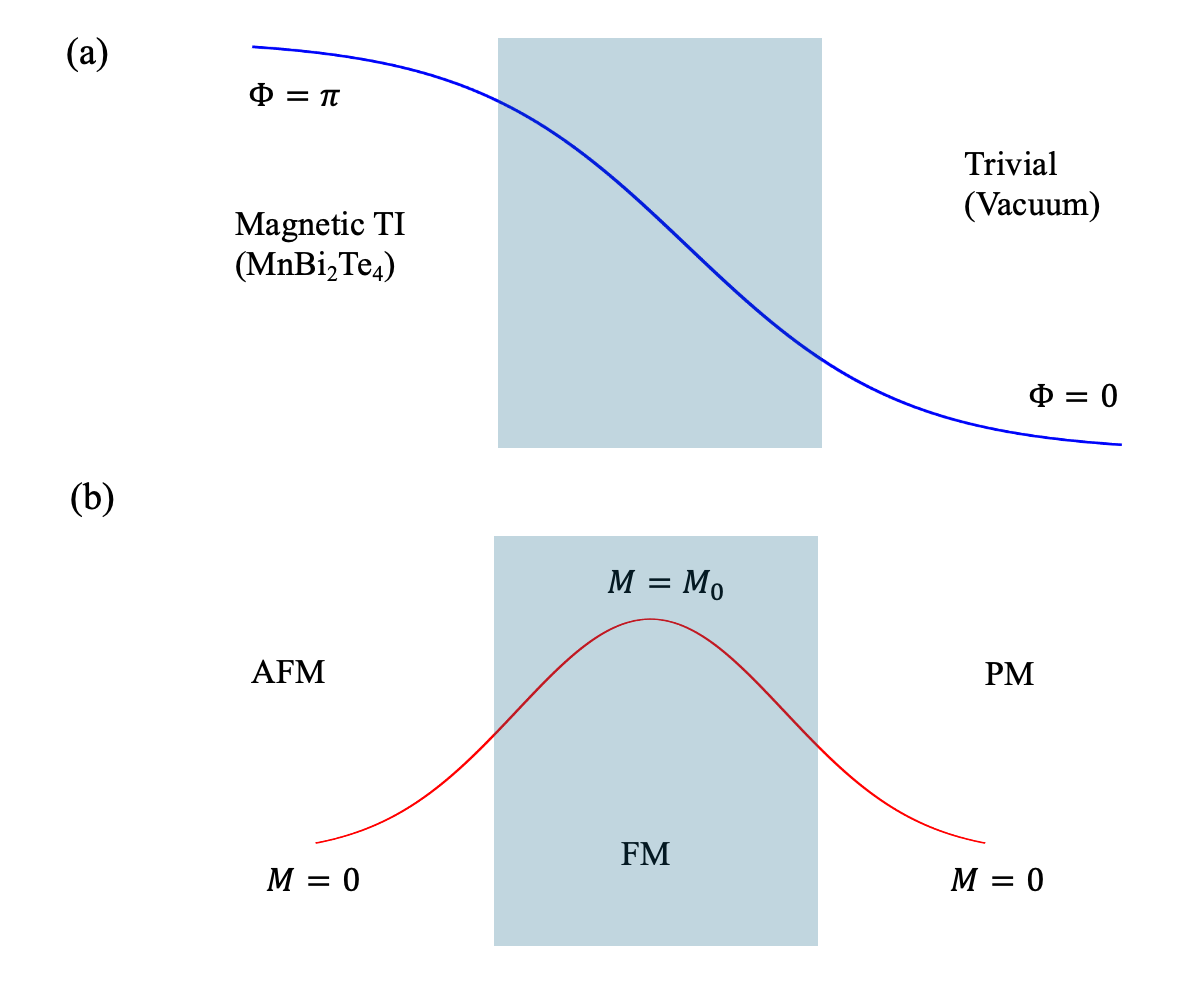}
 \caption{ The adiabatic evolution of (a) the Dirac complex angle $\Phi$ and the (b) magnetization $M$ from the magnetic topological insulator side to the trivial side. }
 \label{fig:Domain Wall Setup}
\end{figure}

In the Hamiltonian (\ref{eqn:HL}), the Dirac mass $m$ is a real number in the bulk due to the anti-unitary $\hat{T}/\hat{S}$ symmetry or $\hat{P}$ symmetry, and is expected to possess opposite signs between topologically non-trivial and trivial phases \cite{zhang2019topological,wang2020dynamical}. However, both the $\hat{P}$ symmetry or the $\hat{T}/\hat{S}$ symmetry is broken at the surface along the z direction in magnetic TI sandwiches and MnBi$_2$Te$_4$, and thus the Dirac mass is generally expected to be a complex number across the surface between magnetic TI sandwiches or MnBi$_2$Te$_4$ and the vacuum. In order to describe this $\hat{T}/\hat{S}$ symmetry breaking at the surface, we introduce $\Phi$ as the phase angle of the complex Dirac mass by rewriting $m \Gamma_0$ in Eq.(\ref{eqn:HL}) as 
\begin{eqnarray}
m \Gamma_0 \rightarrow |m| \Gamma_0 \cos \Phi + |m| \sin \Phi \Gamma_{4}  = |m| \Gamma_0 \left( \cos \Phi + i \sin \Phi \Gamma_{04}  \right) = |m| \Gamma_0 e^{i \Phi \Gamma_{04}} 
\end{eqnarray}
in Eq.(\ref{eqn:HL}). 
Due to both $\hat{P}$ and $\hat{T}/\hat{S}$ symmetries, the complex angle $\Phi$ is restricted to either $\Phi = 0$ (trivial phase) or $\Phi = \pi$ (topological phase) in the bulk \cite{zhang2019topological} However, around the surface/ interface of magnetic TI sandwiches or MnBi$_2$Te$_4$, both $\hat{P}$ and $\hat{S}$ symmetries are broken, and thus the complex mass term is allowed. In order to derive an effective field theory across the surface of magnetic TI sandwiches or MnBi$_2$Te$_4$, we can consider the adiabatic evolution of the $\Phi$ field across the surface of magnetic TI sandwiches or MnBi$_2$Te$_4$, as shown in Fig. \ref{fig:Domain Wall Setup}. We assume the variation of the $\Phi$ field is slow enough, so that we can locally treat the fluctuation of mass angle around a constant value $\Phi_0$ as a perturbation, e.g., $\Phi = \Phi_0 + \delta \Phi$, and the complex mass becomes 
\begin{align}
    |m| \Gamma_0 e^{i \Phi \Gamma_{04}} &= |m| \Gamma_0 e^{i (\Phi_0 + \delta \Phi) \Gamma_{04}} \nonumber \\
    &= |m| \Gamma_0 e^{i \Phi_0 \Gamma_{04}} \left( \cos \delta \Phi  + i \sin \delta \Phi \Gamma_{04} \right) \nonumber \\
     &= |m| \cos \delta \Phi \Tilde{\Gamma}_0 + \sin \delta \Phi \Tilde{\Gamma}_4  
     \nonumber \\
     &\approx |m| \Tilde{\Gamma}_0 + |m| \delta \Phi \Tilde{\Gamma}_4
\end{align}
where 
\begin{equation}
    \Tilde{\Gamma}_0 = \Gamma_0 e^{i \Phi_0 \Gamma_{04}}, \quad \Tilde{\Gamma}_4 = \Gamma_4 e^{i \Phi_0 \Gamma_{04}}
\end{equation}
For the rest of the paper, we will re-define $\Tilde{\Gamma}_0 \rightarrow \Gamma_0, \Tilde{\Gamma}_4 \rightarrow \Gamma_4$, $\delta \Phi \rightarrow \Phi$ and the Dirac mass term $m \Gamma_0$ in Eq.(\ref{eqn:HL}) becomes 
\begin{eqnarray}
    m \Gamma_0 \rightarrow |m| \Gamma_0 + |m|  \Phi \Gamma_4.
\end{eqnarray}

\subsection{Effective action $S_{eff}$} \label{sec:effaction}

From the Hamiltonian $H_{\Gamma}$ in Eq. (\ref{eqn:HL}), we write down the effective action as 
\begin{equation}
    S_{\Gamma} = \int_0^\beta d\tau \int d^3 r \Big[ \psi_{\Gamma}^\dagger \frac{\partial}{\partial \tau} \psi_\Gamma + \psi_L^\dagger H_\Gamma \psi_\Gamma \Big] = S_{\Gamma,0} + S_{\Gamma,1}(\textbf{u})
\end{equation}
where $\tau$ is the imaginary time and
\begin{align}
    S_{\Gamma,0} &= \int_0^\beta d\tau \int d^3 r \Big[ \psi^\dagger_\Gamma \left( \partial_\tau + m\Gamma_0 + v \Gamma_1 (-i \partial_1) + v \Gamma_2 (-i \partial_2) + v_3 \Gamma_3 (-i \partial_3) \right) \psi_\Gamma \Big] \nonumber \\
    S_{\Gamma,1} &= \int_0^\beta d\tau \int d^3 r \Big[ \psi^\dagger_\Gamma \left( C(\textbf{u}) + M(\textbf{u}) \Gamma_0 \right) \psi_\Gamma  +  \frac{1}{2} \psi^\dagger_\Gamma \Delta_a^j(\textbf{u}) \Gamma_a (-i \partial_j \psi_\Gamma) + \frac{1}{2} (i \partial_j \psi_\Gamma^\dagger) \Delta_a^j(\textbf{u}) \Gamma_a \psi_\Gamma + |m|  \psi_\Gamma^\dagger \Phi \Gamma_4  \psi_\Gamma \Big],
\end{align}
where $a=1,2,3$ and $j= 1,2,3$. Here we have taken the replacement $k_i \rightarrow - i\partial_i$ and $k_i \Delta_a^i \rightarrow \frac{1}{2} \{-i\partial_i, \Delta^i_a \}$ to write the action in the real space. We define the Fourier transform as
\begin{align}
  \psi_\Gamma (\textbf{r},\tau) &= \frac{1}{\beta V} \sum_{i \omega, \textbf{k}} e^{i \textbf{k}\cdot \textbf{r}-i\omega \tau} \psi_\Gamma(\textbf{k},i\omega) \nonumber \\
  C (\textbf{r},\tau) &= \frac{1}{\beta V} \sum_{i \nu, \textbf{q}} e^{i \textbf{q}\cdot \textbf{r}-i\nu \tau} C(\textbf{q},i\nu) \nonumber \\
  M (\textbf{r},\tau) &= \frac{1}{\beta V} \sum_{i \nu, \textbf{q}} e^{i \textbf{q}\cdot \textbf{r}-i\nu \tau} M(\textbf{q},i\nu) \nonumber \\
  \Delta_a^j (\textbf{r},\tau) &= \frac{1}{\beta V} \sum_{i \nu, \textbf{q}} e^{i \textbf{q}\cdot \textbf{r}-i\nu \tau} \Delta_a^j(\textbf{q},i\nu)  \nonumber \\
  \Phi(\textbf{r},\tau) &=\frac{1}{\beta V} \sum_{i \nu, \textbf{q}} e^{i \textbf{q}\cdot \textbf{r} - i \nu \tau} \Phi(\textbf{q},i \nu)  
\label{eq:FieldFourier}
\end{align}
and in the momentum space, we have
\begin{equation}
    S_{\Gamma} = S_{\Gamma,0} + S_{\Gamma,1} = \sum_k \psi_k^\dagger \left( - \mathcal{G}_0^{-1}\right) \psi_k + \sum_{k,k'} \psi_k^\dagger \mathcal{\chi}(k,k') \psi_{k'} 
\end{equation}
where \begin{equation}
    \mathcal{G}_0 = \left( i\omega - |m| \Gamma_0 - v (k_x \Gamma_1 + k_y \Gamma_2) - v_z k_z \Gamma_3 \right)^{-1}
\end{equation}
and 
\begin{equation}\label{eq_SM:chi_vertex_form_1}
    \mathcal{\chi}(k,k') = C(q=k'-k) + M(q=k'-k) + \mathcal{T}_i^a(k,k') \Delta_a^i(q=k'-k) + |m| \Phi (q = k'-k) \Gamma_4
\end{equation}
where 
\begin{eqnarray}\label{eq_SM:stress_energy_tensor_1}
    \mathcal{T}_i^a(k,k')=\frac{1}{2}(k_i + k_i') \Gamma_a
\end{eqnarray}
is the stress-energy tensor and $\sum_k = \frac{1}{\beta V} \sum_{i\omega, \textbf{k}}$. We integrate the electron degrees to obtain the effective action, $W(\textbf{u})$, which is defined by 
\begin{eqnarray}
    W(\textbf{u}) = - \text{ln} (\mathcal{Z}/\mathcal{Z}_0), 
\end{eqnarray}
where 
\begin{equation}
    \mathcal{Z} = \int \mathcal{D}\psi^\dagger \mathcal{D} \psi e^ {-S} = \text{Det}\left(-\mathcal{G}_0^{-1}+\mathcal{\chi} \right) 
\end{equation}
and 
\begin{equation}
    \mathcal{Z}_0 = \int \mathcal{D}\psi^\dagger \mathcal{D} \psi e^ {-S_0} = \text{Det}\left(-\mathcal{G}_0^{-1}\right).
\end{equation}
The effective action $W(\textbf{u})$ can be expanded as
\begin{align}
    W &= - \text{ln} \big[ \text{Det} \left(-\mathcal{G}_0^{-1} + \mathcal{\chi} \right) \big] + \text{ln} \big[ \text{Det} \left(-\mathcal{G}_0^{-1} \right) \big] \nonumber \\
    &= -\text{Tr}\big[\text{ln}\left(-\mathcal{G}_0^{-1} + \mathcal{\chi}\right) - \text{ln}\left(-\mathcal{G}_0^{-1} \right)  \big] \nonumber \\
    &= -\text{Tr}\big[ \text{ln} \left( 1- \mathcal{\chi} \mathcal{G}_0 \right) \big] \nonumber \\
    &= \sum_n \frac{1}{n} \text{Tr} \big[ \left( \mathcal{\chi} \mathcal{G}_0 \right)^n \big] = \sum_n W_n, \quad W_n = \frac{1}{n} \text{Tr} \big[ \left( \mathcal{\chi} \mathcal{G}_0 \right)^n \big]. 
\end{align}

\begin{figure*}
 \includegraphics[width=\textwidth]{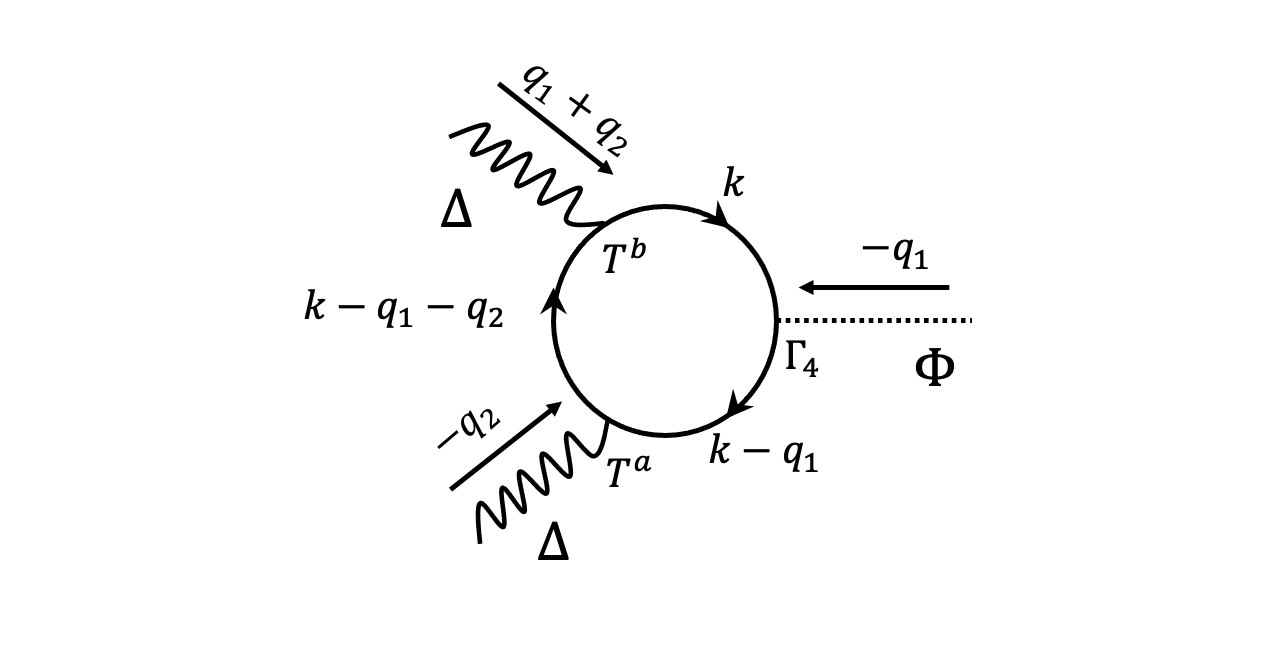}
 \caption{The triangle Feynman diagram, which is the leading order contribution to the interface}
 \label{fig:triangle diagram}
 \end{figure*}
 
We are interested in the terms that involve the mass angle $\Phi$, which varies from $\pi$ inside MnBi$_2$Te$_4$ to $0$ outside MnBi$_2$Te$_4$ - vacuum or trivial insulator around the surface/interface region.  The $\Phi$ field is included in the last term of $\chi$ in Eq.(\ref{eq_SM:chi_vertex_form_1}) and is accompanied by the $\Gamma_4$ matrix. According to the Dirac $\Gamma$-matrix algebra, we have 
\begin{eqnarray} \label{eq_SM:Gamma_matrix_algebra_1}
  \text{Tr}( \Gamma_4) = \text{Tr} (\Gamma_4 \Gamma_\mu) = \text{Tr} (\Gamma_4 \Gamma_\mu \Gamma_\nu) = \text{Tr} (\Gamma_4 \Gamma_\mu \Gamma_\nu \Gamma_\sigma) = 0; \quad
  \text{Tr}(\Gamma_4 \Gamma_\mu \Gamma_\nu \Gamma_\sigma \Gamma_\rho) = 4 \epsilon^{\mu\nu\sigma\rho}, 
\end{eqnarray}
where $\mu,\nu,\rho,\sigma = 0,1,2,3$, and thus the term that involves $\Phi$ field should at least include additional four $\Gamma$ matrices so that the term $\Gamma_4 \Gamma_\mu \Gamma_\nu \Gamma_\sigma \Gamma_\rho$ can appear, which requires at least $n=3$ for $W_n$. The contributions of $W_1$ and $W_2$ do not involve the terms with $\Phi$ and thus merely provide corrections to the bulk phonon dynamics. Thus, we only focus on the third order term $W_3(\Delta,\Phi)$ here, including 
\begin{eqnarray}
    W_3 = W_3^{\Delta} + W_3^{CM} + W_3^{\Delta C} + W_3^{\Delta M},
\end{eqnarray}
where
\begin{eqnarray} \label{eq:W3_Delta}
    && W_3^\Delta = \sum_{q_1,q_2} \Pi_{ij}^{ab}(q_1,q_2) |m| \Phi(-q_1) \Delta_a^i(-q_2) \Delta_b^j(q_1+q_2)\\
    && W_3^{CM} = \sum_{q_1,q_2} \Pi^{CM}(q_1,q_2) |m| \Phi(-q_1) C(-q_2) M(q_1+q_2) \\ 
    && W_3^{\Delta C} = \sum_{q_1,q_2} \Pi^{\Delta C, a}_i(q_1,q_2) |m| \Phi(-q_1) \Delta_a^i (-q_2) C(q_1+q_2)  \\ 
    && W_3^{\Delta M} = \sum_{q_1,q_2} \Pi^{\Delta M, a}_i(q_1,q_2) |m| \Phi(-q_1) \Delta_a^i (-q_2) M(q_1+q_2)
\end{eqnarray}
and
\begin{eqnarray} \label{eq:Pi_Delta}
    && \Pi_{ij}^{\Delta,ab}(q_1,q_2) =  \sum_k \text{Tr} \Big[ \Gamma_4 \mathcal{G}_0(k-q_1) \mathcal{T}_i^a(k-q_1,k-q_1-q_2) \mathcal{G}_0(k-q_1-q_2)\mathcal{T}_j^b(k-q_1-q_2,k) \mathcal{G}_0(k) \Big] \\
    && \Pi^{CM}(q_1,q_2) = \sum_k \text{Tr} \Big[ \Gamma_4 \mathcal{G}_0(k-q_1) I \mathcal{G}_0(k-q_1-q_2) I \mathcal{G}_0(k) \Big] \label{eq:Pi_CM}  \\ 
    && \Pi^{\Delta C, a}_i(q_1,q_2) = \sum_k \text{Tr} \Big[ \Gamma_4 \mathcal{G}_0(k-q_1) \mathcal{T}_i^a(k-q_1,k-q_1-q_2) \mathcal{G}_0(k-q_1-q_2) I \mathcal{G}_0(k) \Big] \label{eq:Pi_Delta_C}  \\ 
    && \Pi^{\Delta M, a}_i(q_1,q_2) = \sum_k \text{Tr} \Big[ \Gamma_4 \mathcal{G}_0(k-q_1) \mathcal{T}_i^a(k-q_1,k-q_1-q_2) \mathcal{G}_0(k-q_1-q_2) I \mathcal{G}_0(k) \Big], \label{eq:Pi_Delta_M}
\end{eqnarray}
with $a, b, i,j = 1,2,3$. We consider the long wavelength limit, so $\Pi^{\Lambda}(q_1,q_2)$ can be expanded in series of $q_1,q_2$, namely, 
\begin{eqnarray}\label{eq_SM:Pi_expansion}
    \Pi^{\Lambda}(q_1,q_2) = \Pi^{\Lambda}(q_1=q_2=0) + {q}_{j,\mu} \cdot \left(\partial_{{q}_{j,\mu}} \Pi^{\Lambda}  \right)_{q_{1} = q_{2} =0} + q_{1,\mu} q_{2,\nu} \left(\partial_{{q}_{1,\mu}} \partial_{{q}_{2,\nu}}\Pi^{\Lambda}  \right)_{q_{1} = q_{2} =0}, 
\end{eqnarray}
up to second order in $q$ ,where $j=1,2$, $\mu,\nu=0,1,2,3$ and $\Lambda$ includes all the indices for $\Pi$, e.g. $\Delta, CM, \Delta C, \Delta M$, $a, b$ and $i, j$. $q_l$ is a 4-vector defined as $q_l = (q_{l,0} = i \omega_1, q_{l,x},q_{l,y},q_{l,z})$with $l=1,2$. The interface dynamics comes from the variation in $\Phi$ i.e., from $\partial_{v} \Phi$, where $v=1,2,3$, which means we must expand to at least one order in $q_{l,v}$. Thus, we neglect the zero-order term $\Pi^{\Lambda}(q_1=q_2=0)$ in Eq.(\ref{eq_SM:Pi_expansion}), which only modifies the bulk elastic modulii. The first order expansion term in Eq.(\ref{eq_SM:Pi_expansion}) provides the contributions with the form $\Phi \Delta_a^i \partial_\mu \Delta_b^j$, $\Phi C \partial_\mu M$ , $\Phi \Delta_a^i \partial_\mu C$ and $\Phi \Delta_a^i \partial_\mu M$ to the effective action $W_3$. However, $\Delta$, $C$, $M$ are the functions of the strain tensor $u_{ij}$ in Eqs.(\ref{eq_SM:C_M_functions}) and (\ref{eq_SM:Delta_functions}), which is even under both inversion symmetry $\hat{P}$ and anti-unitary symmetry $\hat{S}$, while the $\Phi$ field is odd under both $\hat{P}$ and $\hat{S}$, e.g. $\hat{P} \Phi \hat{P}^{-1} = - \Phi$ and $\hat{S} \Phi \hat{S}^{-1} = - \Phi$. Consequently, all these terms either break inversion $\hat{P}$ for $\mu = 0$ or break anti-unitary symmetry $\hat{S}$ for $\mu=1,2,3$, and thus they must vanish as the microscopic Hamiltonian of the system preserves both the symmetries $\hat{P}$ and $\hat{S}$. Therefore, we below focus on the second order term, 
\begin{eqnarray}
    \Pi^\Lambda_{\mu\nu} = \left(\partial_{{q}_{1,\mu}} \partial_{{q}_{2,\nu}}\Pi(q_1,q_2)^\Lambda  \right)_{q_{1} = q_{2} =0}. 
\end{eqnarray}

For the second-order expansion terms, Eqs.(\ref{eq:Pi_Delta}) to (\ref{eq:Pi_Delta_M}) give rise to the terms 
\begin{equation}\label{eq_SM:Pi_Delta_1}
    \Pi_{ij\mu\nu}^{\Delta,ab} = \Pi_{ij0l}^{\Delta,ab (1)} + \Pi_{ij0l}^{\Delta,ab (2)} + \Pi_{ij0l}^{\Delta,ab (3)}
\end{equation}
where \begin{align}\label{eq_SM:Pi_Delta_2}
    \Pi_{ij0l}^{\Delta,ab (1)} &= N^{(1)}\sum_k \text{Tr} \Big[\Gamma_4 \partial_
    {q_{1,0}}\mathcal{G}_0(k-q_1) \mathcal{T}_i^a(k-q_1,k-q_1-q_2)\partial_
    {q_{2,l}} \mathcal{G}_0(k-q_1-q_2)\mathcal{T}_j^b(k-q_1-q_2,k) \mathcal{G}_0(k) \Big]_{q_{1}=q_{2}=0} \nonumber \\
    \Pi_{\Delta,ij0l}^{ab (2)} &= N^{(2)} \sum_k \text{Tr} \Big[ \Gamma_4 \partial_
    {q_{1,0}}\mathcal{G}_0(k-q_1) \partial_
    {q_{2,l}} \mathcal{T}_i^a(k-q_1,k-q_1-q_2) \mathcal{G}_0(k-q_1-q_2)\mathcal{T}_j^b(k-q_1-q_2,k) \mathcal{G}_0(k) \Big]_{q_{1}=q_{2}=0} \nonumber \\
    \Pi_{\Delta,ij0l}^{ab (3)} &= N^{(3)} \sum_k \text{Tr} \Big[ \Gamma_4 \mathcal{G}_0(k-q_1)  \mathcal{T}_i^a(k-q_1,k-q_1-q_2) \partial_
    {q_{1,0}}\partial_
    {q_{2,l}} \mathcal{G}_0(k-q_1-q_2)\mathcal{T}_j^b(k-q_1-q_2,k) \mathcal{G}_0(k) \Big]_{q_{1}=q_{2}=0} 
\end{align}
with $N^{(1,2,3)}$ to be the symmetry factors, arising due to swapping $q_1$ with $q_2$, $\mu = 1,2,3$ with $\mu = 0$ etc. We have $N^{(1)} = 2, N^{(2)} =1 ,N^{(3)} = 2$
and 
\begin{eqnarray}
    & \Pi^{CM}_{\mu \nu} = N^{CM}_1\sum_k \text{Tr} \Big[\Gamma_4 \partial_
    {q_{1,\mu}}\mathcal{G}_0(k-q_1) I \partial_
    {q_{2,\nu}} \mathcal{G}_0(k-q_1-q_2) I \mathcal{G}_0(k) \Big]_{q_{1}=q_{2}=0} \label{eq_SM:Pi_CM_1} \\
    & \Pi^{\Delta C}_{\mu \nu} = N^{\Delta C}_1\sum_k \text{Tr} \Big[\Gamma_4 \partial_
    {q_{1,\mu}}\mathcal{G}_0(k-q_1) \mathcal{T}_i^a(k-q_1,k-q_1-q_2)\partial_
    {q_{2,\nu}} \mathcal{G}_0(k-q_1-q_2) I \mathcal{G}_0(k) \Big]_{q_{1}=q_{2}=0} \label{eq_SM:Pi_Delta_C_1} \\
    & \Pi^{\Delta C, a}_{i,\mu \nu} = N^{\Delta C}_2 \sum_k \text{Tr} \Big[ \Gamma_4 \mathcal{G}_0(k-q_1)  \mathcal{T}_i^a(k-q_1,k-q_1-q_2) \partial_
    {q_{1,\mu}}\partial_
    {q_{2,\nu}} \mathcal{G}_0(k-q_1-q_2) I \mathcal{G}_0(k) \Big]_{q_{1}=q_{2}=0}. 
    \label{eq_SM:Pi_Delta_M_1}
\end{eqnarray}
where $I$ is the $4\times 4$ identity matrix. The Green's function and its first and second derivatives in the above expressions are given by \begin{align} \label{eq:Gderivatives}
    \mathcal{G}_0(k-q_1-q_2)|_{q_{1}=q_{2}=0}&= \frac{\left( i \omega + |m| \Gamma_0 + k_p v_p \Gamma_p\right)}{D^2}  \\
    \partial_{q_{1,0}} \mathcal{G}_0(k-q_1)|_{q_{1}=q_{2}=0} &= \frac{1}{D^4} \left( (i\omega)^2 + \epsilon_0^2 + 2 i\omega \left( |m| \Gamma_0 + vk_x \Gamma_1 + v k_y \Gamma_2 + v_z k_z \Gamma_3 \right) \right) \label{eq:Gderivatives_2} \\
    \partial_{q_{2,l}} \mathcal{G}_0(k-q_1-q_2)|_{q_{1}=q_{2}=0} &= -\frac{v_l \Gamma_l}{D^2} -\frac{2 v_l^2 k_l}{D^4} \left( i \omega + |m|\Gamma_0 + vk_x \Gamma_1 + v k_y \Gamma_2 + v_z k_z \Gamma_3  \right) \label{eq:Gderivatives_3} \\
    \partial_
    {q_{1,0}}\partial_
    {q_{2,l}} \mathcal{G}_0(k-q_1-q_2)|_{q_{1}=q_{2}=0} &= \frac{2 v_l^2 k_l}{D^4} - \frac
    {2 i\omega}{D^4} v_l \Gamma_l - \frac{8 i\omega v_l^2 k_l}{D^6} \left( i \omega + |m|\Gamma_0 + vk_x \Gamma_1 + v k_y \Gamma_2 + v_z k_z \Gamma_3 \right) \label{eq:Gderivatives_4}
\end{align}
where $\epsilon_0^2 = m^2 + v^2 k_{\parallel}^2 + v_z^2 k_z^2 $, $D^2 = (i \omega)^2 - \epsilon_0^2$, $v_{p} = v$ if $p = 1,2$ and $l=1,2,3$. Our task below is to evaluate Eqs.(\ref{eq_SM:Pi_Delta_1}), (\ref{eq_SM:Pi_Delta_2}),(\ref{eq_SM:Pi_CM_1}) (\ref{eq_SM:Pi_Delta_C_1}) and (\ref{eq_SM:Pi_Delta_M_1}). 
We first consider the terms involving $C$ and $M$ in Eqs.(\ref{eq_SM:Pi_CM_1}). To ensure $\Pi^\Lambda_{\mu \nu} \neq 0$, two conditions need to be satisfied. (1) $\Pi^\Lambda_{\mu \nu}$ must be even in $\kk$ and $i\omega$, and otherwise the summation over $\kk$ and $i\omega$ will make it zero; (2) In Eqs.(\ref{eq_SM:Pi_Delta_2})-(\ref{eq_SM:Pi_Delta_M_1}), there is one $\Gamma_4$ matrix together with several $\mathcal{G}_0$'s and $\mathcal{T}_i^a$'s inside the $Tr$ operator, and from Eq.(\ref{eq_SM:Gamma_matrix_algebra_1}), a non-zero $\Pi^\Lambda_{\mu \nu}$ requires that $\mathcal{G}_0$'s and $\mathcal{T}_i^a$'s together provides the product of four different $\Gamma$ matrices, namely $\Gamma_0\Gamma_1\Gamma_2\Gamma_3$, since both $\mathcal{G}_0$'s and $\mathcal{T}_i^a$'s do not contain the $\Gamma_4$ matrix.

We first apply the above two conditions to show that $\Pi^{CM}_{\mu \nu}$ in Eq.(\ref{eq_SM:Pi_CM_1}) must be zero. For a non-zero value of $\Pi^{CM}_{\mu \nu}$, we need four $\Gamma$-matrices $\Gamma_{0,1,2,3}$ from three $\mathcal{G}_0$'s according to the condition (2). However, this is impossible since each $\mathcal{G}_0$ only contains at most one $\Gamma$-matrix in Eqs.(\ref{eq:Gderivatives})-(\ref{eq:Gderivatives_4}). Thus, we must have $\Pi^{CM}_{\mu \nu} = 0$. 

We next evaluate $\Pi^{\Delta C}_{\mu \nu}$, and focus on the $\mu = 0$ component, $\Pi^{\Delta C}_{\mu=0, \nu}$, in which there are three $\mathcal{G}_0$'s and one $\mathcal{T}_i^a$, in addition to $\Gamma_4$ matrix, inside the trace. According to the condition (2), one should choose the terms with $\Gamma$ matrices in all these three $\mathcal{G}_0$'s and one $\mathcal{T}_i^a$ in order to get four $\Gamma$ matrices. From Eq.(\ref{eq:Gderivatives_2}), one can see any term with the $\Gamma$ matrix in $\partial_{q_{1,0}} \mathcal{G}_0(k-q_1)|_{q_{1}=q_{2}=0}$ for $\Pi^{\Delta C}_{\mu=0, \nu}$ must be accompanied with one $i\omega$. To satisfy the condition (1), we require another term with $i\omega$ from the remaining part, namely $\mathcal{T}_i^a(k-q_1,k-q_1-q_2)\partial_{q_{2,\nu}} \mathcal{G}_0(k-q_1-q_2)  \mathcal{G}_0(k)$, inside the trace. However, $\mathcal{T}_i^a(k-q_1,k-q_1-q_2)$ does not contain $i\omega$, while the $i\omega$ terms in $\mathcal{G}_0(k)$ and $\partial_{q_{2,l=1,2,3}} \mathcal{G}_0(k-q_1-q_2)|_{q_{1}=q_{2}=0}$ do not contain any $\Gamma$ matrix, thus violating the condition (2). Therefore, the only possible non-zero term is $\Pi^{\Delta C}_{\mu=0, \nu=0}$. For $\Pi^{\Delta C}_{\mu=0, \nu=0}$, we find that in order to get $\Gamma_0\Gamma_1\Gamma_2\Gamma_3$, we will have odd number of $k_i$ ($i=x,y,z$) and thus still have $\Pi^{\Delta C}_{\mu=0, \nu=0}=0$. This arrives at the conclusion that $\Pi^{\Delta C}_{\mu=0, \nu}=0$ for any $\nu$. 
Similar argument can be applied to the case with $\mu = l = 1,2,3$, and $\Pi^{\Delta C}_{l \nu}$ is odd in $k_i$ for $\nu = 0,1,2,3$, leading to $\Pi^{\Delta C}_{\mu \nu} = 0$. We note from Eqs(\ref{eq:Pi_Delta_C}) and (\ref{eq:Pi_Delta_M}) that $\Pi^{\Delta M, a}_{i,\mu \nu} =  \Pi^{\Delta C, a}_{i,\mu \nu}$. So $\Pi^{\Delta M, a}_{i,\mu \nu} = 0$
. Next we focus on the evaluations of $\Pi_{ij,\mu\nu}^{\Delta,ab}$ in Eq.(\ref{eq_SM:Pi_Delta_1}) and (\ref{eq_SM:Pi_Delta_2}), which corresponds to the one-loop Feynman diagram - also known as the triangle diagram - in Fig. \ref{fig:triangle diagram}. 
Based on similar arguments,  we find $\Pi^{CM}_{\mu \nu} = \Pi^{\Delta C, a}_{i,\mu \nu}=\Pi^{\Delta M, a}_{i,\mu \nu}=0$, and 
$\Pi^{\Delta,ab}_{ij\mu\nu}$ is nonzero only when $\mu =0, \nu =l$ and $\mu =l, \nu =0$. Below we evaluate each term in Eq.(\ref{eq_SM:Pi_Delta_1}) for $\Pi_{ij\mu\nu}^{\Delta,ab}$. We first consider the term $\Pi_{\Delta,ij0l}^{ab(1)}$, which is derived as 
\begin{align}\label{eqn:pi0l1}
    \Pi_{\Delta,ij0l}^{ab(1)} &= \sum_k \text{Tr} \Big[ \Gamma_4 \frac{(i \omega)^2 + \epsilon_0^2 + 2i \omega \left( |m| \Gamma_0 + v_{p_{1}} k_{p_{1}} \Gamma_ {p_{1}} \right) }{D^4} k_i \Gamma_a \left( -\frac{v_l \Gamma_l}{D^2} - \frac{2 v_l^2 k_l}{D^4} \left(  i\omega + |m|\Gamma_0 + v_{p_{2}} k_{p_{2}} \Gamma_{p_{2}}\right) \right) \nonumber \\
    & k_j \Gamma_b \frac{\left( i \omega + |m| \Gamma_0 + k_{p_{3}} v_{p_{3}} \Gamma_{p_{3}}\right)}{D^2}
    \Big] \nonumber \\
    &= -\sum_k \frac{k_ik_j 2 (i \omega)^2 |m| k_l k_p v_p v_l^2}{D^{10}} \text{Tr} \Big[ \Gamma_4 \Gamma_0 \Gamma_a \Gamma_b \Gamma_p\Big] -\sum_k \frac{k_ik_j 2 (i \omega)^2 |m| k_l k_p v_p v_l^2}{D^{10}} \text{Tr} \Big[ \Gamma_4 \Gamma_0 \Gamma_a \Gamma_p \Gamma_b\Big]  \nonumber \\ &-m \sum_k \frac{k_i k_j v_l 2(i \omega)^2 }{D^8} \text{Tr} \Big[ \Gamma_4 \Gamma_0 \Gamma_a \Gamma_l \Gamma_b \Big] - |m| \sum_{k} \frac{k_i k_j \left( (i \omega)^2 + \epsilon_0^2 \right) 2 k_l v_l^2 k_p v_p}{D^{10}} \text{Tr} \Big[ \Gamma_4 \Gamma_a \Gamma_0 \Gamma_b \Gamma_p \Big] \nonumber \\
    & -|m| \sum_k \frac{k_i k_j 4 (i\omega)^2 k_l v_l^2 v_p k_p}{D^{10}}\text{Tr} \Big[  \Gamma_4 \Gamma_p \Gamma_a \Gamma_0 \Gamma_b\Big] - |m| \sum_k \frac{k_i k_j v_l \left( (i \omega)^2 +\epsilon_0^2\right) }{D^8} \text{Tr} \Big[ \Gamma_4 \Gamma_a \Gamma_l \Gamma_b \Gamma_0 \Big] \nonumber \\ &- |m| \sum_{k} \frac{k_i k_j \left( (i \omega)^2 + \epsilon_0^2 \right) 2 k_l v_l^2 k_p v_p}{D^{10}} \text{Tr} \Big[ \Gamma_4 \Gamma_a \Gamma_p \Gamma_b \Gamma_0 \Big]
    -|m| \sum_k \frac{k_i k_j 4 (i\omega)^2 k_l v_l^2 v_p k_p}{D^{10}}\text{Tr} \Big[  \Gamma_4 \Gamma_p \Gamma_a \Gamma_b \Gamma_0\Big] \nonumber \\
&= -\sum_k \frac{k_ik_j 2 (i \omega)^2 |m| k_l k_p v_p v_l^2}{D^{10}}  4 \epsilon^{abp} -\sum_k \frac{k_ik_j 2 (i \omega)^2 |m| k_l k_p v_p v_l^2}{D^{10}} 4 \epsilon^{apb} - |m| \sum_k \frac{k_i k_j v_l 2(i \omega)^2 }{D^8}  4 \epsilon^{alb}  \nonumber \\ &-  |m| \sum_{k} \frac{k_i k_j \left( (i \omega)^2 + \epsilon_0^2 \right) 2 k_l v_l^2 k_p v_p}{D^{10}}  4 \epsilon^{bap} - |m| \sum_k \frac{k_i k_j 4 (i\omega)^2 k_l v_l^2 v_p k_p}{D^{10}}4 \epsilon^{pab} - |m|\sum_k \frac{k_i k_j v_l \left( (i \omega)^2 +\epsilon_0^2\right) }{D^8} 4 \epsilon^{lab} \nonumber \\ &- |m| \sum_{k} \frac{k_i k_j \left( (i \omega)^2 + \epsilon_0^2 \right) 2 k_l v_l^2 k_p v_p}{D^{10}} 4 \epsilon^{pab}
    -|m| \sum_k \frac{k_i k_j 4 (i\omega)^2 k_l v_l^2 v_p k_p}{D^{10}}4 \epsilon^{apb} \nonumber \\
&= -\sum_k \frac{k_ik_j 2 (i \omega)^2 |m| k_l k_p v_p v_l^2}{D^{10}} \left( 4 \epsilon^{abp}  + 4 \epsilon^{apb} \right)  -4 |m| \sum_k \frac{k_i k_j v_l  }{D^8}  \left( 2(i \omega)^2 \epsilon^{alb} + ((i\omega)^2 + \epsilon_0^2)  \epsilon^{lab} \right) \nonumber \\ &-  |m| \sum_{k} \frac{k_i k_j \left( (i \omega)^2 + \epsilon_0^2 \right) 2 k_l v_l^2 k_p v_p}{D^{10}}  \left( 4 \epsilon^{bap} + 4 \epsilon^{pab} \right)  -|m| \sum_k \frac{k_i k_j 4 (i\omega)^2 k_l v_l^2 v_p k_p}{D^{10}}\left( 4 \epsilon^{pab} + 4 \epsilon^{apb} \right) \nonumber \\ & = -4 |m| \sum_k \frac{k_i k_j v_l }{D^8}  \left( ((i\omega)^2 - \epsilon_0^2)  \epsilon^{alb} \right) = -4 |m| \sum_k \frac{k_i k_j  v_l}{D^6}   \epsilon^{alb}
\end{align}
where we have used the identities $\text{Tr}\Big[ \Gamma_4 \Gamma_0 \Gamma_a \Gamma_b \Gamma_c \Big] = 4 \epsilon^{abc}$ and the cyclic and anti-symmetric nature of the Levi-Civita tensor $\epsilon^{pqr} = \epsilon^{rpq} =  -\epsilon^{qpr} $. Next, we have
\begin{align}
    \Pi_{\Delta,ij0l}^{ab(2)} &= \sum_k \text{Tr} \Big[ \Gamma_4 \frac{(i \omega)^2 + \epsilon_0^2 + 2i \omega \left( m \Gamma_0 + v_{p_{1}} k_{p_{1}} \Gamma_{p_{1}}\right) }{D^4} \left( -\frac{1}{2} \delta_{il} \Gamma_a \right) \frac{\left( i \omega + m \Gamma_0 + k_{p_{2}} v_{p_{2}} \Gamma_{p_{2}}\right)}{D^2} \nonumber \\
    &k_j \Gamma_b \frac{\left( i \omega + m \Gamma_0 + k_{p_{3}} v_{p_{3}} \Gamma_{p_{3}}\right)}{D^2}
    \Big] \nonumber \\
    &= -\frac{m}{2} \delta_{il} \sum_k \frac{2 (i \omega)^2}{D^8} k_j k_p v_p \text{Tr} \Big[ \Gamma_4 \Gamma_0 \Gamma_a \Gamma_b \Gamma_p \Big] -\frac{m}{2} \delta_{il} \sum_k \frac{2 (i \omega)^2}{D^8} k_j k_p v_p \text{Tr} \Big[ \Gamma_4 \Gamma_0 \Gamma_a \Gamma_p \Gamma_b \Big] \nonumber \\
    & -\frac{m}{2} \delta_{il} \sum_k \frac{ (i \omega)^2 + \epsilon_0^2}{D^8} k_j k_p v_p \text{Tr} \Big[ \Gamma_4 \Gamma_a \Gamma_0 \Gamma_b \Gamma_p \Big] -\frac{m}{2} \delta_{il} \sum_k \frac{ (i \omega)^2 + \epsilon_0^2}{D^8} k_j k_p v_p \text{Tr} \Big[ \Gamma_4 \Gamma_a \Gamma_p \Gamma_b \Gamma_0 \Big] \nonumber \\
    & -\frac{m}{2} \delta_{il} \sum_k \frac{2 (i \omega)^2 }{D^8} k_j k_p v_p \text{Tr} \Big[ \Gamma_4 \Gamma_p \Gamma_a \Gamma_0 \Gamma_b \Big] -\frac{m}{2} \delta_{il} \sum_k \frac{2 (i \omega)^2 }{D^8} k_j k_p v_p \text{Tr} \Big[ \Gamma_4 \Gamma_p \Gamma_a \Gamma_b \Gamma_0 \Big].
\end{align}
Similarly from $\text{Tr}\Big[ \Gamma_4 \Gamma_0 \Gamma_a \Gamma_b \Gamma_c \Big] = 4 \epsilon^{abc}$ and the cyclic and anti-symmetric nature of the Levi-Civita tensor $\epsilon^{pqr} = \epsilon^{rpq} =  -\epsilon^{qpr} $, we note that $\Pi_{ij0l}^{\Delta,ab(2)} = 0$. Finally, we look at 
\begin{align}
    \Pi_{ij0l}^{\Delta,ab(3)} &= \sum_k \text{Tr} \Big[ \Gamma_4 \left( \frac{2 v_l k_l}{D^4} - \frac
    {2 i\omega}{D^4} v_l \Gamma_l - \frac{8 i\omega k_l}{D^6} \left( i \omega + m\Gamma_0 + vk_x \Gamma_1 + v k_y \Gamma_2 + v_z k_z \Gamma_3 \right) \right)  k_i \Gamma_a \nonumber \\
    &\frac{\left( i \omega + m \Gamma_0 + k_{p_{1}} v_{p_{1}} \Gamma_{p_{1}}\right)}{D^2} k_j \Gamma_b   \frac{\left( i \omega + m \Gamma_0 + k_{p_{2}} v_{p_{2}} \Gamma_{p_{2}}\right)}{D^2}
    \Big] \nonumber \\
    &= -m \sum_k \frac{8 (i \omega)^2}{D^{10}} k_i k_j k_l k_p v_p v_l \text{Tr} \Big[ \Gamma_4 \Gamma_0 \Gamma_a \Gamma_b \Gamma_p \Big] -m \sum_k \frac{8 (i \omega)^2}{D^{10}} k_i k_j k_l k_p v_p v_l \text{Tr} \Big[ \Gamma_4 \Gamma_0 \Gamma_a \Gamma_p \Gamma_b \Big] \nonumber \\
    & -m \sum_k \frac{k_i k_j}{D^4} \left( \frac{2}{D^4} - \frac{8 (i \omega)^2}{D^6}
    \right)  k_l k_p v_p v_l \text{Tr} \Big[ \Gamma_4 \Gamma_a \Gamma_0 \Gamma_b \Gamma_p \Big] -m \sum_k \frac{k_i k_j}{D^4} \left( \frac{2}{D^4} - \frac{8 (i \omega)^2}{D^6}
    \right)  k_l k_p v_p v_l \text{Tr} \Big[ \Gamma_4 \Gamma_a \Gamma_p \Gamma_b \Gamma_0 \Big] \nonumber \\
    & -m \sum_k \frac{8 (i \omega)^2}{D^{10}} k_i k_j k_l k_p v_p v_l \text{Tr} \Big[ \Gamma_4 \Gamma_p \Gamma_a \Gamma_0 \Gamma_b \Big] -m \sum_k \frac{8 (i \omega)^2}{D^{10}} k_i k_j k_l k_p v_p v_l \text{Tr} \Big[ \Gamma_4 \Gamma_p \Gamma_a \Gamma_b \Gamma_0 \Big] \nonumber \\
    & -m \sum_k \frac{2 (i \omega)^2}{D^8}  v_l \text{Tr} \Big[ \Gamma_4 \Gamma_l \Gamma_a \Gamma_b \Gamma_0 \Big] - m \sum_k \frac{2 (i \omega)^2}{D^8}  v_l \text{Tr} \Big[ \Gamma_4 \Gamma_l \Gamma_a \Gamma_0 \Gamma_b \Big]. 
\end{align}
 Again from $\text{Tr}\Big[ \Gamma_4 \Gamma_0 \Gamma_a \Gamma_b \Gamma_c \Big] = 4 \epsilon^{abc}$ and the cyclic and anti-symmetric nature of the Levi-Civita tensor $\epsilon^{pqr} = \epsilon^{rpq} =  -\epsilon^{qpr} $, we also find that $\Pi_{ij0l}^{\Delta,ab(3)} = 0$. Finally, the effective action is given by 
 \begin{equation}
     W_3 = - 4 |m|^2 \sum_k \frac{k_i k_j v_l}{D^6}\sum_{q_1,q_2} \epsilon^{alb} q_{1l} q_{20} \Phi(-q_1) \Delta_a^i(-q_2) \Delta_b^j(q_1+q_2)  
 \end{equation}
 Using the Fourier transform in Eq(\ref{eq:FieldFourier}), we rewrite the effective action in real space and obtain
 \begin{equation}
     S_\Gamma = - |m|^2 v_l \mathcal{I}_{i} \delta_{ij} \int dt d^3r \Phi \epsilon^{alb} \partial_l \Delta_a^i \partial_t \Delta_b^j, \quad \mathcal{I}_{i} = 4  \sum_k \frac{   k_i^2}{\left( (i \omega)^2 - \epsilon_0^2\right)^3}. \label{eq: effSLNY0}
 \end{equation}
Our result matches the effective action derived in Ref. \cite{hughes2011torsional}, which is known as the "Nieh-Yan" term \cite{nieh1982quantized,nieh1982identity,nieh2007torsional,chandia1997topological}. We rescale the momentum $\Tilde{k}_i = v_i k_i $, the frame field $\Tilde{\Delta}^i_a = \Delta^i_a/v_i$, the spatial coordinates $\Tilde{r}_i = r_i/v_i, $, spatial derivatives $\Tilde{\partial}_i = v_i \partial_i$ such that the frame field becomes dimensionless in real space, $\Tilde{\textbf{r}}$ has the dimension of $t$ and $\Tilde{\textbf{k}}$ has the dimension of $\omega$. We rewrite Eq(\ref{eq: effSLNY0}) as 
\begin{equation}
    S_\Gamma = - m^2  \Tilde{\mathcal{I}}_{i} \delta_{ij} \int dt d^3 \Tilde{r} \Phi \epsilon^{alb} \Tilde{\partial}_l \Tilde{\Delta}_a^i \partial_t \Tilde{\Delta}_b^j, \quad \Tilde{\mathcal{I}}_i = v^2 v_z v_i^2 \mathcal{I}_i = 4 v^2 v_z  \sum_k \frac{  v_i^2  k_i^2}{\left( (i \omega)^2 - \epsilon_0^2\right)^3}. \label{eq: effSLNY}
\end{equation}

\subsection{Comparison between Nieh-Yan action and valley Axion terms}\label{sec:compareNYaxion}
In this section, we present a comparison between the derivation of the Nieh-Yan term in magnetic topological insulator sandwiches and MnBi$_2$Te$_4$ and the valley axion term (pseudo-gauge field)  discussed in Ref.\cite{chatterjee2024localized} for Dirac semimetals like Na$_3$Bi. The key difference is the location of the Dirac Hamiltonian. For Dirac semimetals, the Dirac Hamiltonian is located at momenta which are not invariant under TR. Thus, the lowest order coupling term between Dirac electrons and acoustic phonons is independent of momentum. Consequently, acoustic phonons can play the role of a pseudo-gauge field, leading to the valley axion mechanism for localized interfacial acoustic phonons discussed in Ref. \cite{chatterjee2024localized}. In contrast, the Dirac Hamiltonian for the magnetic TI sandwiches or MnBi$_2$Ti$_4$ is located at the $\Gamma$ point, which is an $\mathcal{T}$ or $\mathcal{S}$ invariant momentum for bulk TI materials. As a consequence, for both magnetic TI sandwich structures and MnBi$_2$Ti$_4$, the $k$-independent coupling between Dirac electrons and acoustic phonons is limited to the $C(\textbf{u})I$ term that only shifts the overall energy and the $M(\textbf{u}) \Gamma_0$ term, which is a correction to the mass term $m \Gamma_0$.  Importantly, the acoustic phonons {\it cannot} give rise to the vector potential component of the pseudo-gauge field for the axion term. Thus, we consider the $k$-linear terms for the coupling between Dirac electrons and acoustic phonons. This coupling term can serve as the frame field and give rise to the torsion component of an effective gravitational field for Dirac electrons. As we have shown above, this coupling term can lead to the Nieh-Yan terms in the effective action of phonon dynamics after integrating out Dirac electrons. There is another difference between the Nieh-Yan effective action and axion effective action. When the electron mass $m$ slowly varies from $m<0$ (topological) to $m>0$ (trivial) across the interface of magnetic Ti sandwiches or MnBi$_2$Ti$_4$, the Nieh-Yan effective action $S_\Gamma$ in Eq.(\ref{eq: effSLNY}) varies continuously as a function of $|m|^2$, whereas the valley axion action $S^{ax}$ shows an abrupt change due to the coefficient of the axion term being independent of the Dirac mass $m$ as shown below. 

\subsubsection{Nieh-Yan effective action}

We first evaluate the integral presented in Eq(\ref{eq: effSLNY}). We denote $v_i k_i = \Tilde{k}_i$ for $i=x,y,z$, $\Tilde{k}^2 = v^2(k_x^2 +k_y^2) + v_z^2 k_z^2$, and 

\begin{align}\label{eq:IiNY}
    \Tilde{\mathcal{I}}_i &= 4 v^2 v_z \sum_k \frac{\Tilde{k}_i^2}{\Big[ (i \omega)^2 - m^2 - \Tilde{k}^2 \Big]^3}  = \frac{4}{3 } v^2 v_z \frac{1}{\beta} \sum_{i\omega} \int \frac{d^3k}{(2\pi)^3} \frac{\Tilde{k}^2}{\Big[ (i \omega)^2 - m^2 - \Tilde{k}^2 \Big]^3}  = \frac{4}{3} \frac{1}{\beta} \sum_{i\omega} \int \frac{d^3\tilde{k}}{(2\pi)^3} \frac{\Tilde{k}^2}{\Big[ (i \omega)^2 - m^2 - \Tilde{k}^2 \Big]^3} \nonumber \\
    &=  \frac{4}{3} \frac{1}{\beta} \sum_{i \omega} \int \frac{4 \pi}{(2 \pi)^3} \Tilde{k}^2 d \Tilde{k} \frac{\Tilde{k}^2}{\Big[ (i \omega)^2 - m^2 - \Tilde{k}^2 \Big]^3} =  \frac{2}{3 \pi^2} \frac{1}{\beta} \sum_{i \omega} \int  d \Tilde{k} \frac{\Tilde{k}^4}{\Big[ (i \omega)^2 - m^2 - \Tilde{k}^2 \Big]^3}.
\end{align}

We first consider the Matsubara frequency summation and define $\mathcal{I}_R = \oint_R \frac{dz}{2 \pi i} f(z) \frac{1}{e^{\beta z}+1}$ with $f(z) = \frac{1}{(z^2 - m^2 - \Tilde{k}^2)^3}$ and $n_F(z) = \frac{1}{e^{\beta z}+1}$. The integrand has the poles at $z_n = i \omega = i \frac{(2n+1) \pi}{\beta}$ from the function $\frac{1}{e^{\beta z}+1}$, and the poles at $z=\pm \sqrt{m^2 + \Tilde{k}^2} = \pm z_0$ from $f(z)$. 
We calculate the residue 
\begin{align}
    \text{Res}[f(z) \eta_F(z),z_n] = \lim_{z\rightarrow z_n} (z-z_n) f(z) \eta_F(z) = - \frac{1}{\beta} f(i \omega_n)
\end{align}
and 
\begin{equation}
    \text{Res}[f(z) \eta_F(z),z_0] = -\frac{3}{8 z_0^5} \tanh{\frac{\beta z_0}{2}} + \frac{3 \beta}{8 z_0^4} \frac{1}{1+\cosh{\beta z_0}} + \frac{\beta^2}{8 z_0^3} \frac{\tanh{\beta z_0/2}}{\cosh^2{\frac{\beta z_0}{2}}}. 
\end{equation}
For a large $R$, the integrand vanishes making $\mathcal{I}_R=0$. The integral $\mathcal{I}_R$ is then written in terms of the residues at the poles $z_n$ and $z_0$, which ultimately gives us
\begin{equation}\label{eq:sumfiomega}
    \frac{1}{\beta} \sum_{i \omega_n} f(i \omega_n) = -\frac{3}{8 z_0^5} \tanh{\frac{\beta z_0}{2}} + \frac{3 \beta}{8 z_0^4} \frac{1}{1+\cosh{\beta z_0}} + \frac{\beta^2}{8 z_0^3} \frac{\tanh{\beta z_0/2}}{\cosh^2{\frac{\beta z_0}{2}}} \equiv \mathcal{F}[z_0(\Tilde{k})]
\end{equation}
and
\begin{equation} \Tilde{\mathcal{I}}_i = \frac{2}{3 \pi^2}  \int  d \Tilde{k} \Tilde{k}^4 \mathcal{F}(\Tilde{k}) .
\end{equation}
We first consider the limit of $T \rightarrow 0$, in which $\beta \rightarrow \infty$, $\tanh{\beta z_0/2} \rightarrow 1, \cosh{\beta z_0} \rightarrow \infty, e^{-\beta z_0/2} \rightarrow 0$, so that we have
\begin{equation}\label{eq:fomegasum}
    \frac{1}{\beta} \sum_{i \omega_n} f(i \omega_n) = - \frac{3}{8 (m^2 + \Tilde{k}^2)^{5/2}} ,
\end{equation}
and \begin{equation}
    \Tilde{\mathcal{I}}_i = - \frac{1}{4 \pi^2} \int_0^\infty dx \frac{x^4}{(1+x^2)^{5/2}} ,
\end{equation}
for any value of $i$, where $x=\Tilde{k}/|m|$. We find that $\int_0^\infty dx \frac{x^4}{(1+x^2)^{5/2}}$ has a $\log$ divergence as $x \rightarrow \infty$, so we introduce a cutoff $x_c$, which physically corresponds to the momentum cutoff $\Tilde{k}_c = |m| x_c$ beyond which the physics of the realistic material is no longer described by the Dirac Hamiltonian. We have 
\begin{equation} \label{eq_SM:NiehYan Integral}
 \Tilde{\mathcal{I}}_i = \frac{1}{4 \pi^2}\frac{x_c \left( 3 + 4 x_c^2 \right) }{3 \left( 1 + x_c^2 \right)^{3/2} }  - \frac{1}{4 \pi^2}
 \sinh^{-1}(x_c) .
\end{equation}
In the finite $T$ case, we have 
\begin{equation}\label{eq_SM:FiniteTeq1}
    \Tilde{\mathcal{I}}_i = \frac{2}{3 \pi^2} \int d \Tilde{k} \Tilde{k}^4 \Bigg[ -\frac{3}{8 z_0^5} \tanh{\frac{\beta z_0}{2}} + \frac{3 \beta}{8 z_0^4} \frac{1}{1+\cosh{\beta z_0}} + \frac{\beta^2}{8 z_0^3} \frac{\tanh{\beta z_0/2}}{\cosh^2{\frac{\beta z_0}{2}}}\Bigg] .
\end{equation}
We rescale $\Tilde{k},z_0,m$ with $\beta$  such that $\Bar{k} = \beta \Tilde{k}, \Bar{z}_0 = \beta z_0 = \sqrt{\Bar{k}^2 + \Bar{m}^2}$ with $\Bar{m} = \beta m$. Eq(\ref{eq_SM:FiniteTeq1}) becomes
\begin{equation}
    \Tilde{I}_i = \Tilde{I}_i^{(1)} + \Tilde{I}_i^{(2)} + \Tilde{I}_i^{(3)} ,
\end{equation}
where
\begin{align}
    \Tilde{I}_i^{(1)} &= -\frac{1}{4 \pi^2} \int_0^\infty d \Bar{k} \frac{\Bar{k}^4}{\Bar{z}_0^5} \tanh{\Bar{z}_0/2} \label{eq_SM:I1}\\
    \Tilde{I}_i^{(2)} &= \frac{1}{8 \pi^2} \int_0^\infty d \Bar{k} \frac{\Bar{k}^4}{\Bar{z}_0^4} \frac{1}{(\cosh{\Bar{z}_0/2})^2} \label{eq_SM:I2}\\
    \Tilde{I}_i^{(3)}  &= \frac{1}{24 \pi^2} \int_0^\infty d\Bar{k} \frac{\Bar{k}^4}{\Bar{z}_0^3} \frac{\tanh{\Bar{z}_0/2}}{(\cosh{\Bar{z}_0/2})^2} .  \label{eq_SM:I3}
\end{align}
Eqs(\ref{eq_SM:I2},\ref{eq_SM:I3}) can be evaluated numerically from $0$ to $\infty$, and are converging, but Eq(\ref{eq_SM:I1}) diverges at $\Bar{k} \rightarrow \infty$. So, we choose 2 cut-offs $\Lambda_0$ and $\Lambda_c$ such that $\tanh{\Lambda_0/2} \sim \tanh{\Lambda_c/2} \sim 1$ and in this limit, Eq(\ref{eq_SM:I1}) can be evaluated numerically. We evaluate $\Tilde{I}_i$ numerically as a function of $|m|$ for different $\Lambda_c$ in Fig. 2a of the main text. Therefore,    
\begin{equation} \label{eq_SM:S_Gamma_eff_1}
     S_\Gamma =  \eta_0(|m|) \delta_{ij} \int d t d^3 \Tilde{r} \Phi \epsilon^{alb} \Tilde{\partial}_l \Tilde{\Delta}_a^i \partial_t \Tilde{\Delta}_b^j, 
\end{equation}
which is a continuous function when the Dirac mass $m$ changes sign, in line with the vanishing viscoelastic response of trivial insulators. Moreover, $\eta_0$ is independent of $T$ and is treated as a tuning parameter for the rest of the paper. 

We consider another limit of Eq(\ref{eq_SM:I1}), where we first take $T \rightarrow 0$ and then $m \rightarrow 0$. For $T \rightarrow 0$,  Eq(\ref{eq_SM:I1}) becomes 
\begin{equation} \label{eq_SM:I1limitm}
    \Tilde{I}_i^{(1)} = - \frac{1}{4 \pi^2} \int d \Tilde{k} \frac{\Tilde{k}^4}{z_0^5} \tanh{ \beta z_0 /2} = - \frac{1}{4 \pi^2} \int_\epsilon^{\Lambda_c} d \tilde{k} \frac{\Tilde{k}^4}{z_0^5},
\end{equation}
where $\tanh{\beta \Tilde{k}/2} \rightarrow 1$. In the limit $m \rightarrow 0$, Eq(\ref{eq_SM:I1limitm}) becomes
\begin{equation}
    \Tilde{I}_i^{(1)} = - \frac{1}{4 \pi^2} \int_\epsilon^{\Lambda_c} \frac{d \Tilde{k}}{\Tilde{k}} = - \frac{1}{4 \pi^2} \ln \left(\Lambda_c /\epsilon \right)  .
\end{equation}
We find a $log$-divergence for both infrared (IR) and ultraviolet (UV) sectors in this limit. The UV divergence is cut-off by $\Lambda_c$ which is determined by microscopic details (lattice constant). In our model, $\Lambda_c$ is set by the quadratic term $m_1 k^2$. The IR cut-off $\epsilon$ is determined by the sample size $L$, e.g., $\epsilon \sim 1/L$, and thus we have
\begin{equation}
    \Tilde{I}_i^{(1)} (L) \approx -\frac{1}{4 \pi^2} \ln (L \Lambda_c),
\end{equation}
which leads to 
\begin{equation}
    \frac{d \ln \Tilde{I}_i^{(1)}(L)}{d \ln L} = -\frac{1}{4 \pi^2} \frac{1}{\Tilde{I}_i^{(1)}(L)} = \beta (\Tilde{I}_i^{(1)}). 
\end{equation}
We note this equation describes the scaling of the parameter $\Tilde{I}_i^{(1)}$ with respect to the sample size $L$ and is similar to the scaling equation of weak localization in two dimensions localization physics \cite{abrahams1979scaling,rammer2018quantum}. 

\subsubsection{Valley axion mechanism}
When the acoustic phonons couple as a valley axion field $A_{\mu,a}$ where $a=1,2$ correspond to the valleys, the effective action becomes \cite{yu2021dynamical,chatterjee2024localized}
\begin{equation}
     S^{ax} = m^2 I^{ax} \sum_{a} \int d t d^3r \Phi_a \epsilon^{\mu \nu \rho \sigma} \partial_\mu A_{\nu,a} \partial_\rho A_{\sigma,a}, \quad I^{ax} = v^2 v_z \sum_k \frac{1}{\left( (i \omega)^2 - \epsilon_0^2\right)^3}. 
 \end{equation}
Using Eq(\ref{eq:fomegasum}), we find \begin{align}
     I^{ax} = \frac{3}{32  \pi^2 m^2  } \int_0^\infty dx \frac{x^2}{(1+x^2)^{5/2}}  = \frac{3}{32  \pi^2 m^2 } \frac{\Gamma(3/2) \Gamma(1)}{2 \Gamma(5/2)} = \frac{1}{32 \pi^2} \frac{1}{m^2}
 \end{align}
The effective action in the valley axion mechanism becomes  
\begin{equation}\label{eq_SM:axioneffaction}
     S^{ax} = \frac{1}{32 \pi^2}   \int d t d^3 \Tilde{r} \Phi \epsilon^{\mu \nu \rho \sigma} \partial_\mu A_{\nu} \partial_\rho A_{\sigma}
 \end{equation}
 which is independent of Dirac mass $m$ and has a jump when Dirac mass $m$ changes sign because $\Phi$ has a jump from $0$ to $\pi$ and vice versa.

\subsection{Effective action $S_{eff}$ as a surface phonon Hall viscosity} \label{sec:effactiontotal}
In this section, we will show that the effective action $S_\Gamma$ can give rise to the surface phonon Hall viscosity. Firstly, we will show when $\Phi$ is a constant, the effective action $S_\Gamma$ in Eq.(\ref{eq_SM:S_Gamma_eff_1}) is a total derivative. When $\Phi$ is a constant, we have
\begin{align}
    S_\Gamma &= \eta_0 \delta_{ij} \int dt d^3 r \Phi \epsilon^{alb} \partial_l \Delta_a^i \partial_t \Delta_b^j \nonumber \\ &= \eta_0 \delta_{ij} \int dt d^3 r \Phi \epsilon^{alb} \partial_l \left( \Delta_a^i \partial_t \Delta_b^j \right) - \eta_0 \delta_{ij} \int dt d^3 r \Phi \epsilon^{alb}  \Delta_a^i \partial_l \partial_t \Delta_b^j. \label{eq_SM:S_Gamma_eff_2}
\end{align}
Since
\begin{align}
    & \eta_0 \delta_{ij} \int dt d^3 r \Phi \epsilon^{alb}  \Delta_a^i \partial_l \partial_t \Delta_b^j = - \eta_0 \delta_{ij} \int dt d^3 r \Phi \epsilon^{alb}  \partial_t \Delta_a^i \partial_l  \Delta_b^j \nonumber \\ &= - \eta_0 \delta_{ij} \int dt d^3 r \Phi \epsilon^{bla}  \partial_t \Delta_b^i \partial_l  \Delta_a^j =  \eta_0 \delta_{ij} \int dt d^3 r \Phi \epsilon^{alb}  \partial_t \Delta_b^i \partial_l  \Delta_a^j \nonumber \\ &= - \eta_0 \delta_{ij} \int dt d^3 r \Phi \epsilon^{alb}  \Delta_a^j \partial_l \partial_t \Delta_b^i, 
\end{align}
which makes the second term in $S_\Gamma$ of Eq.(\ref{eq_SM:S_Gamma_eff_2})  vanish, i.e. $\eta_0 \delta_{ij} \int dt d^3 r \Phi \epsilon^{alb}  \Delta_a^i \partial_l \partial_t \Delta_b^j = 0$. Here we used the periodic boundary condition for the time domain in the second and last steps. Thus, we have shown 
\begin{eqnarray}
    S_\Gamma &= \eta_0 \delta_{ij} \int dt d^3 r \Phi \epsilon^{alb} \partial_l \left( \Delta_a^i \partial_t \Delta_b^j \right)  
\end{eqnarray}
which is a total derivative of the spatial dimension. 

We next consider the interface to be the $xy$-plane with the $\Phi$ field only depending on the z direction, e.g. $\Phi = \Phi(z)$. In this case, Eq(\ref{eq: effSLNY}) can be written in the form
\begin{eqnarray} \label{eq_SM:Actionetaijkl}
    S_\Gamma = - \int dt d^3 r \frac{\partial \Phi}{\partial z} \eta_{ijmn} u_{ij} \Dot{u}_{mn},
\end{eqnarray}
where $\eta_{ijmn}$ is the Hall viscosity tensor \cite{avron1995viscosity,avron1998odd} and satisfies 
\begin{eqnarray}\label{eq_SM:symmetry_eta}
    \eta_{ijmn} = \eta_{jimn} = \eta_{ijnm} = -\eta_{mnij}.
\end{eqnarray}
In the specific context of acoustic phonons, it is referred to as phonon Hall viscosity \cite{barkeshli2012dissipationless, shapourian2015viscoelastic}. 
Due to Eqs.(\ref{eq_SM:symmetry_eta}) and the $D_{3d}$ symmetry, we find three independent phonon Hall viscosity coefficients $\eta_1, \eta_2, \eta_3$ defined as
\begin{align} \label{eq:eta symmetry}
    & \eta_1 \equiv \eta_{xxxy} = -\eta_{yyxy} = -\eta_{xyxx} = \eta_{xyyy}  \nonumber \\
    & \eta_2 \equiv \eta_{xzyz} = -\eta_{yzxz} \nonumber \\
    & \eta_3 \equiv = \eta_{xyyz}/2 = -\eta_{xxxz} = \eta_{yyxz} = -\eta_{yzxy}/2 = \eta_{xzxx} = -\eta_{xzyy}. 
\end{align}
We note that other symmetries of the $D_{3d}$ such as $\Hat{C}_{2x}$ and spatial inversion $\Hat{P}$ do not lead to additional constraints on $\eta_{ijkl}$. Using Eq(\ref{eq_SM:Delta_functions}), we have
\begin{align} \label{eq:eta values}
    \eta_{1} &= \eta_0 \left( -2  B_4^2 + 4 D_2^2 \right) \nonumber \\ \eta_{2} &= \eta_0 \left( B_3^2 - 2 D_1^2 \right) \nonumber \\
    \eta_{3} &= \eta_0 \left( B_3 B_4/2 -  D_1 D_2 \right)
\end{align}
and other $\eta_{ijmn}$'s can be related to Eqs.(\ref{eq:eta values}) via the relation (\ref{eq_SM:symmetry_eta}) Given the independent components of phonon Hall viscosity, we find the effective action of the surface phonon Hall viscosity is 
\begin{align}\label{eq:SPHV}
    S_{PHV} = -  \int d^3 r dt \left( \frac{\partial \Phi}{\partial z} \right) \Big[ \eta_1 \left( (u_{xx}-u_{yy}) 2\dot{u}_{xy} - 2 u_{xy} ( \dot{u}_{xx} -\dot{u}_{yy}) \right) + \eta_2 \left( u_{xz} \dot{u}_{yz} - u_{yz} \dot{u}_{xz} \right) \nonumber \\ + \eta_3 \left( 2u_{xy} \dot{u}_{yz} - (u_{xx} - u_{yy}) \dot{u}_{xz} - u_{yz} 2 \dot{u}_{xy} + u_{xz} (\dot{u}_{xx} - \dot{u}_{yy}) \right) \Big].
\end{align}
As compared with the Hall viscosity term derived in Refs. \cite{barkeshli2012dissipationless} and  \cite{shapourian2015viscoelastic} for the 2D and 3D isotropic systems, we find the $\eta_1$ term is isotropic, while $\eta_2$ and $\eta_3$ terms are new independent Hall viscosity coefficients for materials with the $D_{3d}$ group. In Sec. \ref{sec:bulkaction} we will show both $\eta_2$ and $\eta_3$ terms can bring new physical phenomena.


\subsection{Bulk action} \label{sec:bulkaction}
We consider the bulk elastic wave in a cubic crystal lattice with symmetry group $D_{3d}$. Generally, the effective action of the elastic waves has the form
\begin{equation}
    S_0 = \int dt d^3r \Big[ \frac{1}{2} \rho_{ij} \partial_t u_i \partial_t u_j - F_0  \Big] ; \ \ F_0 = \frac{1}{2} \lambda_{ijkl} u_{ij} u_{kl}. \label{S0general}
\end{equation}
We assume the bulk to be uniform, i.e. $\rho_{ij} = \rho \delta_{ij}$. For magnetic TI materials with the $D_{3d}$ group, the bulk free energy is given by \cite{landau2012theory} 
\begin{equation}\label{eq:F0}
    \begin{split}
        F_0 &= \frac{1}{2} (a+b) \left( u_{xx}^2 + u_{yy}^2 \right)   + (a-b)  u_{xx} u_{yy} + 2 b  u_{xy}^2 + c \left( u_{xx} + u_{yy} \right) u_{zz} + 2 d \left( u_{xz}^2 + u_{yz}^2 \right)  \\ & +  \frac{1}{2} f u_{zz}^2 + 2 g \Big[\left( u_{xx} - u_{yy} \right)) u_{yz} + 2 u_{xy} u_{xz} \Big] 
    \end{split}. 
\end{equation}
The parameters of elastic modulii are given in Table \ref{table:material}. 


\subsection{Equation of motion with the Nieh-Yan term}\label{sec:EOMNiehYan}

In this section, we derive the full equation of motion with the Nieh-Yan contribution $S_{PHV}$, given in Eq(\ref{eq_SM:Actionetaijkl}), which we refer to as the surface phonon Hall viscosity term. The equation of motion can be obtained from $\frac{\delta S}{\delta u_i} = 0$ with $S = S_0 + S_{\text{PHV}}$ where $S_0$ is given in Eq(\ref{S0general}) and $S_{\text{PHV}}$ is given by Eq(\ref{eq_SM:Actionetaijkl}), which leads to 
\begin{equation}\label{eq:EOMgeneral}
    \rho \ddot{u_i}  = \partial_j \left( \lambda_{ijkl} u_{kl} - 2 \eta_{ijkl} \frac{\partial \Phi}{\partial z} \dot{u}_{kl} \right) 
\end{equation} 
where non-zero $\eta_{ijkl}$'s are defined in Eqs(\ref{eq:eta values},\ref{eq:eta symmetry}). We point out a key distinction between the Nieh-Yan equation of motion and the Maxwell-Chern-Simons equation of motion, obtained as the electromagnetic response of a topological insulator. The Maxwell-Chern-Simons equation of motion gaps out the electromagnetic spectrum i.e. it is a topological \textit{massive} gauge theory \cite{deser2000topologically}, whereas in the Nieh-Yan and the Axion response of acoustic phonons, the phonons remain gapless at $\textbf{k}=0$.

\section{Interface phonon modes}\label{sec:AppAnal}

In the main text, we considered the surface phonon modes on each surface of a slab configuration of a magnetic TI sandwich structure or MnBi$_2$Te$_4$ using stress-free boundary conditions \cite{landau2012theory,benedek2013surface}. In this section, we will derive phonon modes localized at the interface of a magnetic TI and a trivial TI with the same elastic modulii.

\subsection{Interface modes}\label{sec:bulkinf}

We consider periodic boundary conditions along $x$ and $y$ directions and an interface at the origin along the $z$ direction, so $\Phi(z) = \pi \Theta(-z)$ where $\Theta(z)$ is the Heaviside step function. Here we consider an interface between MnBi$_2$Te$_4$ and a trivial insulator with the same symmetry group and bulk elastic modulii such that the existence of localized phonon modes purely originates from the change in $\Phi$. We assume the solution with the plane wave in the in-plane (xy) direction and spatial variation in the z-direction, so we choose the ansatz 
\begin{eqnarray}\label{eq_SM:ansatz_elasticwave}
\textbf{u}(\textbf{r},t) = \textbf{f}(z)e^{i \textbf{k}_{\parallel}\cdot\textbf{r}_{\parallel}}
\end{eqnarray}
with $\textbf{r}_{\parallel} = (x,y)$ and $\textbf{k}_{\parallel} = (k_x,k_y)$. Substituting the ansatz Eq(\ref{eq_SM:ansatz_elasticwave}) into Eq.(\ref{eq:EOMgeneral}), we find the eigen equation
\begin{eqnarray}\label{eq_SM:eigen_elasticwave}
    H_{ph}(\omega) \textbf{f} = \omega^2 I \textbf{f}
\end{eqnarray}
where $I$ is an identity matrix and
\begin{equation}\label{eq:H2iw}
    H_{ph}(\omega) = H_0 + 2 i \omega  H_{PHV}
\end{equation}
with
\begin{equation}
    H_0  = \begin{pmatrix} 
     (a+b) k_x^2 + b k_y^2 - d \partial_z^2 -2i g k_y \partial_z  & a k_x k_y -i 2g k_x \partial_z  &  2g k_x k_y -i (c+d) k_x \partial_z \\
    a k_x k_y - i 2g k_x \partial_z  & b k_x^2 +  (a+b) k_y ^2 - d \partial_z^2 + 2g i k_y \partial_z & g (k_x^2 - k_y^2) -i (c+d) k_y \partial_z  \\
     2g k_x k_y - i (c+d) k_x \partial_z & g (k_x^2 - k_y^2) - i (c+d) k_y \partial_z & d (k_x^2 + k_y^2)  - f \partial_z^2 
    \end{pmatrix},\label{eq:H0}
\end{equation}
and
\begin{equation}
    H_{PHV}  =  \begin{pmatrix} 
     0  &  \frac{\partial \Phi}{\partial z}  \left( \eta_1 (k_x^2 + k_y^2) - \eta_2 \partial_z^2 \right) & -\frac{\partial \Phi}{\partial z}  \left(  \eta_2 (i k_y) \partial_z + \eta_3 (k_x^2- k_y^2) \right)\\ 
       -\frac{\partial \Phi}{\partial z}  \left( \eta_1 (k_x^2 + k_y^2) - \eta_2 \partial_z^2 \right) & 0 & \frac{\partial \Phi}{\partial z} \left(  \eta_2 (i k_x) \partial_z + 2 \eta_3 k_x k_y \right) \\
     \frac{\partial \Phi}{\partial z}  \left(  \eta_2 (i k_y) \partial_z + \eta_3 (k_x^2-k_y^2) \right)  & -\frac{\partial \Phi}{\partial z}  \left(  \eta_2 (i k_x) \partial_z + 2\eta_3 k_x k_y \right)  & 0
    \end{pmatrix}. \label{eq:HNY}
\end{equation}
Below our task is to solve the eigen-equation (\ref{eq_SM:eigen_elasticwave}), which is not trivial as $H$ depends on $\omega$. In Sec. \ref{sec:isotropic approx}, we will first discuss a simplified situation, in which analytical solution for the domain wall modes can be obtained. In Sec. \ref{sec_SM:Numerical results}, we will present the results of full numerical calculations.

\subsection{Isotropic approximation }\label{sec:isotropic approx}
In this section, we first consider
the isotropic approximation, where 
\begin{equation} \label{eq:isoparameters}
    a =  c_l^2 - c_t^2, b = c_t^2, c = (c_l^2 - 2 c_t^2), d = c_t^2, f = c_l^2, g=0. 
\end{equation}
We further assume 
\begin{equation} \label{eq:isoctclc0parameters}
    c_t = c_l = c_0, \eta_2 = 0. 
\end{equation} 
Even though this limit is unrealistic, the analytical solution in this limit will provide us insight into the localized phonon modes. The full Hamiltonian in the eigen-equation (\ref{eq_SM:eigen_elasticwave}) has the form
\begin{eqnarray}
    H_{ph}(\omega) = H_0 + 2 h_1 \delta(z)
\end{eqnarray} 
with
\begin{equation} \label{eq:isomatrix}
H_{0} = \begin{pmatrix}
    c_0^2 k^2 - c_0^2 \partial_z^2 -\omega^2 & 0 & 0\\
    0  & c_0^2 k^2 - c_0^2 \partial_z^2 -\omega^2  & 0 \\
    0 & 0 &  c_0^2 k^2 - c_0^2 \partial_z^2 -\omega^2 
    \end{pmatrix},
\end{equation}
and
\begin{equation} \label{eq:isomatrixh1}
h_1 = i \omega  \begin{pmatrix}
    0 & \eta_1  k^2
    & -\eta_3 (k_x^2 - k_y^2) \\
    -\eta_1  k^2  & 0 & 2 \eta_3 k_x k_y \\
    \eta_3 (k_x^2 - k_y^2) & -2 \eta_3 k_x k_y &  0
    \end{pmatrix}. 
\end{equation}
Due to the $\delta$-function in front of $h_1$, we have two regions I and II for $z>0$ and $z<0$, respectively, and choose the ansatz of exponentially localized phonon modes around $z=0$ such that they vanish at infinity, i.e. 
\begin{equation}
    \textbf{u} = \begin{cases}
        \sum_{\tau=1}^{3} A_\tau e^{-\zeta_\tau z} \textbf{u}_0^\tau(z,\zeta_\tau) &  z>0 \\
        \sum_{\tau=1}^{3} B_\tau e^{\xi_\tau z} \textbf{v}_0^\tau(z,\xi_\tau) &  z<0
    \end{cases}
\end{equation}
where $\textbf{u}_0^\tau$ and $\textbf{v}_0^\tau$ are eigenmodes indexed by $\tau=1,2,3$ and given by $u_i^\tau(z) = e^{-\zeta_\tau z} \delta_{i\tau}$ and $v_i^\tau(z) = e^{\xi_\tau z} \delta_{i\tau}$ with the penetration depth ( or the inverse localization length) defined as
\begin{equation} \label{eq_SM:zetaequation}
    \zeta_\tau = \xi_\tau = \sqrt{k^2 - \omega^2/c_0^2} \equiv \zeta  > 0. 
\end{equation}
The bulk modes occur when $\zeta = 0$. The boundary conditions of the interface at $z=0$ are captured by (1) continuity of the wavefunction $\textbf{u}$ at $y=0$ and (2) integrating the equation of motion (Eqs.(\ref{eq:isomatrix}),\ref{eq:isomatrixh1}) around $y=0$ for a small spatial interval, from which we obtain 6 equations for 6 variables $A_{1,2,3}$ and $B_{1,2,3}$, given by (for $i=x,y,z$)
\begin{align}
    &\sum_{\tau=1}^3 u_i^\tau(0^+,\zeta_\tau) A_\tau - \sum_{\tau=1}^3 v_i^\tau (0^-,\xi_\tau) B_\tau = 0 \nonumber \\
    &\sum_{\tau=1}^3 \Big[ \sum_j (h_1)_{ij} u_j^\tau(0,\zeta_\tau) + c_0^2 \zeta_\tau u_i^a (0,\zeta_\tau)\Big] A_\tau + \sum_{\tau=1}^3 \Big[ \sum_j (h_1)_{ij} v_j^\tau(0,\xi_\tau) + c_0^2 \xi_\tau v_i^\tau(0,\xi_\tau) \Big] B_\tau =0. \label{case1eq}
\end{align}

\begin{figure*}
\includegraphics[width=\textwidth]{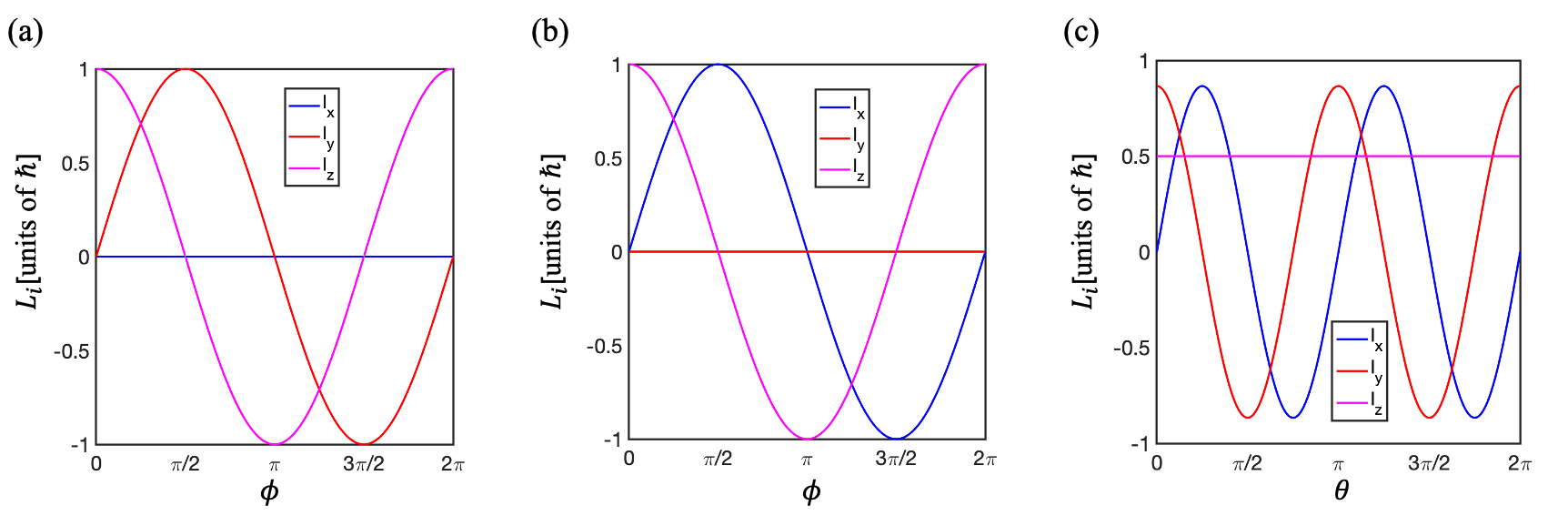}
 \caption{(a) The phonon angular momentum $L_{x,y,z}$ of the surface phonon mode as a function of $\phi$ when $\theta=0$ i.e. along $k_x$ ($k_y=0$). (b) The phonon angular momentum $L_{x,y,z}$ of the surface phonon mode as a function of $\phi$ when $\theta=\pi/4$ i.e. along $k_x = k_y$. (c) The phonon angular momentum $L_{x,y,z}$ of the surface phonon mode as a function of $\theta$ when $\phi=\pi/3$. 
}
 \label{fig:PAM_iso}
 \end{figure*}

Eq(\ref{case1eq}) can be re-written in a compact form as
\begin{equation}\label{MA1}
    M_{6\times6} \begin{pmatrix}
        A \\
        B
    \end{pmatrix} = 0,
\end{equation}
where
\begin{equation}
    M = \begin{pmatrix}
       \mathbb{I}_{3\times 3} & -\mathbb{I}_{3\times 3} \\
       h_1 + c_0^2 \sqrt{k^2 - \omega^2/c_0^2} \mathbb{I}_{3\times 3} & h_1 + c_0^2 \sqrt{k^2 - \omega^2/c_0^2} \mathbb{I}_{3\times 3}
    \end{pmatrix}
\end{equation}
with a 3-by-3 identity matrix $\mathbb{I}_{3\times 3}$, $A = (A_1, A_2, A_3)^T$ and $B = (B_1, B_2, B_3)^T$. A nontrivial solution exists if the secular equation, $\text{Det} (M) = 0$ is satisfied, which gives
\begin{equation}\label{case1dipersion}
   2 c_0^6 \left( k^2 - \omega^2/c_0^2\right) = \Big[ 8 c_0^2 \eta_3^2 k_x^2 k_y^2 + 2 \eta_1^2 c_0^2 k^4 + 2 \eta_3^2 c_0^2 \left(k_x^2 - k_y^2 \right)^2\Big] \omega^2.
\end{equation}
The frequency of the surface model is given by
\begin{equation}\label{eq_SM:isoc0frequency}
    \omega = \frac{c_0^2 k}{\sqrt{ c_0^2 + \Bar{\eta}^2 k^4}} \approx c_0 k - \frac{\Bar{\eta}^2}{2 c_0}k^5,
\end{equation}
with $|\Bar{\eta}| = \sqrt{\eta_1^2 + \eta_3^2}$, 
which has a lower energy than the bulk mode. 
From Eq(\ref{eq_SM:zetaequation}), we find the penetration depth of the interface phonon modes is 
\begin{eqnarray}
    \zeta =\frac{1}{2} |\Bar{\eta}| k^3 /\sqrt{c_0^2 + \Bar{\eta}^2 k^4}. 
\end{eqnarray}
From Eq(\ref{MA1}), we can solve the eigen-vector as
\begin{align}
    A_1 = B_1 &= - \left( \sin^2 \phi \sin 2\theta \cos 2\theta + i \cos \phi \right) \nonumber \\
    A_2 = B_2 &= 1 - \sin^2 \phi \cos^2 2 \theta \nonumber \\
    A_3  = B_3 &= -\sin \phi \left( \cos\phi \cos 2\theta - i \sin 2\theta \right) 
\end{align}
where $\eta_1/|\Bar{\eta}| = \cos\phi, \eta_3/|\Bar{\eta}| = \sin \phi, k_x = k \cos\theta, k_y = k \sin \theta$ and $k^2 = k_x^2 + k_y^2$. The surface phonon displacement field has the eigenvector form
\begin{equation}\label{eq_SM:isodisplacement}
    \textbf{u} = 
        N e^{-\lambda z} e^{i \textbf{k}\cdot \textbf{r}} \begin{pmatrix}
            - \left( \sin^2 \phi \sin 2\theta \cos 2\theta + i \cos \phi \right) \\
            1 - \sin^2 \phi \cos^2 2 \theta \\
            -\sin \phi \left( \cos\phi \cos 2\theta - i \sin 2\theta \right)
        \end{pmatrix}.
\end{equation}
The normalization condition $\int_{-\infty}^0 dz \textbf{u}^\dagger \textbf{u} = 1$ gives $N = (1/2 \pi)\sqrt{\lambda/2 (1-\sin^2 \phi \cos^2 2\theta)}$. The phonon angular momentum is defined by 
\begin{eqnarray}
    l_i = \hbar \textbf{u}_0^{\dagger} M_i \textbf{u}_0, \quad \textbf{u}_0 = \frac{1}{\sqrt{2(1-\sin^2 \phi \cos^2 2\theta) }} \begin{pmatrix}
        - \left( \sin^2 \phi \sin 2\theta \cos 2\theta + i \cos \phi \right) \\
            1 - \sin^2 \phi \cos^2 2 \theta \\
            -\sin \phi \left( \cos\phi \cos 2\theta - i \sin 2\theta \right)
    \end{pmatrix}.
\end{eqnarray}
where $i=x,y,z$ and $(M_i)_{jk} = (-i) \epsilon_{ijk}$  \cite{zhang2014angular,hamada2018phonon}. Specifically,
\begin{align} \label{eq_SM:PAMiso}
    \textbf{l} = (l_x,l_y,l_z) =  \hbar \left(  \sin \phi \sin 2\theta,  \sin \phi \cos 2\theta,  \cos \phi \right).
\end{align}
In Fig. \ref{fig:PAM_iso}, we plot the phonon angular momentum of the surface phonon mode, given in Eq(\ref{eq_SM:PAMiso}) as a function of $\phi$ - the relative strength of $\eta_1$ and $\eta_3$ and $\theta$ - the polar angle in the $k_x-k_y$ plane. The above eigen-function of surface phonon modes (Eq(\ref{eq_SM:isodisplacement})) is generally elliptically polarized e.g. in \ref{fig:PAM_iso}a when $\phi = \pi/4$, neither $L_y$ nor $L_z$ is quantized. It becomes circularly polarized in certain limits. We illustrate three examples. (1) From Eq(\ref{eq_SM:PAMiso}), when $\phi=0$, we have $\textbf{l} =(0,0, \hbar)$ indenpend of $\theta$. $\phi = 0$ corresponds to the case when $\eta_3 =0$ and $\eta_1 \neq 0$. (2) From Fig. \ref{fig:PAM_iso}a, when $\phi=\pi/2$, we have $\textbf{l} =(0,\hbar,0)$. This occurs when $\theta=0, \pi, 2\pi$, $\eta_1 = 0$ and $\eta_3 \neq 0$. (3) From  
Fig. \ref{fig:PAM_iso}b, when $\phi = \pi/2$ i.e. $\eta_1 = 0,  \eta_3 \neq 0$, we have $\textbf{l} =(\hbar,0,0)$. This occurs at $\theta=\pi/4$ which corresponds to the $k_x=k_y$ line. In Fig. \ref{fig:PAM_iso}c, we plot the phonon angular momentum at $\phi = \pi/3$, i.e. $\eta_3 = \sqrt{3} \eta_1$ as a function of $\theta$, e.g. along different directions in the $k_x-k_y$ plane, which confirms our claim that for a generic $\phi$ and $\theta$, the surface phonon mode is elliptically polarized.

\subsection{Numerical Method and Results}\label{sec_SM:Numerical results}
In this section, we outline the numerical iterative method used to solve eigen equations Eq(\ref{eq_SM:eigen_elasticwave}) with Eqs(\ref{eq:H0},\ref{eq:HNY}). We solve for $\omega$ iteratively, i.e., we first find the eigenvalues of $H_0$, denoted as $\omega_1$ (first iteration). We then find the eigenvalues of $H_0 + 2(i \omega_1) H_{PHV} $, denoted as $\omega_2$ and we proceed until the convergence condition $|\omega_n - \omega_{n-1}| < \epsilon \omega_1 $, where $\epsilon = 10^{-10}$ is the relative tolerance. The flow chart of the self-consistent calculations is summarized as

\begin{eqnarray}
    & H_0 \rightarrow \omega_1 \nonumber \\
    & H_0 + 2 i \omega_1 H_{PHV} \rightarrow \omega_2
    \nonumber \\
    & H_0 + 2 i \omega_2 H_{PHV} \rightarrow \omega_3 \nonumber \\
    . \nonumber \\
    . \nonumber \\
    . \nonumber \\
    & H_0 + 2 i \omega_{n-1} H_{PHV} \rightarrow \omega_n
\end{eqnarray}
with $|\omega_n - \omega_{n-1}| < \epsilon \omega_1 $.

\begin{figure}
\includegraphics[width=\textwidth]{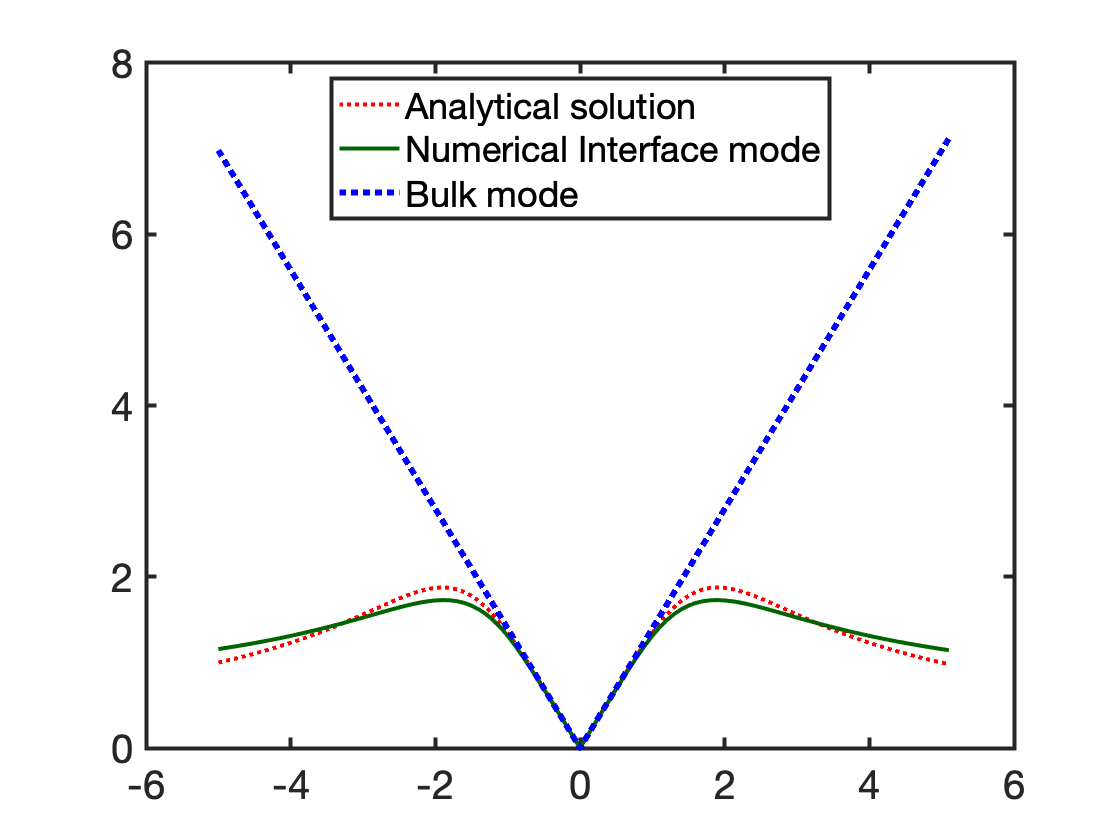}
 \caption{The comparison between the analytically obtained surface mode dispersion (in red) and the numerically obtained surface mode dispersion (in green). The blue lines depict the lowest bulk mode. We use $c_0 = 1500$ m/s, $\eta_1 = 0.16, \eta_2 = 0, \eta_3 = 0.17$ in units of m$^3$/s$^2$.
}
 \label{fig:Compareiso}
 \end{figure}

\begin{figure*}
 \includegraphics[width=\textwidth]{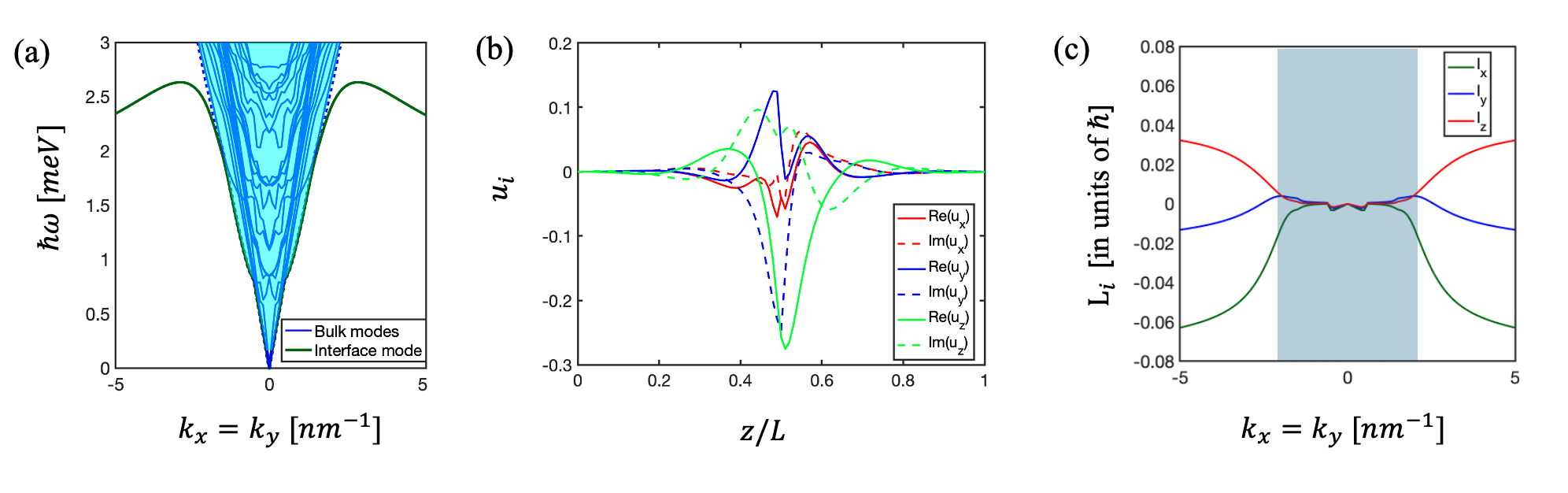}
 \caption{(a) The phonon energy $\hbar \omega$  along the $k_x=k_y$ line in the 2D Brillouin zone. The surface mode (green) frequency is lower than those of the bulk modes (in blue) (b) The spatial variation of the phonon displacement field $\textbf{u}$ of the surface mode at $k_x = k_y = 2.5$ nm $^{-1}$. The surface mode is localized at $z/L \sim 0.5$, which is the location of the $\Phi$ domain wall. (c) The phonon angular momentum of the lowest surface phonon mode along $k_x=k_y$. All the directions of the phonon angular momentum ($l_{x,y,z}$) are non-zero, which makes the surface phonon mode elliptically polarized. The middle region is where the surface modes cannot be separated from the bulk modes and therefore the phonon angular momentum calculation isn't accurate. We use $\eta_1 = 0.16, \eta_2 = 0.11, \eta_3 = 0.17$ in units of $0.005$ m$^3$/s$^2$.  
}
 \label{fig:012}
 \end{figure*}

\begin{figure*}
\includegraphics[width=\textwidth]{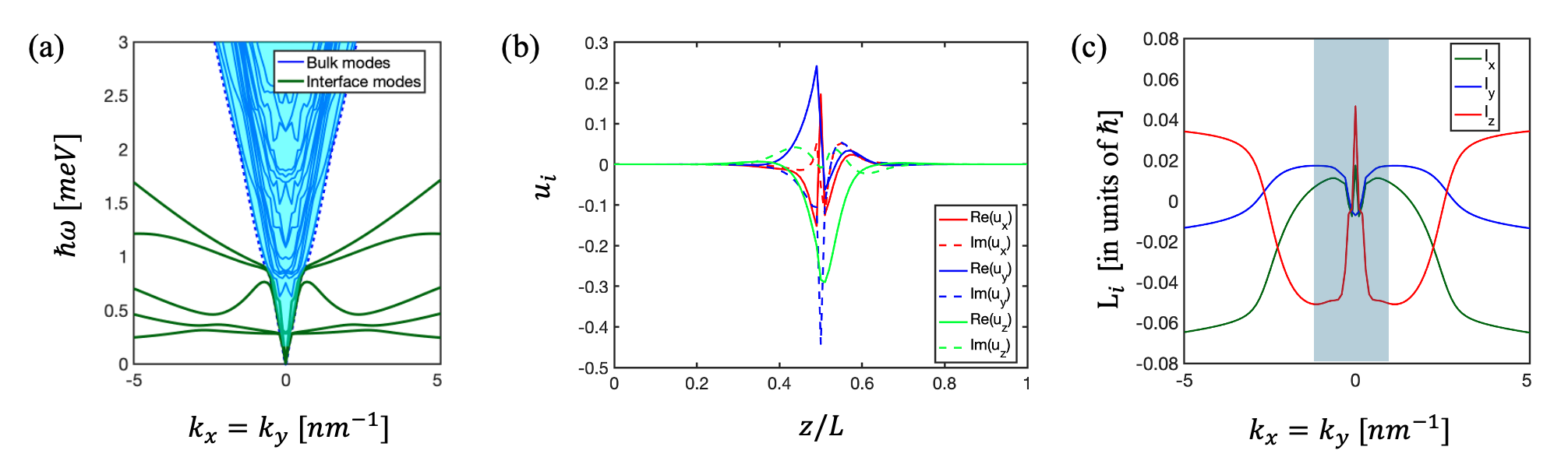}
 \caption{(a) The phonon energy $\hbar \omega$  along the $k_x=k_y$ line in the 2D Brillouin zone. There are several surface modes (green) when the electron-acoustic coupling strength is increased to $\eta_1 = 1.6, \eta_2 = 1.1, \eta_3 = 1.7$ in units of $0.005$ m$^3$/s$^2$ . The surface phonon bands remain flat for a large range of momenta and do not lead to lattice instabilities i.e. $\omega>0$. (b) The spatial variation of the phonon displacement field $\textbf{u}$ of the lowest frequency surface mode at $k_x = k_y = 2.5$ nm $^{-1}$. The surface mode is highly localized across a few lattice sites along $z$. (c) The phonon angular momentum of the lowest surface phonon mode along $k_x=k_y$. 
}
 \label{fig:120}
 \end{figure*}

Based on the above numerical method, we next present our numerical results. First, in Fig. \ref{fig:Compareiso}  we show that using the isotropic approximation described in Eqs(\ref{eq:isoparameters},\ref{eq:isoctclc0parameters}) of Sec. \ref{sec:isotropic approx}, there exists a surface phonon mode (in green), whose frequency is lower than the lowest bulk mode (in blue). Furthermore, in Fig. \ref{fig:Compareiso}, we show that the numerically obtained surface mode frequency (in green) matches reasonably well with the surface mode frequency obtained in Eq(\ref{eq_SM:isoc0frequency}) (in red). 

Next, we turn to our realistic system of a magnetic TI material. We use the elastic modulii of Bi$_2$Te$_3$, as given in Ref. \cite{jenkins1972elastic}, due to the close connection between these two crystals and the lack of complete list for the elastic modulii of MnBi$_2$Te$_4$ in literature \cite{chai2024thermoelectric,bartram2022ultrafast}. We plot the phonon dispersion in Fig. \ref{fig:012}a for non-zero surface phonon Hall viscosity, where we observe a surface phonon mode (in green), whose frequency is below the frequencies of the bulk modes (in blue). In Fig. \ref{fig:012}b, we plot the spatial variation of the surface phonon displacement field $\textbf{u}$ and we show that this surface phonon mode is localized around the interface $z \sim L/2$ and the amplitude of the displacement field decays exponentially away from the interface. In Fig. \ref{fig:012}c, we plot the phonon angular momentum of the surface phonon mode and show that each of the phonon angular momentum components i.e., $l_{x,y,z}$ is non-zero, which signifies 3D polarization of the surface mode. The middle region in Fig. \ref{fig:012}c, around $k_x \sim [-2,2] nm^{-1}$ is the region where the surface mode overlaps with the bulk modes and it is nearly impossible to separate them. In Fig. \ref{fig:120}, we increase the surface phonon Hall viscosity and find multiple surface modes (in green) as seen in Fig. \ref{fig:120}a. In Fig. \ref{fig:120}b, we plot the spatial variation of the displacement field of the lowest frequency surface mode. We observe that this surface mode is highly localized across a few sites near the interface. The phonon angular momentum of this surface phonon mode, plotted in  Fig. \ref{fig:120}c shows a slightly higher magnitude of phonon angular momentum compared to Fig. \ref{fig:012}c. Again as in Fig. \ref{fig:012}c, the middle region of Fig. \ref{fig:120}c corresponds to the surface modes overlapping with the bulk and they are inseparable.

\section{Effective theory of acoustic phonon modes in magnetic TI films}
In this section, we will discuss the physical phenomena induced by acoustic phonon modes, and their interplay with surface Dirac electrons and surface magnons, in magnetic TI sandwiches and MnBi$_2$Te$_4$ films. We will first outline our numerical calculations of acoustic phonon modes in magnetic TI films in Sec. \ref{sec_SM:acousticphononsinFilms}. In Sec. \ref{sec:SurfaceHamiltonian}, we develop the effective Hamiltonian to describe the coupling between electrons, strain tensor and magnetization at both surfaces of magnetic TI films using symmetry considerations. By integrating out gapped electronic surface states, we obtain the 2D effective theory for in-plane acoustic phonon modes in magnetic TI films in Sec.\ref{sec:nonreciprocal}, from which we can derive analytical results for the phonon thermal Hall effect in Sec.\ref{sec:thermal Hall}. In Sec.\ref{sec: surfacemagnonpolarons}, we discuss consider the coupling between strain tensor and magnetization and develop the theoretical formalism for magnon-polaron in magnetic TI films.

\subsection{Acoustic phonon modes in Magnetic TI films}\label{sec_SM:acousticphononsinFilms}

In this section, we will obtain the surface phonon modes for a slab of thickness $L_0$ with stress-free boundary conditions \cite{landau2012theory,benedek2013surface}. The acoustic phonons in a slab configuration are described by the Hamiltonian given in Eqs.(\ref{eq:H2iw},\ref{eq:H0},\ref{eq:HNY}) with appropriate $\frac{\partial\Phi}{\partial z}$ values at the top ($z=L_0$) and bottom surfaces ($z=0$) of the slab. As illustrated in Fig. 1(d) and (e) of the main text, we consider two configurations depending on whether the magnetization on the top and bottom surfaces are parallel or anti-parallel. The parallel and anti-parallel configurations are referred to as the surface ferromagnetic configuration (FM) and anti-ferromagnetic (AFM) configurations, respectively. For the FM case, the variation of the complex Dirac mass is given by \begin{equation} \label{eq_SM:FMPhi}
    \frac{\partial \Phi}{\partial z} = \begin{cases}
        +\pi, &z = 0 \\
        0, &z \in (0,L_0) \\
        +\pi, &z = L_0 ,
    \end{cases}
\end{equation} whereas for the AFM case,
\begin{equation} \label{eq_SM:AFMPhi}
    \frac{\partial \Phi}{\partial z} = \begin{cases}
        + \pi, &z = 0 \\
        0 , &z \in (0,L_0) \\ 
        - \pi, &z = L_0 .
    \end{cases}
\end{equation}Using Eqs.(\ref{eq_SM:FMPhi},\ref{eq_SM:AFMPhi}) and the stress-free boundary conditions on each surface ($z=0,L_0$), we solve Eqs.(\ref{eq:H2iw},\ref{eq:H0},\ref{eq:HNY}) numerically to obtain the acoustic phonon dispersion and their chiral and nonreciprocal properties for the FM and AFM cases, as shown in Fig. 2 of the main text.

\subsection{Effective surface Dirac Hamiltonian coupled to phonons and magnons} \label{sec:SurfaceHamiltonian}
Since the PHV term only occurs at the surface, we may consider the effective theory to describe the coupling between electrons, strain tensor and magnetization on both the top and bottom surfaces of magnetic TI sandwiches. For TIs, the surface Dirac electrons are highly localized, and its penetration depth is normally within a few nano-meters away from the surface layer \cite{zhang2010crossover}. Thus, for magnetic TI sandwiches with the thickness of tens or hundreds nano-meters, two surface states are well decoupled. Similarly, the magnetization on the top and bottom surfaces are also decoupled by the middle pure TI layer. Thus, we treat surface electrons and magnetization on the top and bottom surfaces, separately. Below we denote the surface electron Hamiltonian as $H^a$ and magnetization as $\mm^a$ with $a=t, b$ for top and bottom surfaces, separately. In contrast, the surface acoustic wave has a penetration depth around tens of microns \cite{landau2012theory,gerus1974amplification,gerus1975amplification}, much larger than the thickness of magnetic TI sandwiches. Thus, the surface acoustic waves on the top and bottom surfaces are strongly coupled, and the acoustic phonons in magnetic TI sandwiches should be treated as two dimensions. Here we only consider the in-plane acoustic phonon modes in magnetic TI sandwiches, and the corresponding strain fields for acoustic phonon modes are uniform in the z direction (film growth direction). Thus, both the top and bottom surface electrons will coupled to one strain field $\uu$ for the acoustic phonons in magnetic TI sandwiches. 

Below we first construct the effective electron Hamiltonian at the surface $a = t, b$ from the magnetic symmetry group $3m'$, including symmetries $\Hat{C}_{3z}$ and $\Hat{m}_x \Hat{T}$ \cite{yu2019magnetic,garate2010inverse} and the Hamiltonian reads
\begin{equation}
    H^a (\kk) = H^a_{\text{e}} + H^a_{\text{e-ph}} + H^a_{\text{e-m}} ,  
\end{equation} 
where $H^a_{\text{e}}, H^a_{\text{e-ph}}, H^a_{\text{e-m}} $ are the electron, electron-strain and electron-magnon Hamiltonians respectively, given by
\begin{align}
    & H^a_{\text{e}} = v_f^a \left( k_y \sigma_x - k_x \sigma_y \right) + m_z^a \sigma_z \label{eq_SM:Heonly} \\
    & H^a_{\text{e-ph}} = \Tilde{C}^a_0  (u_{xx} + u_{yy}) + \Tilde{C}^a_1 u_{zz} + \Tilde{B}_1^a \Big[ \left(u_{xx} - u_{yy} \right) \sigma_y + 2 u_{xy}  \sigma_x \Big] + \Tilde{B}^a_2 \left( u_{xz} \sigma_x + u_{yz} \sigma_y  \right)  \label{eq_SM:Hephonly} \\
    & H^a_{\text{e-m}} = g^a ( m^a_x \sigma_x + m^a_y \sigma_y), \label{eq_SM:Hemonly}
\end{align}
where $k_{x,y}$ are the surface electron momenta and $a=t,b,$ denotes the top and bottom surfaces. At the surface $a$,  $v_f^a$ is the Fermi velocity, $m_z^a$ is $z$-direction static magnetization and the magnon field $\mm_{\parallel}^a = (m_x^a,m_y^a)$ describes surface magnon dynamics at the surface $a$. We assume $|m^a_{x,y}|\ll m^a_z$ so magnetization is along the $z$-axis. The strain field $\uu$ describes phonon dynamics. We also neglect the electron tunneling between two surface states and the coupling of magnetization between two surfaces in the above Hamiltonian. The Hamiltonian $H^a$ can be written as   
\begin{align} \label{eq:surfDiracH}
    H^a (\kk) = A^a_0 I + v_f^a\left( k_y + A^a_y \right) \sigma_x - v_f^a \left( k_x + A_x^a\right) \sigma_y + m^a_z \sigma_z
\end{align}
where the pseudo-gauge field ${\bf A}^a$ is given by
\begin{align} \label{eq:surfgaugefield}
    A^a_0 & =  \Tilde{C}^a_0  (u_{xx} + u_{yy}) + \Tilde{C}^a_1 u_{zz}, \nonumber \\ A^a_x &= -\frac{1}{v_f^a} \Big[  \Tilde{B}^a_1 (u_{xx} - u_{yy}) + \Tilde{B}^a_2 u_{yz}
    + g^a m_y^a \Big], \nonumber \\
    A^a_y &= \frac{1}{v_f^a} \Big[ \Tilde{B}^a_1 2 u_{xy} + \Tilde{B}^a_2 u_{xz} + g^a m_x^a \Big].
\end{align} 

The top and bottom surfaces are related by inversion symmetry $\mathcal{P}$ in the FM case and $\mathcal{PT}$ in the AFM case, and thus we require
\begin{eqnarray}\label{eq_SM:FMRelation}
    D_{\mathcal{P}} H^t(\kk; \mm^t_\parallel, \uu) D^{-1}_{\mathcal{P}} = H^b( \mathcal{P}\kk; \mathcal{P} \mm^b_\parallel, \mathcal{P} \uu)
\end{eqnarray}
for the FM case and 
\begin{eqnarray}\label{eq_SM:AFMRelation}
    D_{\mathcal{PT}} H^t(\kk; \mm^t_\parallel, \uu) D^{-1}_{\mathcal{PT}} = [H^b(\mathcal{PT}\kk; \mathcal{PT} \mm^b_\parallel, \mathcal{PT} \uu)]^*
\end{eqnarray}
for the AFM case, where $D_{\mathcal{P}}$ and $D_{\mathcal{PT}}$ are the representation matrices for $\mathcal{P}$ and $\mathcal{PT}$, respectively, and $\mm^a_\parallel$ labels the in-plane magnetization fluctuation, which is treated as a dynamical field ($m^a_z$ is treated as an external parameter) 

For the FM case, the inversion $\mathcal{P}$ reverses the momentum $\kk$ but keep the form of spin $\sigma$, magnetization $\mm_{\parallel}^a$ and the strain tensor $\uu$, and thus the representation matrix is given by 
\begin{eqnarray}
    D_{\mathcal{P}} = \sigma_0
\end{eqnarray}
and 
\begin{eqnarray}\label{eq_SM:FMRelation2}
    \mathcal{P}\kk = -\kk, \quad \mathcal{P}\mm^t_\parallel = \mm^b_\parallel, \quad \mathcal{P} \uu = \uu.
\end{eqnarray}
In contrast, for the AFM case, the $\mathcal{PT}$ symmetry reverses the magnetization $\mm^a$ and spin $\sigma$, but keep the sign of $\kk$ and $\uu$, and thus
\begin{eqnarray}
    D_{\mathcal{PT}} = i \sigma_y
\end{eqnarray}
and
\begin{eqnarray} \label{eq_SM:AFMRelation2}
    \mathcal{PT}\kk = \kk, \quad \mathcal{PT}\mm^t_\parallel = -\mm^b_\parallel, \quad \mathcal{PT} \uu = \uu.
\end{eqnarray} 

Based on the above symmetry transformation of Eqs(\ref{eq_SM:FMRelation},\ref{eq_SM:FMRelation2}), for the FM case, we find
\begin{align}
   & \Big[ \Tilde{C}^t_0  (u_{xx} + u_{yy}) + \Tilde{C}^t_1 u_{zz} \Big]  I + \Big[  v_f^t k_y + \Tilde{B}^t_1 2 u_{xy} + \Tilde{B}^t_2 u_{xz} + g^t m_x^t \Big]  \sigma_x  + \Big[ -v_f^t  k_x + \Tilde{B}^t_1 (u_{xx} - u_{yy}) + \Tilde{B}^t_2 u_{yz}
    + g^t m_y^t  \Big] \sigma_y + m^t_z \sigma_z \nonumber \\
    &= \Big[ \Tilde{C}^b_0  (u_{xx} + u_{yy}) + \Tilde{C}^b_1 u_{zz} \Big]  I + \Big[  -v_f^b k_y + \Tilde{B}^b_1 2 u_{xy} + \Tilde{B}^b_2 u_{xz} + g^b m_x^t \Big]  \sigma_x  + \Big[  v_f^b  k_x + \Tilde{B}^b_1 (u_{xx} - u_{yy}) + \Tilde{B}^b_2 u_{yz}
    + g^b m_y^t  \Big] \sigma_y +  m^b_z \sigma_z
\end{align}
which implies 
\begin{eqnarray}\label{eq_SM:FMTopBottom}
   v_f^t = - v_f^b,\quad m^t_z = m^b_z,\quad g^t = g^b ,\quad \Tilde{C}^t_0 = \Tilde{C}^b_0,\quad  \Tilde{C}^t_1 = \Tilde{C}^b_1,\quad  \Tilde{B}^t_1 = \Tilde{B}^b_1,\quad \Tilde{B}^t_2 = \Tilde{B}^b_2.
\end{eqnarray}
Similarly, using Eqs(\ref{eq_SM:AFMRelation},\ref{eq_SM:AFMRelation2})   for the AFM case, 
\begin{align}
   & \Big[ \Tilde{C}^t_0  (u_{xx} + u_{yy}) + \Tilde{C}^t_1 u_{zz} \Big]  I - \Big[  v_f^t k_y + \Tilde{B}^t_1 2 u_{xy} + \Tilde{B}^t_2 u_{xz} + g^t m_x^t \Big]  \sigma_x  + \Big[ -v_f^t  k_x + \Tilde{B}^t_1 (u_{xx} - u_{yy}) + \Tilde{B}^t_2 u_{yz}
    + g^t m_y^t  \Big] \sigma_y - m^t_z \sigma_z \nonumber \\
    &= \Big[ \Tilde{C}^b_0  (u_{xx} + u_{yy}) + \Tilde{C}^b_1 u_{zz} \Big]  I + \Big[  v_f^b k_y + \Tilde{B}^b_1 2 u_{xy} + \Tilde{B}^b_2 u_{xz} - g^b m_x^t \Big]  \sigma_x  - \Big[ - v_f^b  k_x + \Tilde{B}^b_1 (u_{xx} - u_{yy}) + \Tilde{B}^b_2 u_{yz}
    - g^b m_y^t  \Big] \sigma_y + m^b_z \sigma_z,
\end{align}
which implies
\begin{equation}\label{eq_SM:AFMTopBottom}
   v_f^t = - v_f^b, \quad m^t_z = -m^b_z, \quad g^t = g^b , \quad \Tilde{C}^t_0 = \Tilde{C}^b_0, \quad  \Tilde{C}^t_1 = \Tilde{C}^b_1, \quad  \Tilde{B}^t_1 = - \Tilde{B}^b_1, \quad \Tilde{B}^t_2 = -\Tilde{B}^b_2.
\end{equation}

\subsection{In-plane acoustic phonon modes in magnetic TI sandwiches} \label{sec:nonreciprocal}
In this section, we first focus on in-plane phonon dynamics and thus ignore the magnons by setting $m_{x} = m_{y} = 0$ in Eqs.(\ref{eq:surfDiracH},\ref{eq:surfgaugefield}). We integrate out the Dirac electrons in Eq.(\ref{eq:surfDiracH}) and obtain the effective of the surface phonon modes as 
\begin{equation}\label{eq_SM:Effaction2D}
    S_{2D,a}^{PHV} = \frac{\text{sgn}(m_z^a)}{8 \pi} \int dx dy dt  \epsilon^{\mu \nu \rho} A^a_\mu \partial_\nu A^a_{\rho}, \quad \mu, \nu,\rho = 0,x,y .
\end{equation}
Substituting Eq.(\ref{eq:surfgaugefield}) into Eq.(\ref{eq_SM:Effaction2D}) leads to 
\begin{align}\label{eq_SM:2Dphononaction}
    S_{2D,a}^{PHV} & =  \int dx dy dt \mathcal{L}_{2D,a}^{PHV} = \int dx dy dt \Bigg[ \eta^a_{S_{1}} \Big[ \left( u_{xx} - u_{yy} \right) 2 \dot{u}_{xy} - \left( \dot{u}_{xx} - \dot{u}_{yy} \right) 2 u_{xy}  \Big] + \eta^a_{S_{2}} \Big[ u_{xz} \Dot{u}_{yz} - u_{yz} \Dot{u}_{xz}  \Big]  \nonumber \\
    &+ \eta^a_{S_{3}} \Big[ u_{xz} \left( \Dot{u}_{xx} - \Dot{u}_{yy} \right) - 2 u_{xy} \Dot{u}_{yz} -  \left( u_{xx} - u_{yy} \right) \Dot{u}_{xz} + u_{yz} 2 \Dot{u}_{xy}  \Big]
   \nonumber \\ 
   &+ 2 \xi^a_{S_{1}} \Big[ \left( u_{xx} + u_{yy} \right) \partial_x \left( 2 u_{xy}  \right) + \left( u_{xx} + u_{yy} \right) \partial_y \left( u_{xx} - u_{yy} \right) \Big] + 2 \xi^a_{S_{2}} \Big[ u_{zz} \partial_x \left( 2 u_{xy}  \right) +  u_{zz} \partial_y \left( u_{xx} - u_{yy} \right) \Big] \Bigg] 
\end{align}
where $
\eta^a_{S_1}, \eta^a_{S_2}$ and $\eta^a_{S_3}$ terms are  the phonon Hall viscosity coefficients at the surface $a$, given by
\begin{equation}
    \eta^a_{S_1} = \frac{\text{sgn}(m^a_z) (\Tilde{B}^a_1)^2}{8 \pi (v_f^a)^2}, \quad \eta_{S_2} = \frac{\text{sgn}(m^a_z) (\Tilde{B}^a_2)^2}{8 \pi (v_f^a)^2}, \quad \eta_{S_3} = \frac{\text{sgn}(m^a_z) \Tilde{B}^a_1 \Tilde{B}^a_2}{8 \pi (v_f^a)^2}, 
\end{equation}
and $\xi^a_{S_1}$ and $\xi^a_{S_2}$ terms are the flexo-elastic modulii \cite{ren2025nonreciprocal} at the surface $a$, given by
\begin{equation}
    \xi^a_{S_1} = \frac{\text{sgn}(m^a_z) \Tilde{C}^a_0 \Tilde{B}^a_1}{8 \pi v_f^a}, \quad  \xi^a_{S_2} = \frac{\text{sgn}(m^a_z) \Tilde{C}^a_1 \Tilde{B}^a_1}{8 \pi v_f^a}.
\end{equation}
As demonstrated in the main text, the surface phonon Hall viscosity alone is sufficient to induce nonreciprocity and chiral properties of acoustic phonons in magnetic TI films. As a result, we neglect the flexo-elastic modulii. The surface phonon Hall viscosity terms $\eta_{S_1}, \eta_{S_2}$ and $\eta_{S_3}$ in Eq.(\ref{eq_SM:2Dphononaction}) is equivalent to S$_{PHV}$ in Eq.(\ref{eq:SPHV}), whch is derived from bulk action. By choosing $\left(\frac{\partial \Phi}{\partial z} \right) $ to be a $\delta$-function at $z=0$, we obtain 
\begin{equation}
    \eta_{S_1} = - \eta_1, \quad \eta_{S_2} = - \eta_2, \quad \eta_{S_3} =  -\eta_3 .
\end{equation}
After establishing the connection between the bulk and surface actions, we will discuss the physical phenomena induced by surface Hall viscosity. We focus on the case with $\eta^a_2 = \eta^a_3 = 0$, so we will drop the terms with $\eta^a_{S_{2,3}}$ for simplicity. For the remainder of the paper, we redefine $\eta^a_{S} \equiv \eta^a_{S_1}$, which makes
\begin{align}\label{eq_SM:2Dphononaction2}
    S_{2D,a}^{PHV} & =\int dx dy dt \Bigg[ \eta^a_{S} \Big[ \left( u_{xx} - u_{yy} \right) 2 \dot{u}_{xy} - \left( \dot{u}_{xx} - \dot{u}_{yy} \right) 2 u_{xy}  \Big] .
\end{align}
The PHV effective action is given by
the sum of the top and bottom surfaces
\begin{equation}
    S_{2D}^{PHV} = \sum_{a = t,b} S_{2D,a}^{PHV}
\end{equation}
for magnetic TI films. For the FM case, using Eqs.(\ref{eq_SM:FMTopBottom}), we can determine $\eta_S^t = \eta_S^b$ and for the AFM case, using Eqs.(\ref{eq_SM:AFMTopBottom}) we have $\eta_S^t = - \eta^b_S$. Therefore, the PHV parameter in $S_{2D}^{PHV}$ is just twice for the parameter of the effective action of each surface for the FM case and $S_{2D}^{PHV} = 0$ for the AFM case, which is consistent with our discussion from the numerical calculations of the slab model in the main text. Next, we derive the equation of motion from $\frac{\partial S_{2D}}{\partial u_i} = 0$ where $i=x,y$ and $S_{2D} = S_{2D}^0 + S_{2D}^{PHV}$ with $S_{2D}^0$ given by
\begin{equation}
    S_{2D}^0 = \int dx dy dt\mathcal{L}_{2D}^0 = \int dx dy dt \Bigg[ \frac{1}{2} \left( \Dot{u}^2_{x} + \Dot{u}^2_{y}\right) - \frac{1}{2} \left(a+b\right) \left(u_{xx}^2 + u_{yy}^2 \right) - \left( a- b \right) u_{xx} u_{yy} + 2 b u_{xy}^2 \Bigg],
\end{equation}
for magnetic TI films. Here we only focus on in-plane strain tensor so we would expect two acoustic phonon modes in this effective theory.
We choose an ansatz
\begin{equation} \label{eq:surfansatz}
    \textbf{u}(k,t) = \textbf{f}_0 e^{i \textbf{k}\cdot \textbf{r} - i \omega t}, 
\end{equation}
where $\textbf{k} = (k_x,k_y)$ and $\textbf{r} = (x,y)$. Using the ansatz Eq.(\ref{eq:surfansatz}), the equation of motion is transformed to an eigen-problem
\begin{equation}
       H_{2D} \textbf{f}_0 = \omega^2 \textbf{f}_0, \quad H_{2D} = H_{2D}^0 + H_{2D}^{\text{e-ph}} ,
\end{equation}
where 
\begin{equation} \label{eq_SM:H2DPhonon}
    H_{2D}^0 = \begin{pmatrix}
        (a+b)k_x^2+b k_y^2 & a k_x k_y \\
        a k_x k_y & b k_x^2 + (a+b) k_y^2
    \end{pmatrix}, 
\end{equation}
and 
\begin{equation}\label{eq_SM:H2DPhonon2}
     H_{2D}^{\text{e-ph}} = \begin{pmatrix}
        0 &  2i \omega \eta_{S} k^2  \\
        -2 i \omega \eta_{S} k^2 & 0
    \end{pmatrix} ,
\end{equation}
where we have redefined $2 \eta^t_S \rightarrow \eta_S$ for the remainder of the SM.



\subsection{Phonon thermal Hall effect}  \label{sec:thermal Hall}
\begin{figure*}
\includegraphics[width=\textwidth]{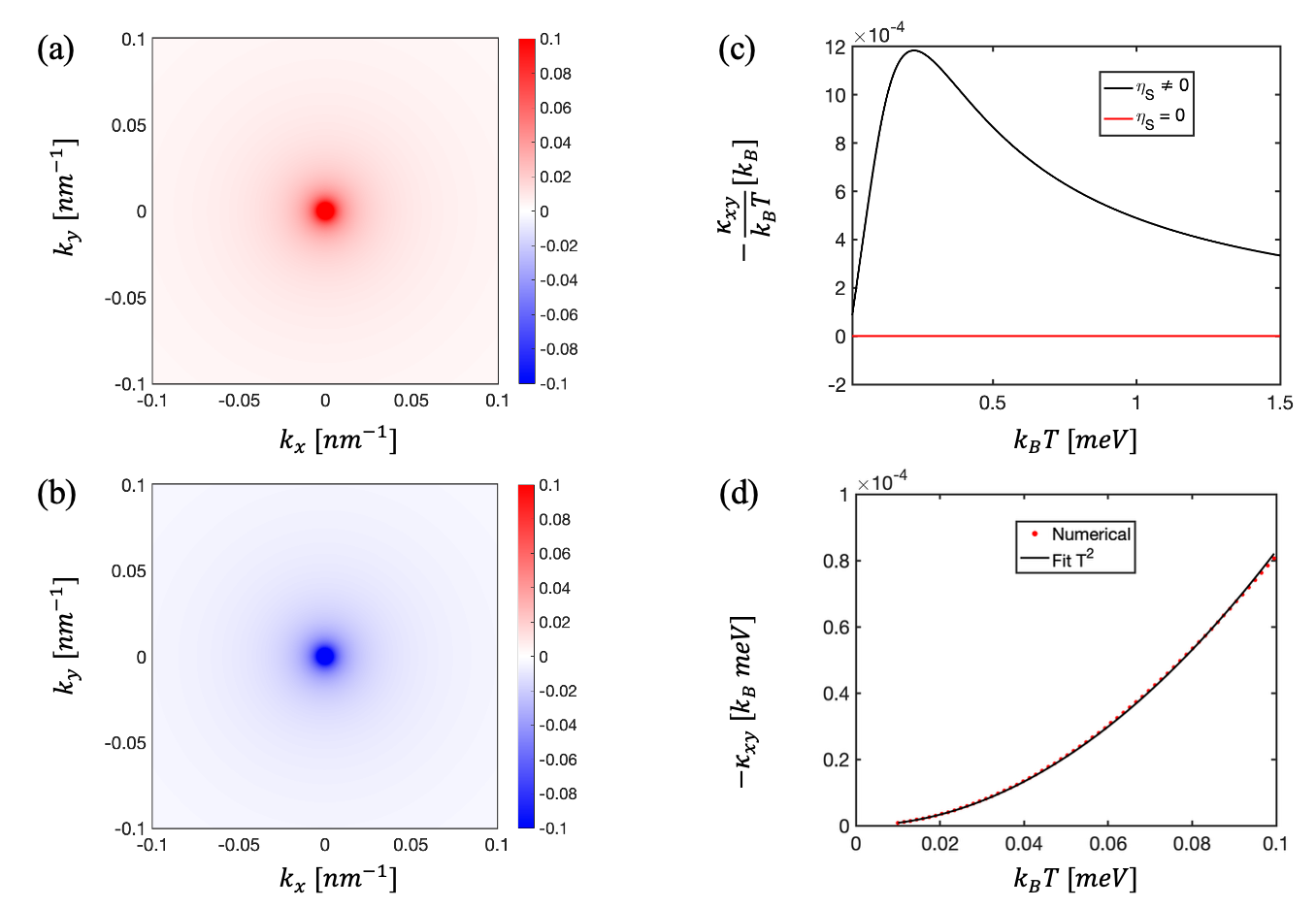}
 \caption{ (a) and (b) The phonon Berry curvature distribution of the two phonon bands in the $k_x-k_y$ plane. (c) The thermal Hall conductivity due to the two phonon bands (d) The thermal Hall conductivity due to the phonon modes fitted to $T^2$.
}
 \label{fig:PhononBC}
 \end{figure*}

In this section, we first discuss the formalism of the phonon thermal Hall effect and then evaluate the phonon thermal Hall conductivity for magnetic TI sandwiches in the FM configuration. We start by defining the Lagrangian density from the total action given in Eqs.(\ref{eq_SM:2Dphononaction},\ref{eq_SM:2Dphononaction2})
\begin{equation}
    S_{2D}^0 + S_{2D}^{PHV} = \int dt d^3r \mathcal{L}_{2D}. 
\end{equation}
We only consider the FM case as the PHV term, as well as the phonon thermal Hall effect, vanishes for the AFM case, as shown by our numerical calculations in the main text. We perform a Legendre transformation to go from the Lagrangian formalism to the Hamiltonian formalism,
\begin{equation}
    H_{2D} = p_i \dot{u}_i - \mathcal{L}_{2D}, \quad p_i = \frac{\partial \mathcal{L}_{2D}}{\partial \Dot{u}_i}.
\end{equation}
We obtain 
\begin{equation}
    p_i = \dot{u}_i - \mathcal{A}_i, \quad i = x,y
\end{equation}
where 
\begin{align}
    \mathcal{A}_x &= \eta_S  \Big[ -\partial_y 
    \left( u_{xx} - u_{yy} \right) + 2 \partial_x u_{xy}\Big] \nonumber \\
    \mathcal{A}_y &= \eta_S \Big[ -\partial_x \left( u_{xx} - u_{yy} \right)  - 2 \partial_y u_{xy}  \Big].
\end{align}
Therefore, the Hamiltonian operator is given by
\begin{align}
    H_{2D} &= \frac{1}{2}\dot{u}_i^2 + \frac{1}{2} u_{i} \left( H_{2D}^{1}\right)_{ij} u_j \nonumber \\
    &= \frac{1}{2} \left(p_i + \mathcal{A}_i \right)^2 + \frac{1}{2} u_{i} \left( H_{2D}^{1}\right)_{ij} u_j
\end{align}
where $i=x,y$ and
\begin{equation}
    H_{2D}^1 = \begin{pmatrix}
        (a+b)k_x^2+b k_y^2 & a k_x k_y \\
        a k_x k_y & b k_x^2 + (a+b) k_y^2
    \end{pmatrix}.
\end{equation}
Using Hamilton's equations of motion \begin{equation}
    \dot{u}_i = \frac{\partial H}{\partial p_i}, \quad \dot{p}_i = - \frac{\partial H}{\partial u_i},
\end{equation}
we have 
\begin{equation}\label{eq:H6}
   \mathcal{H} \psi = \omega \psi , \quad \mathcal{H} = i\begin{pmatrix}
         H_{\text{2D}}^1 & - I \\
         H_{2D}^1 - \left({H_{2D}^2}\right)^2 &  H_{2D}^{2}
    \end{pmatrix}, \quad \psi = (\textbf{u},\textbf{p})^T
\end{equation}
where
\begin{equation}
    H_{2D}^{2} = \begin{pmatrix}
        0 &  \eta_{S} k^2  \\
        - \eta_{S} k^2 & 0
    \end{pmatrix}. 
\end{equation}
Since $\mathcal{H}$ is non-Hermitian, we define the right and left eigenvectors $\psi $ and $\Bar{\psi}$, where $\Bar{\psi} \mathcal{H} = \Bar{\psi} \omega $. The formalism of phonon thermal Hall effect has been developed in Ref.\cite{qin2011energy,qin2012berry} and the phonon thermal Hall effect has been calculated in magnetic insulators in the presence of a PHV term \cite{ye2021phonon}, in which the phonon Hall viscosity presented in Ref. \cite{ye2021phonon} is the bulk PHV and arises from the electron-spin interaction. In our system, the PHV only exists at the surface and the bulk PHV vanishes due to the presence of the time reversal symmetry $\mathcal{T}$ in magnetic TI sandwiches. The Hamiltonian used in Ref. \cite{qin2012berry} is of the form
\begin{equation}
    \Tilde{\mathcal{H}} = i\begin{pmatrix}
        0 & I \\
        -  H_{2D}^{1} &  H_{2D}^{2}
    \end{pmatrix}
\end{equation}
and therefore the left and right eigenvalues are related by $\Bar{\psi} = \psi^\dagger H_{2D}^{1}$ which is important for calculating the phonon thermal Hall effect. For the Hamiltonian described in Eq(\ref{eq:H6}), $\Bar{\psi} = \psi^\dagger H_{2D}^{1}$ iff $[H_{2D}^1, H_{2D}^2] = 0 $, which is not generally true. But a unitary transformation $U$ relates $\mathcal{H}$ to $\mathcal{\Tilde{H}}$ as
\begin{equation} \label{eq_SM:Unitarytransform}
    \Tilde{\mathcal{H}} = U^{-1} \mathcal{H} U = i \begin{pmatrix}
        0 & I \\
        -H_{2D}^{1} & 2 H_{2D}^{2}
    \end{pmatrix}, \quad U = \begin{pmatrix}
        I & 0 \\
        H_{2D}^{2} & -I
    \end{pmatrix}. 
\end{equation}
Therefore, we can apply the formalism presented in Ref.\cite{qin2012berry} to $\Tilde{\mathcal{H}}$ in Eq.(\ref{eq_SM:Unitarytransform}) and calculate the phonon thermal Hall conductivity $\kappa_{xy}^{tr}$ as
\begin{equation}\label{eq:kappaxy}
   \kappa_{xy}^{tr} = - \frac{1}{T} \int d \epsilon \epsilon^2 \sigma_{xy} (\epsilon) \frac{d n (\epsilon)}{d \epsilon}  
\end{equation}
where 
\begin{equation} \label{eq: sigmaxyZph}
    \sigma_{xy} (\epsilon) = -\frac{1}{V \hbar} \sum_{\hbar \omega_{\textbf{k},n} \leq \epsilon} \Omega^z_{\textbf{k},n}. 
\end{equation}
$\Omega_{\textbf{k}n} = -\text{Im}[ \frac{\partial \Bar{\psi}_{\textbf{k}n}}{\partial \textbf{k}} \times \frac{\partial \psi_{\textbf{k}n}}{\partial \textbf{k}} ]$ is the phonon Berry curvature distribution of band $n$ and $n(\epsilon) = \frac{1}{e^{\beta \epsilon} -1}$ is the Bose distribution. In Eq.(\ref{eq:kappaxy}), we have dropped the topological term $\mathcal{Z}_{ph}$ considered in Ref.\cite{qin2012berry} because of the lack of a global phonon topology in our system i.e., $\mathcal{Z}_{ph} = 0$. In Fig.\ref{fig:PhononBC}(a) and (b), we calculate the Berry curvature for the two phonon bands and observe that the Berry curvature is peaked around $\Gamma$. Next, we plot the phonon thermal Hall conductivity $\kappa_{xy}$ due to both the phonon modes as a function of temperature $k_B T$ in Fig. \ref{fig:PhononBC}(c), where we find a peak at low temperatures, consistent with the numerical results of the phonon thermal Hall conductivity for a slab configuration in the main text. We also note that in the absence of the surface PHV ($\eta_{S}=0$), the Berry curvature distribution is negligible in the $k_x-k_y$ plane. Therefore, $\kappa_{xy}$ vanishes when the surface phonon Hall viscosity is zero as seen from the red line in Fig. \ref{fig:PhononBC}(c). By fitting the low temperature behavior of the phonon thermal Hall conductivity in Fig. \ref{fig:PhononBC}(c), we find that $\kappa_{xy} \sim T^2$. 

Next, we carry out a non-Hermitian perturbation theory calculation to derive an analytical form of the phonon Berry curvature and the phonon thermal Hall conductivity, from which we can show the temperature dependence of $\kappa_{xy}$ analytically. The zeroth order Hamiltonian becomes
\begin{equation}
    \mathcal{H}_0 = \begin{pmatrix}
        0 & I \\
        -H_{2D}^{1} & 0
    \end{pmatrix}
\end{equation}
and we consider the surface PHV term as a perturbation, i.e.
\begin{equation}
    \mathcal{V}_0 = \begin{pmatrix}
        0 & 0 \\
        0 & 2 H_{2D}^{2}
    \end{pmatrix}.
\end{equation}
We find no correction at the linear $\eta_{S}$ order to the phonon frequencies. To second order in $\eta_S$, the frequencies of the phonon bands are given by
\begin{align}
    \omega_{1}^{\pm} &= \pm \sqrt{a+b} k \pm 2 \frac{\sqrt{a+b}}{a} \eta_{S}^2 k^3  \nonumber \\
    \omega_{2}^{\pm} &= \pm \sqrt{b} k \mp 2 \frac{\sqrt{b}}{a} \eta_{S}^2 k^3 .
\end{align}
The corresponding phonon mode wavefunctions at the linear order in $\eta_{S}$ are
\begin{align}
& \psi_{1}^{+} =
\begin{bmatrix}
\dfrac{i k_x}{\sqrt{2}\sqrt{a+b}(k_x^2+k_y^2)} 
    - \dfrac{\sqrt{2}\eta_S k_y }{a\sqrt{k_x^2+k_y^2}} \\[0.8em]
\dfrac{i k_y}{\sqrt{2}\sqrt{a+b}(k_x^2+k_y^2)}
    + \dfrac{\sqrt{2}\eta_S k_x }{a\sqrt{k_x^2+k_y^2}} \\
\dfrac{k_x}{\sqrt{2}\sqrt{k_x^2+k_y^2}}
    + \dfrac{i\sqrt{2}\sqrt{a+b}\eta_S k_y }{a} \\[0.8em]
\dfrac{k_y}{\sqrt{2}\sqrt{k_x^2+k_y^2}}
    - \dfrac{i\sqrt{2}\sqrt{a+b}\eta_S k_x }{a}
\end{bmatrix},
\quad \quad \quad
\psi_{1}^- \;=\;
\begin{bmatrix}
-\dfrac{i k_x}{\sqrt{2}\sqrt{a+b}(k_x^2+k_y^2)}
    - \dfrac{\sqrt{2}\eta_S k_y }{a\sqrt{k_x^2+k_y^2}} \\
-\dfrac{i k_y}{\sqrt{2}\sqrt{a+b}(k_x^2+k_y^2)}
    + \dfrac{\sqrt{2}\eta_S k_x }{a\sqrt{k_x^2+k_y^2}} \\
\dfrac{k_x}{\sqrt{2}\sqrt{k_x^2+k_y^2}}
    - \dfrac{i\sqrt{2}\sqrt{a+b}\eta_S k_y }{a} \\[0.8em]
\dfrac{k_y}{\sqrt{2}\sqrt{k_x^2+k_y^2}}
    + \dfrac{i\sqrt{2}\sqrt{a+b}\eta_S k_x }{a}
\end{bmatrix},
\\ 
& \psi_{2}^+ \;=\;
\begin{bmatrix}
-\dfrac{i k_y}{\sqrt{2}\sqrt{k_x^2+k_y^2}\sqrt{b(k_x^2+k_y^2)}}
    + \dfrac{\sqrt{2}\eta_S k_x }{a\sqrt{k_x^2+k_y^2}} \\
\dfrac{i k_x}{\sqrt{2}\sqrt{k_x^2+k_y^2}\sqrt{b(k_x^2+k_y^2)}}
    + \dfrac{\sqrt{2}\eta_S k_y }{a\sqrt{k_x^2+k_y^2}} \\
-\dfrac{k_y}{\sqrt{2}\sqrt{k_x^2+k_y^2}}
    - \dfrac{i\sqrt{2}\eta_S k_x\sqrt{b(k_x^2+k_y^2)} }{a\sqrt{k_x^2+k_y^2}} \\
\dfrac{k_x}{\sqrt{2}\sqrt{k_x^2+k_y^2}}
    - \dfrac{i\sqrt{2}\eta_S k_y\sqrt{b(k_x^2+k_y^2)} }{a\sqrt{k_x^2+k_y^2}}
\end{bmatrix},
\quad \quad \quad
\psi_{2}^- \;=\;
\begin{bmatrix}
\dfrac{i k_y}{\sqrt{2}\sqrt{k_x^2+k_y^2}\sqrt{b(k_x^2+k_y^2)}}
    + \dfrac{\sqrt{2}\eta_S k_x }{a\sqrt{k_x^2+k_y^2}} \\
-\dfrac{i k_x}{\sqrt{2}\sqrt{k_x^2+k_y^2}\sqrt{b(k_x^2+k_y^2)}}
    + \dfrac{\sqrt{2}\eta_S k_y }{a\sqrt{k_x^2+k_y^2}} \\
-\dfrac{k_y}{\sqrt{2}\sqrt{k_x^2+k_y^2}}
    + \dfrac{i\sqrt{2}\eta_S k_x\sqrt{b(k_x^2+k_y^2)} }{a\sqrt{k_x^2+k_y^2}} \\
\dfrac{k_x}{\sqrt{2}\sqrt{k_x^2+k_y^2}}
    + \dfrac{i\sqrt{2}\eta_S k_y\sqrt{b(k_x^2+k_y^2)} }{a\sqrt{k_x^2+k_y^2}}
\end{bmatrix}
\end{align}

From the phonon wavefunctions, we can calculate the phonon Berry curvature distribution in momentum space as 
\begin{align}
    \Omega_{1}^{\pm}(k) &= \pm \eta_{S} \frac{3a + 4b}{a \sqrt{a+b}} \frac{1}{k}\nonumber \\
    \Omega_{2}^{\pm}(k) &= \mp \eta_{S} \frac{a + 4b}{a \sqrt{b}} \frac{1}{k},  
\end{align}
and $\sigma_{xy}(\epsilon)$ is given by
\begin{equation}
    \sigma_{xy} (\epsilon) = -  \sum_{i=1,2} \int \frac{d^2 q}{(2 \pi)^2} \Omega_{i} (q) \Theta(\epsilon - \omega_{q,i}) .
\end{equation}
At low temperatures, $\omega_{i} \approx v_i q$, where $v_1 = \sqrt{a+b}, v_2 = \sqrt{b}$. Therefore, $\sigma_{xy}$ for the FM case becomes 
\begin{align}\label{eq_SM:phonon_Berry_curvature}
    \sigma_{xy}(\epsilon) &= - \sum_{i=1,2} \int \frac{d q}{2 \pi} q \Omega_{i}(q) \Theta(\epsilon - v_i q) \nonumber \\ 
    &= - \sum_{i=1,2} \int_0^{\epsilon/v_i} \frac{d q}{2 \pi} q \Omega_{i}(q)\nonumber \\
   &= -\frac{\eta_S}{ 2\pi} \frac{1}{a^2 (a+b)} \epsilon \left( \frac{a+4b}{b} - \frac{3a+4b}{a+b} \right) \nonumber \\
    &= -\frac{\eta_S}{2 \pi} \frac{a+2b}{a(a+b)} \epsilon .
\end{align}
Since, $\Omega_i(q) \sim 1/q$, the integrand $q \Omega_i(q)$ in $\sigma_{xy}(\epsilon)$ is independent of $q$ and the $q$-integral leads to $\sigma_{xy}(\epsilon) \sim \epsilon$. 
The thermal Hall conductivity becomes
\begin{align}
    \kappa_{xy}^{tr} &= - \frac{1}{T} \int d \epsilon \epsilon^2 \sigma_{xy}(\epsilon) \frac{dn}{d \epsilon} \nonumber \\
    &=  -\frac{\eta_S}{ 2 \pi } \frac{a+2b}{a(a+b)} \frac{1}{T^2} \int d \epsilon \epsilon^3 \frac{e^{\epsilon/k_B T}}{(e^{\epsilon/k_B T}-1)^2} \nonumber \\
    &= -\frac{\eta_S}{ 2 \pi } \frac{a+2b}{a(a+b)} k_B^3 T^2 \int d x x^3 \frac{e^{x}}{(e^{x}-1)^2}, \quad x = \epsilon/k_B T \nonumber \\
    &\propto \eta_{S} T^2
\end{align}
which is exactly the $T^2$ behavior that we find numerically in Fig. \ref{fig:PhononBC}(d). We note that this behavior originates from $\sigma_{xy}(\epsilon) \propto \epsilon$ in Eq.(\ref{eq_SM:phonon_Berry_curvature}). 


\subsection{Magnon polarons in magnetic TI films} \label{sec: surfacemagnonpolarons}

In this section, we include non-zero magnonic excitations $m^a_{x,y} \neq 0$ in Eq.(\ref{eq:surfDiracH},\ref{eq:surfgaugefield}). We assume that the magnons on the top and bottom surface are decoupled in the magnetic TI sandwiches, and only consider the phonon-magnon interaction term at each surface and neglect the interaction of the surface magnon with the bulk phonons. It should be noted that the discussion below assumes the decoupling between the magnetization at the top and bottom surfaces, which cannot be applied to MnBi$_2$Te$_4$. The phonon-magnon coupling in MnBi$_2$Te$_4$ is beyond the scope of this work. 

In magnetic TI sandwiches, the magnon Lagrangian includes the free magnon contribution and acoustic phonon-magnon interaction, given by
\begin{align}
    \mathcal{L}_{2D,a}^m =  \mathcal{L}_{2D,a}^{m_0} + \mathcal{L}_{2D,a}^{m-ph}
\end{align}
where the free magnon Lagrangian $\mathcal{L}_{2D,a}^{m_0}$ is given by \cite{herring1951theory,kittel1958interaction}
\begin{align}\label{eq_SM:Lagrangianm0}
    \mathcal{L}^{m_0}_{2D,a} = \frac{m^a_x \dot{m}^a_y}{\omega_s} - \frac{\gamma_A}{2} \left( (\nabla m^a_x)^2 + (\nabla m^a_y)^2  \right) - \frac{\gamma_K}{2} \left( (m_x^a)^2 + (m_y^a)^2 \right) 
\end{align}
with $\gamma_A,\gamma_K$ as the material parameters and the magnetization axis is along the $z$-axis. The phonon-magnon interaction term is given by
\begin{align}\label{eq_SM:lagrangianmph}
    \mathcal{L}_{2D,a}^{m-ph} = &\zeta^a_1\Big[ \left(  u^a_{xx} - u^a_{yy} \right) \dot{m}^a_x - m^a_x \left( \dot{u}^a_{xx} - \dot{u}^a_{yy} \right) + m^a_y 2 \dot{u}^a_{xy} - 2 u^a_{xy} \dot{m}_y^a + m^a_y \dot{m}^a_x - m_x^a \dot{m}^a_y \Big] \nonumber \\
    & + 2 \zeta^a_2 \Big[ \left( u^a_{xx} + u^a_{yy} \right) \left( \partial_x m^a_x + \partial_y m^a_y\right) \Big] ,
\end{align}
where 
\begin{equation}\label{eq_SM:zetatopbottom}
    \zeta^a_1 = \frac{\text{sgn}(m^a_z) g^a \Tilde{B}^a_1 }{8 \pi (v_f^a)^2}, \quad \zeta^a_2 = \frac{\text{sgn}(m^a_z) g^a \Tilde{C}^a_0}{8 \pi v_f^a}.
\end{equation} 
We note that 
since $\zeta^a_1 \left( m_y \dot{m}_x - m_x \dot{m}_y \right)$ only provides a correction to $\omega_s$, we can absorb this term by redefining $\omega_s$. From Eq(\ref{eq_SM:zetatopbottom}) and using the relations in Eq.(\ref{eq_SM:FMTopBottom}), we determine $\zeta_1^t = \zeta_1^b, \zeta_2^t = - \zeta_2^b$ for the FM configuration and using Eq.(\ref{eq_SM:AFMTopBottom}) we get $\zeta_1^t = \zeta_1^b, \zeta_2^t = \zeta_2^b$ for the AFM configuration. 

In our slab model of the main text, we include the magnon modes described by Eqs.(\ref{eq_SM:Lagrangianm0},\ref{eq_SM:lagrangianmph}) as interaction terms on the top and bottom surfaces of the magnetic TI slab configuration and calculate the magnon-polaron dispersion and the magnon-polaron thermal Hall in Fig. 4 of the main text.
